\documentclass[preprint,11pt]{imsart}

\pdfoutput=1

\RequirePackage[OT1]{fontenc}
\RequirePackage{amsthm,amsmath}
\RequirePackage[numbers]{natbib}
\RequirePackage[colorlinks=true, citecolor=blue, filecolor=black,  urlcolor=blue]{hyperref}
\usepackage{hyperref}

\usepackage[utf8]{inputenc} %
\usepackage[T1]{fontenc}    %

\usepackage{amsfonts}
\usepackage{amssymb}
\usepackage{bbm}
\usepackage{bm}
\usepackage{nicefrac} %
\usepackage{url}

\usepackage{color}
\usepackage{enumerate}
\usepackage{enumitem}
\usepackage{graphicx}
\usepackage{mathrsfs}
\usepackage{mathtools}
\usepackage[textsize=tiny]{todonotes}
\usepackage{ifpdf}
\usepackage{wrapfig}
\usepackage{float}
\usepackage{subfig} 
\usepackage{bbm}
\usepackage[letterpaper,centering,top=1.5in,bottom=1.5in]{geometry}

\usepackage{algorithm}
\newcommand{\spalg}{1.2}

\usepackage{caption}
\usepackage[noend]{algpseudocode}

\usepackage{varwidth}
\usepackage{dsfont}
\usepackage{multirow}

\usepackage{tikz}
\usetikzlibrary{shapes,arrows,decorations.pathmorphing,backgrounds,positioning,fit,petri}
\usetikzlibrary{calc}
\usetikzlibrary{matrix}
\usetikzlibrary{backgrounds}
\usepackage{pgfplots}
\pgfplotsset{compat=newest}
\usepgfplotslibrary{groupplots}
\tikzset{
	default/.style={circle,draw=blue!50,fill=blue!20,thick,
		inner sep=0pt,minimum size=6mm},
	defaultOrange/.style={circle,draw=orange!50,fill=orange!20,thick,
		inner sep=0pt,minimum size=6mm},		
	defaultGreen/.style={circle,draw=ForestGreen!50,fill=ForestGreen!20,thick,
		inner sep=0pt,minimum size=6mm},
	defaultGray/.style={circle,draw=black,fill=Gray!10,thick,
		inner sep=0pt,minimum size=6mm},			
	defaultCyan/.style={circle,draw=cyan!50,fill=cyan!20,thick,
		inner sep=0pt,minimum size=6mm},				
	defaultBlank/.style={circle,draw=white,fill=white,thick,
		inner sep=0pt,minimum size=6mm},	
	paramSetNodes/.style={circle,draw=blue!50,fill=blue!20,thick,
		inner sep=0pt,minimum size=11mm},	
	dataNode/.style={circle,draw=red!50,fill=red!20,thick,
		inner sep=0pt,minimum size=6mm},		
	dataSetNodes/.style={circle,draw=red!50,fill=red!20,thick,
		inner sep=0pt,minimum size=11mm},	
	default2/.style={circle,draw=black,fill=blue!20,thick,
		inner sep=0pt,minimum size=6mm}, 
	edgeStyle/.style={ semithick},
	entriesMatrixZero/.style={rectangle,draw=white,fill=white, 
		inner sep=0pt,minimum size=3mm}, 
	entriesMatrixZeroDashed/.style={rectangle,draw=black,fill=white, dashed,
		inner sep=0pt,minimum size=3mm},
	entriesMatrix/.style={rectangle,draw=black,fill=blue!60!white, 
		inner sep=0pt,minimum size=3mm},
	entriesMatrixRed/.style={rectangle,draw=black,fill=red!60!white, 
		inner sep=0pt,minimum size=3mm},	
	entriesMatrixOrange/.style={rectangle,draw=black,fill=orange!60!white, 
		inner sep=0pt,minimum size=3mm},				
        shadedNode/.style={circle,very thick,draw=orange!50,fill=orange!20,inner sep=0pt,minimum size=6mm},	background/.style={ draw, fill=yellow!30, align=right}												
}

\newcommand{\colorA}{red!40}  
\newcommand{\colorB}{yellow!50} 
\newcommand{\colorS}{orange!40}

\pgfdeclarelayer{background}
\pgfdeclarelayer{bBack}
\pgfsetlayers{bBack,background,main}
\usetikzlibrary{fit}

\newcommand{\Ex}{\mathbb{E}}       
\newcommand{\mapSi}{S_{\bf Id}}
\newcommand{\mapSih}{\widehat{S}_{\bf Id}}

\newcommand{\Var}{ {\mathbb{V}\mathrm{ar}}}   %
\newcommand{\Cov}{ {\mathbb{C}\mathrm{ov}}}   %
\newcommand{\Gauss}{\mathcal{N}}                                      %

\DeclareMathOperator\erf{erf}
\newcommand{\push}{_\sharp}                                      %
\newcommand{\orth}{ \perp\!\!\!\perp }  %

\newcommand{\re}{\mathbb{R}}
\newcommand{\ra}{\rightarrow}

\newcommand{\Bc}{\mathcal{B}}

\newcommand{\Vc}{\mathcal{V}}

\newcommand{\Ec}{\mathcal{E}}

\newcommand{\Ac}{\mathcal{A}}
\newcommand{\Sc}{\mathcal{S}}

\newcommand{\Tc}{\mathcal{T}}
\newcommand{\Mc}{\mathcal{M}}

\newcommand{\Xib}{\boldsymbol{\Xi}}
\newcommand{\Wcb}{\boldsymbol{\mathcal{W}}}
\newcommand{\Xb}{\boldsymbol{X}}
\newcommand{\xb}{\boldsymbol{x}}

\newcommand{\Yb}{\boldsymbol{Y}}
\newcommand{\yb}{\boldsymbol{y}}

\newcommand{\Zb}{\boldsymbol{Z}}
\newcommand{\zb}{\boldsymbol{z}}

\newcommand{\Hb}{\boldsymbol{H}}

\newcommand{\Rb}{\boldsymbol{R}}
\newcommand{\Sb}{\boldsymbol{S}}

\newcommand{\Ab}{\boldsymbol{A}}

\newcommand{\db}{\boldsymbol{d}}

\newcommand{\Lb}{\boldsymbol{L}}
\newcommand{\Sigmab}{\boldsymbol{\Sigma}}
\newcommand{\sigmab}{\boldsymbol{\sigma}}

\newcommand{\xib}{\boldsymbol{\xi}}

\newcommand{\Hbrm}{\boldsymbol{\mathrm{H}}}

\newcommand{\Gcb}{\boldsymbol{\mathcal{G}}}

\newcommand{\Hcb}{\boldsymbol{\mathcal{H}}}
\newcommand{\Ecb}{\boldsymbol{\mathcal{E}}}
\newcommand{\Ycb}{\boldsymbol{\mathcal{Y}}}
\newcommand{\Xcb}{\boldsymbol{\mathcal{X}}}

\newcommand{\mfrak}{\mathfrak{m}}

\newcommand{\ufrak}{\mathfrak{u}}
\newcommand{\ufrakb}{\boldsymbol{\ufrak}}

\newcommand{\neigh}{ {\rm Nb} }

\newcommand{\Aset}{ \Ac }
\newcommand{\Bset}{ \Bc }
\newcommand{\Sset}{ \Sc }

\newcommand{\Dkl}{\mathcal{D}_{\rm KL}}         %

\newcommand{\pull}{^\sharp}

\ifpdf
  \usepackage{epstopdf} 
\fi
\PassOptionsToPackage{numbers}{natbib}

\startlocaldefs
\numberwithin{equation}{section}

\newtheorem{exmp}{Example} %
\newtheorem{remark}{Remark} %

\begin{document}

\begin{frontmatter}

\title{Coupling techniques for nonlinear ensemble filtering}

\runtitle{Couplings for nonlinear ensemble filtering}

\begin{aug}
\author{\fnms{Alessio} \snm{Spantini}%
\ead[label=e1]{spantini@mit.edu}},
\author{\fnms{Ricardo} \snm{Baptista}%
\ead[label=e2]{rsb@mit.edu}},
\and
\author{\fnms{Youssef} \snm{Marzouk}%
\ead[label=e3]{ymarz@mit.edu}
\ead[label=u1,url]{http://uqgroup.mit.edu}
}

\runauthor{Spantini, Baptista, Marzouk}

\address{Massachusetts Institute of Technology\\
Cambridge, MA 02139 USA\\
\printead{e1},
\printead*{e2},
\printead*{e3} 
}
\end{aug}

\begin{abstract}
We consider filtering in high-dimensional non-Gaussian state-space
models with intractable transition kernels, nonlinear and possibly
chaotic dynamics, and {sparse} observations in space and time.  We
propose a novel filtering methodology that harnesses
 transportation of measures, convex optimization, and ideas from probabilistic graphical
models to yield robust ensemble approximations of the filtering
distribution in high dimensions. Our approach can be understood as the
natural generalization of the ensemble Kalman filter (EnKF) to
\emph{nonlinear} updates, using stochastic or deterministic
couplings.  The use of nonlinear updates can reduce the intrinsic bias
of the EnKF at a marginal increase in computational cost.  We avoid
any form of importance sampling and %
introduce %
non-Gaussian
localization approaches for dimension scalability.  Our framework
achieves state-of-the-art tracking performance on challenging
configurations of the Lorenz-96 model in the chaotic regime.
\end{abstract}


\end{frontmatter}

\setcounter{tocdepth}{2}
\tableofcontents

\section{Introduction}

State-space models formalize the probabilistic description of a
time dependent latent process---the state---observed indirectly at
discrete times \cite{durbin2012time}. These models can approximate a
wide variety of stochastic processes, ranging from the evolution of
atmospheric variables in meteorology to the volatility of financial
assets.  This paper is concerned with the problem of (discrete-time)
{\it filtering}, i.e., characterizing the sequence of conditional
distributions of the latent field at observation times, given all
available measurements up to that time.  Filtering arises in virtually
every discipline that seeks an online integration of models
with data, e.g., imaging, pharmacology, atmospheric sciences, and
oceanography. In geophysical applications, filtering and several
closely related inference problems (e.g., smoothing, sequential
parameter inference) fall under the broad label of \emph{data
assimilation}.

Despite its importance and ubiquity, filtering remains a challenging
task, particularly when (i) the state is high dimensional; (ii) the
state transition dynamic is nonlinear, expensive to simulate, and
intractable, e.g., if it involves the integration of a chaotic partial
differential equation; and (iii) observations of the state are
{sparse}, both in space and time
\cite{majda2012filtering,reich2015probabilistic}.  These constraints
reflect typical challenges faced in numerical weather prediction or
geophysical data assimilation, and pose severe challenges for {\it
consistent} sequential Monte Carlo (SMC) algorithms, which invariably
face particle degeneracy or impoverishment
\cite{snyder2008obstacles,doucet2009tutorial}.
In such scenarios, state-of-the-art results\footnote{At least in terms
of time-averaged errors %
in {point estimates} of the state 
\cite{law2012evaluating}.
}  are typically
obtained with the ensemble Kalman filter (EnKF), which is the
workhorse of modern ensemble-based data assimilation
\cite{evensen2007data}.
The EnKF implements a two-step Monte Carlo approximation of the
classical Kalman recursions.  In the {\it forecast} step, a particle
approximation of the filtering distribution is propagated
through the transition dynamic to yield a ``forecast'' ensemble at the
next observation time.  In the {\it analysis} step, the forecast
ensemble is updated via the action of a {\it linear} 
transformation to yield an empirical approximation of the new
filtering marginal.  The linear transformation is estimated under
Gaussian assumptions.
Hence, the EnKF cannot yield consistent estimators of the filtering
distribution for non-Gaussian models \cite{mandel2011convergence}.
In essence, the EnKF trades consistency of the estimators for lower
variance, and thus robustness in high dimensions. Yet the intrinsic
bias of the EnKF---due both to the linear transformation and the way
it is estimated---implies that increasing the ensemble size beyond a
certain threshold does {\it not} improve accuracy; 
increasing computational effort does not yield better inference.  
We wish to address these limitations.

In this work, we introduce \emph{non-Gaussian generalizations of the EnKF} by
considering {nonlinear} transformations derived from %
 couplings.
Couplings provide a link between distributions.  That is, for a pair
of distributions $(\pi_1,\pi_2)$, a coupling is defined by a pair of
random variables 
$(\Xb_1, \Xb_2)$, which admit $\pi_1$ and $\pi_2$ as
marginal distributions \cite{villani2008optimal}. 
We consider couplings that are induced by a continuous transformation
$T$ such that $\Xb_2 = T(\Xb_1)$.  The transformation can be either
deterministic or stochastic.  When $T$ is deterministic, we call it a
{\it transport map} {and say that the coupling is deterministic (see~\cite[Definition 1.2]{villani2008optimal})}.  The transformation allows one to sample $\pi_2$
by evaluating $T$ at samples from $\pi_1$.

We interpret the analysis step of the EnKF as a problem of coupling
the forecast distribution $\pi_{\Xb}$ with the filtering
distribution $\pi_{\Xb \vert \Yb}$.  That is, given samples
$\xb^1,\ldots,\xb^M$ from $\pi_{\Xb}$ and a likelihood function
$\pi_{\Yb \vert \Xb}$, we seek a transformation $T$ that yields
samples $\smash{T(\xb^1),\ldots,T(\xb^M)}$ from $\pi_{\Xb \vert \Yb}$.
Couplings are not unique.  We are particularly interested in
transformations that can be estimated efficiently without resorting to
importance sampling---perhaps using only convex optimization---and
that are easy to ``localize'' in high dimensions.  We want to avoid
the use of weights and thus issues of particle degeneracy.  Moreover,
the computation of the transformation should {\it not} become
increasingly challenging as the variance of the observation noise
decreases.  The latter is a typical concern of filtering algorithms
that rely on some attributes---e.g., moments---of the bootstrap particle
filter approximation.  \emph{This paper proposes two new algorithms}, called the
stochastic and deterministic map filters.  The former represents a
non-Gaussian generalization of the EnKF with ``perturbed
observations'' \cite{burgers1998analysis}, while the latter is a
nonlinear extension of the square-root EnKF
\cite{tippett2003ensemble}.

The \emph{stochastic map filter} seeks a non-deterministic
transformation from forecast to filtering distribution.  This
transformation is given by a transport map that pushes forward the
joint distribution of state \emph{and} data at a given observation time,
$\pi_{\Xb, \Yb}$, to the filtering distribution.  We target a
specific transport map derived from the Knothe--Rosenblatt (KR)
rearrangement that pushes forward $\pi_{\Xb, \Yb}$ to a ``reference''
distribution with independent components, e.g., a standard normal.
The KR rearrangement is the unique monotone triangular transport map
that defines a deterministic coupling between two distributions
\cite{bogachev2005triangular},
and can be estimated within a finite-dimensional function space using
only {\it convex} optimization and samples from $\smash{\pi_{\Xb,
\Yb}}$ \cite{parno2014transport}.  These samples are easy to obtain
given a forecast ensemble $(\xb^i)_{i \leq M}$: it suffices to simulate the
likelihood $\pi_{\Yb \vert \Xb}$ at each particle $\smash{\xb^i}$. If
we restrict the estimator of the rearrangement to be linear, then we
recover the stochastic EnKF \cite{evensen2007data}.  If we consider
nonlinear parameterizations of the estimator, however, we obtain a
whole new class of nonlinear filtering algorithms, which rely on fast
and robust convex optimization.

In the \emph{deterministic map filter}, we adopt a complementary
strategy and seek a deterministic transformation from the 
forecast to
the 
filtering distribution.  We view the transformation as a composition
of two functions.  The first function is a KR rearrangement that
pushes forward the forecast $\pi_{\Xb}$ to a reference distribution.
This rearrangement can be estimated via convex optimization given only
forecast samples, and yields an implicit approximation of the forecast
density, $\widehat{\pi}_{\Xb}$.  The second function is a KR
rearrangement that pushes forward the reference to an approximation of
the filtering distribution given by $\widehat{\pi}_{\Xb \vert \Yb}
\propto \pi_{\Yb \vert \Xb} \, \widehat{\pi}_{\Xb}$, which can be
evaluated in closed form up to a normalizing constant.  The estimation
of the latter rearrangement leads to a variational problem, first
proposed by \cite{el2012bayesian}, that is in general non-convex.
This is an important difference with respect to the stochastic map
filter, which relies entirely on convex optimization, at the cost of
estimating higher-dimensional rearrangements.  Yet, in the special
case of local observations that are conditionally independent, we show
how to implement the deterministic map filter using {\it only} convex
optimization (Appendix \ref{sec:detmap_local}).  The resulting scheme
has a flavor similar to the multivariate rank histogram filter \cite{metref2014non}.  For a
linear parameterization of the estimators {\it and} Gaussian
likelihoods, the deterministic map filter reduces to a (deterministic)
square-root EnKF \cite{tippett2003ensemble}; otherwise, it is far more
general.

In principle, both the stochastic and 
deterministic map filters 
can
approximate a nonlinear update 
of arbitrary complexity:
it suffices to enrich the function 
space for the estimator of the KR
rearrangements.
In practice, however, there is a tradeoff.
Depending on the 
number of forecast particles $M$,
overly complex parameterizations 
can lead to estimators with low bias but
unacceptable variance, and
vice-versa.
Skillful filtering requires a careful
balance between bias and variance.
A key ingredient
of the
functional framework proposed in
this paper
is the ability to 
depart {\it gradually} from the linear ansatz.
For example, 
in our numerical experiments we
consider maps whose components are additively separable, as sums of univariate nonlinear functions.
These parameterizations 
represent a 
natural extension of linear functions
in terms of complexity,
and lead to filtering algorithms that can outperform the EnKF
at a marginal increase in computational cost
and
ensemble requirements (Section \ref{sec:numerics}).
Of course, more
general parameterizations are also 
possible.
In essence, we provide
a natural tradeoff between 
computational cost and statistical accuracy---a tradeoff which is very limited in the EnKF, due to the
restrictive linear ansatz.

In high-dimensional
problems with limited ensemble sizes, we must further
regularize the estimation of the
rearrangements.
In this paper, we introduce a notion of
``localization'' for nonlinear updates, 
by dropping some variable dependence from each component of the transport map.
We show how {\it sparsity} of the KR
rearrangement---a nonlinear function---is linked to %
properties
of the filtering distribution, such as %
decay of correlation and (approximate)
conditional independence \cite{spantini2017inference}.
Intuitively, these sparse transformations 
 approximate 
the projection of the filtering 
distribution 
onto a manifold of %
 non-Gaussian Markov random fields.

 \subsection{Related work} The idea of transporting measures via couplings has
a long history \cite{monge1781memoire,villani2008optimal},
and has found applications in fluid dynamics, economics, statistics, 
machine learning \cite{douglas1999applications,kantorovich1965best,el2012bayesian,goodfellow2014generative}, and many other fields.
One of the %
first filtering algorithms to rely on the %
explicit construction of (optimal) couplings is 
the ensemble transform particle filter \cite{reich2013nonparametric,acevedo2017second}, which
uses an
importance sampling
approximation of the filtering distribution---the same given by the 
bootstrap particle filter \cite{gordon1993novel}---to
obtain %
a consistent {particle} approximation of Monge's optimal 
map \cite{villani2008optimal} %
(cf.\ Section \ref{sec:analysis}). 
Earlier work on particle flows 
investigated the use of transport maps
induced by flows of ODEs \cite{daum2008particle}. %
Related ideas include the feedback particle filter 
\cite{yang2013feedback}
and the ``Gibbs flow'' \cite{heng2015gibbs}.
In particular, \cite{heng2015gibbs} defines an approximation
of the KR rearrangement between an input and a
target measure via the solution of
an ODE whose drift term depends only on the full conditionals of the 
target distribution.
This approximation can be used as a proposal density for sequential Monte Carlo (SMC) methods.
Also, the implicit sampling algorithm of \cite{chorin2009implicit,morzfeld2012random} 
defines a proposal density using %
an %
approximate
transport map, which can be evaluated
by 
solving a one-dimensional optimization problem: in this case,
optimization is used to define the action of a single map,
not to search for the best rearrangement within a class of candidate maps 
(see %
\cite[Sec.\ 6]{marzouk2016introduction} for more details).
Among the flow methods,
the variational mapping particle filter implements a particle
flow given by 
a %
particular form of 
functional gradient descent on a %
reproducing kernel Hilbert space \cite{pulido2018kernel,liu2016stein}.
We clarify its connection with the deterministic map
filter 
in Section \ref{sec:square-root}.

In our filters,
we use a variational approximation of the KR rearrangement to Gaussianize
a collection of particles.
The idea of Gaussianizing multivariate data is
well rooted in 
statistics \cite{box1964analysis}, and
is known as Gaussian anamorphosis (GA)
in the geostatistics literature \cite{wackernagel1996multivariate}.
Usually, these approaches Gaussianize %
marginals of the data
by  estimating one-dimensional cumulative distribution functions.
A typical application of GA within the context of non-Gaussian extensions of the
EnKF is described in \cite{amezcua2014gaussian}:
the idea is to (1) estimate a nonlinear transformation that
Gaussianizes joint samples from the distribution of state {\it and} data
at a given observation %
time, (2) derive an equivalent nonlinear
observation operator in the transformed space, (3) apply EnKF formulas
for nonlinear observation operators in the transformed space, and (4) map the analysis ensemble
back to the original space.
As explained in \cite{amezcua2014gaussian}, this approach cannot
recover the exact filtering distribution for non-Gaussian models, even if we
had the {\it exact} nonlinear transformation.
The problem is that we insist on using EnKF formulas in the transformed space.
In the stochastic map filter introduced %
in this paper, on the other hand, we show how to bypass this
issue by designing a nonlinear update that depends only on the
nonlinear transformation that Gaussianizes the data.
If we had the exact  transformation, then the stochastic map
filter would be exact. 

Yet another measure transport approach to filtering
is that of \cite{spantini2017inference}, which 
uses the idea of decomposable couplings to
yield 
a %
recursive 
approximation of the full Bayesian solution to the sequential
inference
problem, 
using purely variational techniques (i.e., no sampling). {See also~\cite{houssineau2018multilevel} for a multilevel generalization of this variational method.}
Outside the coupling literature,
an early attempt to devise non-Gaussian generalizations of the
EnKF %
is the
Gaussian mixture filter of
\cite{bengtsson2003toward}. 
Here, the idea is to learn the forecast density via a
regularized mixture of
Gaussians.
{Additional efforts at generalizing the EnKF to nonlinear updates include the hierarchical methods of~\cite{katzfuss2020ensemble}, which complement the EnKF with a consistent Bayesian method, e.g., a particle filter, in a low-dimensional subspace of the variables of interest. The stochastic and deterministic map filters introduced in this paper %
could also be adapted to this construction, after identifying a suitable decomposition of the variables. 
\cite{del2017stability} analyzes stability properties of a class of (nonlinear) EnKFs that converge to the extended Kalman--Bucy filter as the number of particles goes to infinity. }
Finally, we note that complementary approaches to %
non-Gaussian high dimensional
filtering include hybrid filters  \cite{frei2013bridging}
and local weights
particle filters \cite{poterjoy2016localized}.
These algorithms 
provide an alternative route to the regularization of 
 SMC methods, which %
are
usually
consistent but %
plagued by %
large estimation variance in high dimensions. 
See \cite{van2018particle} for a recent review.

\subsection{Organization of the paper} The rest of the paper is organized as 
follows.
In Section \ref{sec:state-space}, we review
the notion of state-space models.
In Section \ref{sec:gen_enkf}, we describe
the
EnKF and discuss %
possible
non-Gaussian generalizations.
In Section \ref{sec:analysis}, we interpret the analysis step
of the EnKF as a problem of coupling the forecast distribution
with the filtering distribution, and 
analyze 
the difference between stochastic and
deterministic couplings.
In Section \ref{sec:KR}, we 
recall properties of the KR 
rearrangement---a %
key coupling for our analysis.
In Section \ref{sec:pert_obs}, we introduce
the %
stochastic map filter.
In Section \ref{sec:est_KR}, we
address estimation of the %
KR rearrangement via convex optimization, while in
Section \ref{sec:reg_map} we
introduce regularization ideas 
for high dimensions.
Specific parameterizations of the
KR rearrangement are discussed in Appendix \ref{sec:param_tri}.
In Section \ref{sec:rem_pert_obs},
we collect a few remarks on the
stochastic map filter, including the
connection with the stochastic EnKF and
additional localization ideas.
In Section
\ref{sec:square-root}, 
we introduce the deterministic map
filter.
We report on our numerical experiments with chaotic dynamical systems in 
Section \ref{sec:numerics}, evaluating both state estimation error and
fidelity to the underlying Bayesian solution, among other performance metrics.
Section \ref{sec:discussion}
discusses open issues and
outline directions
for future work.
Code and 
numerical examples are available online.\footnote{\url{https://github.com/map-filters}}

\subsection{Remarks on notation} %
For a pair of functions $f$ and $g$, we denote their composition by
$f \circ g$.
We denote by $\partial_k f$ the partial derivative of $f$ with
respect to its $k$th input variable.
We use boldface capital letters, e.g., $\Xb$, to 
denote random variables on
$\re^n$,
while we write 
scalar-valued
random variables as $X$.
For a density $\pi$,
$\Xb \sim \pi$ means that
$\Xb$ is distributed according to $\pi$.
For all $n>0$, we let $\mathbb{N}_n=\{1,\ldots,n\}$ denote the set
of the first $n$ integers.
If $\Xb=(X_1,\ldots,X_p)$ 
is a collection of 
random variables and $\Ac \subset \mathbb{N}_p$, then 
$\Xb_{\Ac}=(X_i, i\in\Ac)$ denotes a subcollection of $\Xb$.
In the same way, for $j < k$, $\Xb_{j:k}=(X_j,X_{j+1},\ldots, X_k)$.
The same index notation applies to dummy variables $\xb \in \re^n$.
If $\Xb=(X_1,\ldots,X_p)$ has joint
density $\pi$ and $\Ac \subset \mathbb{N}_p$,  
we denote by $\pi_{\Xb_{\Ac}}$ the marginal of $\pi$ along $\Xb_{\Ac}$.
If %
$\pi$  %
is the density of 
$\Zb=(\Xb, \Yb)$, we denote by 
$\pi_{\Xb \vert \Yb}$
the density
of $\Xb$ given $\Yb$.
We denote independence of a pair of random variables $\Xb,\Yb$ by 
  $\Xb \orth \Yb$. In the same way, $\Xb \orth \Yb \vert \Rb$ means
  that $\Xb$ and $\Yb$ are independent given a third random variable $\Rb$.

\section{{Background on nonlinear filtering}}
\subsection{State-space models}
\label{sec:state-space}

We consider a  
nonlinear, non-Gaussian
state-space model given by a pair of discrete-time stochastic processes,
$(\Zb_k,\Yb_k)$ for $k\ge 1$, where
$\Zb_k$ denotes the unobserved state of a Markov chain taking values in 
$\smash{\re^n}$, while $(\Yb_k)$ refers to an observed process in $\re^d$ that is conditionally independent given the states at all times.
In filtering, we want to leverage realizations $(\yb_k)_{k \ge 1}$ of the observed process to infer the distribution of the latent state conditioned on all available measurements, i.e., we want to estimate the sequence of  conditionals
\begin{equation}
    \pi_{\Zb_k \vert \yb_{1:k}}(\zb) \coloneqq \pi_{\Zb_k \vert \Yb_{1},\ldots,\Yb_k}(\zb \vert \yb_1,\ldots,\yb_k),
\end{equation}
recursively in time, for all $k \ge 1$.

The joint law of the process $(\Zb_k,\Yb_k)$ is fully specified by the sequence of transition kernels 
$(\pi_{\Zb_{k+1}\vert \Zb_{k}})$
and likelihood functions
$(\pi_{\Yb_k \vert \Zb_k})$,
together with a distribution
on the initial conditions of the latent process.
We can think of $\Yb_k$ as an indirect and noisy observation of $\Zb_k$, e.g.,
\begin{equation}
  \Yb_k = h(\Zb_k, \Ecb_k)
\end{equation}
for some mapping $h$ and ``noise'' random variable $\Ecb_k$.
In particular, the likelihood function $\pi_{\Yb_k \vert \Zb_k}$ need {\it not} be Gaussian.
Moreover, 
in many cases $(\Zb_k)_{k \ge 1}$ represents a subcollection of random variables
from a larger state process---e.g., $(\Zb_{j/\Delta})_{j \in \mathbb{N}}$ for some integer $\Delta$---that is only observed at certain times 
$k = j/\Delta \in \mathbb{N}$ 
(see Figure \ref{fig:dataAssg}).
We thus consider settings with {\it sparse} measurements, both in space and time.

\begin{exmp}[State-space model]
\label{ex:state}
A possible dynamic for the state is given by $\Zb_0 \sim \Gauss( {\bf 0} , {\bf I})$, and by a set of stochastic difference equations of the form
\begin{equation} \label{eq:example_state_recursion}
  \Zb_{j + \frac{1}{\Delta}} = f\left(\Zb_{j} \right) + 
  \Wcb_{j},\qquad \forall \, j = 0, \,1/\Delta, \,2/\Delta, \ldots,
\end{equation}
for some deterministic forward model $f:\re^n \ra \re^n$ and i.i.d.\
$(\Wcb_j) \sim \Gauss( {\bf 0}, {\bf I}/\Delta )$. 
Sampling from $\pi_{\Zb_{k+1} \vert \Zb_k}(\cdot \vert \zb)$
requires 
iterating over
\eqref{eq:example_state_recursion} for $\Delta$ times
with initial conditions $\Zb_0 = \zb$. We can then consider that the state is observed through $(\Yb_k)$ only at integer times. %
\end{exmp}
We assume that we can %
{\it sample} the transition kernel $\pi_{\Zb_{k+1} \vert \Zb_k}$, but not evaluate its density, which might 
not even %
exist when dealing with deterministic dynamics, e.g., $\Wcb_{j}=0$ in 
\eqref{eq:example_state_recursion}.
\begin{figure}[!ht]
\centering
  \newcommand{\wStarGraph}{1cm} %
\newcommand{\scaleStarGraph}{.8} %
\newcommand{\minsize}{9mm} %
\newcommand{\varState}{\Zb} %
\newcommand{\varRef}{\Xb} %

\subfloat{
\begin{tikzpicture}[transform shape, scale = \scaleStarGraph]
	\node[default,minimum size=\minsize]  (nd1) 										{$\varState_k$};
 	\node[default,minimum size=\minsize, label = {\footnotesize $\varState_{k+\frac{1}{\Delta}}$}]	 (nd2)		  		[right=of nd1]		
  {} 
 		edge[edgeStyle] (nd1);
 		
 	\node[default,minimum size=\minsize, label = {\footnotesize
  $\varState_{k+\frac{\Delta - 1}{\Delta}}$}]	 (nd5)		  		[right=of nd2]		
  {} 
 		edge[edgeStyle,dashed] (nd2); 	

  	\node[default,minimum size=\minsize]	 (nd6)		  		[right=of nd5]		{$\varState_{k+1}$} 
 		edge[edgeStyle] (nd5); 			

  	\node[dataNode,minimum size=\minsize]	 (data1)		  		[below=of nd1]		{$\Yb_k$} 
 		edge[edgeStyle] (nd1);	
  	\node[dataNode,minimum size=\minsize]	 (data5)		  		[below=of nd6]		{$\Yb_{k+1}$} 
 		edge[edgeStyle] (nd6);

      \node[default,minimum size=\minsize, label = {\footnotesize 
      $\varState_{(k-1)+\frac{\Delta - 1}{\Delta}}$}]   (nd7)          [left=of nd1]    
  {} 
    edge[edgeStyle] (nd1); 
      \node[default,minimum size=\minsize, label = {\footnotesize $\varState_{(k-1)+\frac{1}{\Delta}}$}]   (nd8)          [left=of nd7]    
  {} 
    edge[edgeStyle,dashed] (nd7); 
    \node[default,minimum size=\minsize]  
    (nd9)          [left=of nd8]    {$\varState_{k-1}$}  
  {} 
    edge[edgeStyle] (nd8);
        \node[dataNode,minimum size=\minsize]  (data1)          [below=of nd9]    {$\Yb_{k-1}$} 
    edge[edgeStyle] (nd9);
\end{tikzpicture}
} 	
\caption{
Markov structure for a state-space model consisting of a latent
field $(\Zb_{ \frac{j}{\Delta}})_{j \in \mathbb{N}}$ which is observed
every $\Delta$ time steps via a second stochastic process 
$(\Yb_k)_{k \in \mathbb{N}}$.
}
\label{fig:dataAssg}
\vspace{-10pt}
\end{figure}

\subsection{Two-step ensemble filtering}
\label{sec:gen_enkf}
\begin{figure}[!ht]
\centering
\includegraphics[width=0.80\textwidth]{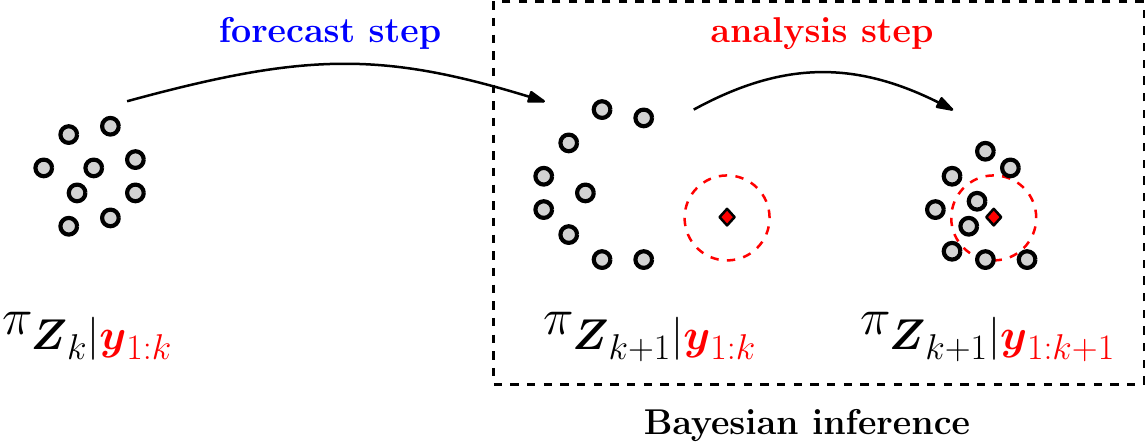}
\caption{
A typical ensemble filtering algorithm first propagates a
 particle approximation of the filtering distribution at time $k$ until the next
 observation time $k+1$ using the transition dynamic ({\it forecast step}).
 Then, it %
 updates the ensemble via a transformation that incorporates information from
 the most recent observation $\yb_{k+1}$ ({\it analysis step}).
 The analysis step can be regarded as a problem of ``static'' Bayesian inference
}
\label{fig:schemeAlg}
\vspace{-10pt}
\end{figure}

We work with a general class of ensemble filtering algorithms, wherein the filtering
distribution is approximated by a collection of particles %
that are
updated recursively over time via a two-step process consisting of a {\it forecast} and an {\it analysis} step (Figure \ref{fig:schemeAlg}).
The recursion takes a particle approximation  
$\zb^1,\ldots,\zb^M$ 
of
the filtering distribution
 $\pi_{\Zb_{k}\vert\yb_{1:k}}$ at  time $k$, and turns it into a particle approximation of the filtering distribution at the next observation time,
 $\pi_{\Zb_{k+1}\vert\yb_{1:k+1}}$, %
 by
 considering the following identity, 
 \begin{equation}
  \pi_{\Zb_{k+1}\vert\Yb_{1:k+1}} \propto
  \pi_{\Yb_{k+1}  \vert \Zb_{k+1}} \,
  \int 
  \pi_{\Zb_{k+1}\vert \Zb_k} \,
  \pi_{\Zb_{k}\vert\Yb_{1:k}}\,{\rm d}{\scriptstyle \Zb_k},
 \end{equation}
 which links filtering marginals at consecutive observation times. 
In the 
{forecast} step, 
 each particle 
$\zb^j$ is updated independently with a sample
from $\pi_{\Zb_{k+1} \vert \Zb_k}(\cdot \vert \zb^j)$, obtained by simulating the %
state dynamic
until the next observation time.
The resulting ensemble $\zb^1_{f},\ldots,\zb^M_{f}$ yields a particle approximation of the forecast (or predictive)
distribution
$\pi_{\Zb_{k+1}\vert\yb_{1:k}}$---i.e., the distribution of the state at time
$k+1$ conditioned on all observations available up to time $k$.
The forecast step might involve the evaluation of an
expensive forward model, like a  %
partial differential equation, and represents the computational
bottleneck for high dimensional and complex models, like
those employed in weather forecasting.
Yet, the forecast step is of little statistical
concern, since
we can usually sample {\it exactly} from
$\pi_{\Zb_{k+1} \vert \Zb_k}$ {up to the discretization of the
  forward model.}\footnote{{In this paper, for simplicity, we take the discretized model to be an exact representation of the forward dynamics.}}
The analysis step, however, 
is more challenging, since it involves
the approximation of 
an inference task.
We can think of the
forecast 
distribution 
$\pi_{\Zb_{k+1}\vert\yb_{1:k}}$ %
as the prior of a  
Bayesian inverse problem, where the likelihood function and posterior distribution are given by 
$\pi_{\yb_{k+1}\vert \Zb_{k+1}}$
and 
$\pi_{\Zb_{k+1}\vert\yb_{1:k+1}}$,
respectively.
In the analysis step, we seek a particle approximation of the posterior $\pi_{\Zb_{k+1}\vert\yb_{1:k+1}}$
in a setting where %
the prior
density $\pi_{\Zb_{k+1}\vert\yb_{1:k}}$ 
is not known explicitly, and 
where
we only have
finitely many samples
$\zb^1_{f},\ldots,\zb^M_{f}$ from the prior.

The analysis step is thus a problem of approximate Bayesian inference, and can be implemented in
various ways.
For instance, the EnKF yields a particle
approximation of the filtering distribution 
by pushing forward the forecast 
ensemble through a {\it linear}
transformation estimated %
under Gaussian assumptions.
In this paper, %
we leverage ideas from measure transport to generalize the EnKF %
by estimating %
{\it nonlinear} %
transformations in %
function space.
Our goal is to reduce the estimation bias 
in the EnKF,
while retaining applicability and robustness of the %
methodology in high dimensions.

\section{{The stochastic and deterministic map filters}}

{In this section we introduce two novel methodologies for addressing the analysis step in nonlinear filtering problems. We start by framing the analysis step as a problem of coupling the forecast and the analysis distributions (in Section~\ref{sec:analysis}) and propose two solutions based on either stochastic (Section~\ref{sec:pert_obs}) or deterministic (Section~\ref{sec:square-root}) couplings.}

\subsection{Analysis step with couplings}
\label{sec:analysis}
The analysis step of the ensemble 
algorithms described in Section
\ref{sec:gen_enkf}
can be formulated as an
abstract Bayesian inference problem, defined by a pair
of random variables $(\Xb,\Yb)$ 
on $\re^n \times \re^d$,
which denote,
respectively, the
unobserved state variable
and the data at the current assimilation time.
The joint law 
$\pi_{\Yb, \Xb}$ %
is specified by the prior marginal $\pi_{\Xb}$, which corresponds to the forecast distribution, and by the
likelihood function $\pi_{\Yb \vert \Xb}$.
For simplicity, we assume that $\pi_{\Yb,\Xb}$ is 
fully supported 
on
$\re^d \times \re^n$.
In filtering, we are only given %
$M$ samples 
$\xb^1,\ldots,\xb^M$ from the prior $\pi_{\Xb}$---the so-called forecast ensemble---and 
we wish to sample the posterior (i.e., the filtering distribution)
\begin{equation}
  \pi_{\Xb \vert \yb^*}(\xb) = 
  \pi_{\Xb \vert \Yb}(\xb \vert \yb^*)
\end{equation}
for some realization $\yb^* \in \re^d$ of the data.

A possible approach to posterior sampling 
is to construct a
{\it coupling} between  prior and posterior
distributions using a transformation
$T$ such that $\Zb \coloneqq T(\Xb)$ follows
the posterior law
\cite{villani2008optimal}.
This transformation and the associated coupling may or may not be
deterministic.

\begin{figure}[!ht]
\centering
\includegraphics[width=0.40\textwidth]{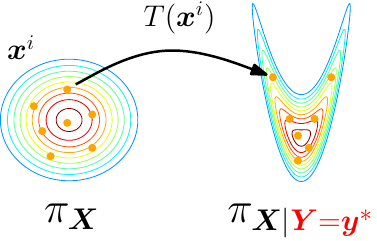}
\caption{
Illustration of a transport map $T$ that pushes forward the prior $\pi_{\Xb}$ to the
posterior $\pi_{\Xb \vert \Yb = \yb^*}$. By the definition of $T$, if $(\xb^i)$ are i.i.d.\ samples from
$\pi_{\Xb}$, then $(T(\xb^i))$ are also i.i.d.\ samples from $\pi_{\Xb \vert \Yb = \yb^*}$.
}
\label{fig:det_coupling}
\vspace{-10pt}
\end{figure}

In a deterministic coupling, $\Zb = T(\Xb)$ for some 
deterministic 
transformation $T:\re^n \ra \re^n$ called a 
{\it transport map}.
In this case, we say that $T$ pushes forward 
the prior to the posterior distribution.
For any $\Xb\sim \pi_{\Xb}$ and map $T$, the pushforward 
of $\pi_{\Xb}$ by $T$
is given by the law of $T(\Xb)$ and is denoted by
$T\push \pi_{\Xb}$.
The existence of a map
$T$ such that $T(\Xb) \sim \pi_{\Xb \vert \yb^*}$ 
is guaranteed under very mild conditions.
For instance, it suffices to have a prior distribution without atoms \cite{villani2008optimal}.
In fact, there are infinitely many ways to push forward one distribution to another, some of which minimize a notion of integrated cost. Monge's optimal map, for example, minimizes the Euclidean $L_2$ cost  
$\Ex[ \| T(\Xb) - \Xb \|^2 ]$ over all admissible transformations 
\cite{monge1781memoire}.
Transport maps are interesting because they enable
direct posterior sampling.
That is, if $\xb^1,\ldots,\xb^M$ are i.i.d.\ samples from the prior, then $T(\xb^1),\ldots,T(\xb^M)$ are also i.i.d.\  samples from the posterior 
(cf.~Figure \ref{fig:det_coupling}).
We will see in Section
\ref{sec:square-root} that deterministic couplings
are associated with nonlinear generalizations of the square-root EnKF \cite{tippett2003ensemble}.
\begin{remark}[Ensemble transform particle filter]
In his pioneering approach to filtering \cite{reich2013nonparametric}, Reich proposes a nonparametric estimator for %
Monge's %
map, by solving the Kantorovich relaxation of a discrete optimal transport problem
between the empirical measure given by the prior ensemble and 
the empirical approximation of the 
posterior distribution 
obtained by {\it reweighting} %
prior
particles according to their likelihood. %
In this paper, we 
consider continuous rather than discrete transport, and look for 
prior-to-posterior transformations---not necessarily optimal with respect to any cost---that can
be estimated {\it without} resorting to importance sampling, thus avoiding 
the root cause of particle
degeneracy.
\end{remark}
A coupling need not be deterministic.
For instance, we can have %
$\Zb = T(\Xib, \Xb)\sim \pi_{\Xb \vert \yb^*}$ for some deterministic transformation $T$ and  additional random variable $\Xib$.
As a function of $\Xb$ alone, 
we can think of $T$ 
as a ``stochastic'' map.
In this 
work,
we %
consider 
non-deterministic couplings induced by a 
map
$T:\re^d \times \re^n \ra \re^n$ such that $\Zb = T(\Yb, \Xb)$, where
$\Yb$ represents the data random variable of the Bayesian inference problem.
These
couplings can be used for
posterior sampling like their deterministic
counterpart:
if $(\yb^1,\xb^1),\ldots,(\yb^M,\xb^M)$ are i.i.d.\ samples from the joint distribution $\pi_{\Yb,\Xb}$, then
$T(\yb^1,\xb^1),\ldots, T(\yb^M,\xb^M)$ are also
i.i.d.\ samples from the posterior.
In fact, the transformation $T$ %
is
a transport map
that pushes forward $\pi_{\Yb,\Xb}$ to
the
posterior.
Inference with non-deterministic couplings is interesting because we can build %
estimators for 
$T$ using only
{\it convex} optimization.
We will see in Section
\ref{sec:pert_obs} that non-deterministic couplings 
are linked to nonlinear generalizations of the stochastic EnKF \cite{burgers1998analysis}.

There is no unique way to couple prior and posterior distributions in
a Bayesian inference problem.
In the context of filtering, 
we need couplings that can be estimated efficiently in high dimensions from finitely many prior samples.
In Sections \ref{sec:pert_obs} and 
\ref{sec:square-root},
we discuss specific choices for these couplings, which lead to novel 
nonlinear filtering algorithms.
Before that, however, we need to discuss the
properties of an important transport for our analysis,
the Knothe--Rosenblatt rearrangement \cite{rosenblatt1952remarks}.

\subsection{Knothe--Rosenblatt rearrangement}
\label{sec:KR}
Given any pair of positive densities $\pi, \eta$ on $\re^n$, there exists a 
unique monotone triangular transport map $S$---called the
Knothe--Rosenblatt (KR) rearrangement---that
pushes forward one of the corresponding distributions to the other \cite{bogachev2005triangular}, i.e., if $\Zb \sim \pi$, then
$S(\Zb)\sim \eta$.
A (lower) triangular map $S:\re^n\ra\re^n$ is a multivariate 
function whose $k$th component, $S^k$, depends only on the first $k$ input variables, i.e.,
\begin{equation}   
S( \zb ) = \left[\begin{array}{l}
S^1(z_1)\\[2pt]
S^2(z_1,z_2)\\ 
\vdots \\ 
S^n(z_1,z_2,\dots z_n)
\end{array}\right].%
\label{eq:lowerTri}
\end{equation}
Monotonicity of the multivariate transformation
is intended with respect to the 
lexicographic order on $\re^n$, which is 
equivalent to 
each slice
\begin{equation} \label{eq:componentMap}
  \xi \mapsto S^k( z_1 ,\ldots, z_{k-1}, \xi )
\end{equation}
being an increasing function.
For $\pi,\eta>0$, %
each function %
\eqref{eq:componentMap} is also strictly increasing and the KR rearrangement becomes an invertible map.
The existence of the KR rearrangement stem from
intuitive factorization properties of any density, e.g., if $\Zb=(Z_1,\ldots,Z_n)\sim \pi$, then
$\pi = \prod_k \pi_{Z_k \vert \Zb_{1:k-1}}$.
In fact, perhaps the most important property 
for our analysis is that 
the  rearrangement
provides an implicit
characterization of  
the marginal conditionals of $\pi$, whenever
$\eta = \Gauss({\bf 0}, {\bf I}_n)$.
In this case, one can show that %
each one dimensional mapping 
\eqref{eq:componentMap} 
pushes forward
the marginal conditional
$\xi \mapsto \pi_{Z_k\vert\Zb_{1:k-1}}(\xi \vert \zb_{1:k-1})$ to a one dimensional standard normal,
for all $\zb_{1:k-1}\in \re^{k-1}$ %
\cite{santambrogio2015optimal}.
We leverage this property in the next section, where we discuss our first filtering algorithm, while we
defer questions pertaining to the estimation of 
the rearrangement to Section \ref{sec:est_KR}.

\subsection{Stochastic map filter} %
\label{sec:pert_obs}
In this section, we outline an
algorithm to carry out the analysis step 
at a given observation time,
based on the 
construction of non-deterministic couplings
between forecast and filtering distributions.
We refer to the notation of Section
\ref{sec:analysis} for an abstract Bayesian inverse problem. 

\begin{figure}[!ht]
\centering
\includegraphics[width=0.50\textwidth]{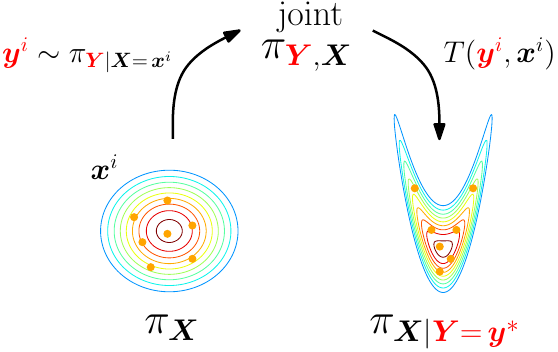}
\caption{
Illustration of the analysis step in the stochastic map filter.
For each prior particle $\xb^i$, we generate a sample $\yb^i$ from
the likelihood 
function $\pi_{\Yb \vert \Xb}(\cdot \vert \xb^i)$.
The pairs $(\yb^i, \xb^i)$ represent i.i.d.\ samples from the joint
distribution $\pi_{\Yb,\Xb}$ and are used to estimate a transport map $T$
that pushes forward $\pi_{\Yb,\Xb}$ to the posterior $\pi_{\Xb \vert \Yb = \yb^*}$.
By the definition of $T$, $(T(\yb^i, \xb^i))$ gives the
desired samples from the posterior
distribution.
}
\label{fig:pert_obs_alg}
\vspace{-10pt}
\end{figure}

Our goal is to find a map  
$T:\re^d \times \re^n \ra \re^n$ such that 
$T(\Yb, \Xb)\sim \pi_{\Xb \vert \yb^*}$.
If we had such a map, then we could easily sample
the posterior $\pi_{\Xb \vert \yb^*}$ by pushing forward samples from
the joint distribution $\pi_{\Yb, \Xb}$ through $T$.
See Figure \ref{fig:pert_obs_alg} for an illustration.
Samples from $\pi_{\Yb, \Xb}$ are also easy to obtain:
given a collection of prior (or forecast) samples $(\xb^i)\sim \pi_{\Xb}$, %
let $\yb^i$ be a sample from 
$\pi_{\Yb\vert\Xb}(\cdot \vert \xb^i)$, for
all $i=1,\ldots,M$.
The resulting pairs $(\yb^i,\xb^i)$ define
samples from $\pi_{\Yb,\Xb}$.
We are then left with the problem of
finding a suitable $T$.
We call $T$ the ``analysis'' map.

Let $S$ be the KR rearrangement that pushes forward
$\pi_{\Yb,\Xb}$ to $\Gauss({\bf 0}, {\bf I}_{d+n})$.
Since the rearrangement is a triangular map,
we can always partition it as
\begin{equation}   \label{eq:blockS}
S( \yb, \xb ) = \left[\begin{array}{l}
S^{\Ycb}(\yb)\\[2pt]
S^{\Xcb}(\yb,\xb)
\end{array}\right]%
\end{equation}
for some functions $S^{\Ycb}: \re^d \ra \re^d$ and
$S^{\Xcb}: \re^d \times \re^n \ra \re^n$,
where $\yb \in \re^d$ and $\xb \in \re^n$.
By the definition of KR rearrangement, we have that the function $S^{\Xcb}$ pushes forward
$\pi_{\Yb,\Xb}$ to a standard normal on $\re^n$, i.e.,
\begin{equation}
  S^{\Xcb}(\Yb,\Xb)\sim \Gauss( {\bf 0}, {\bf I}_n ).
\end{equation}
Moreover, since
the KR rearrangement implicitly characterizes the
conditionals of $\pi_{\Xb,\Yb}$ (cf.~Section \ref{sec:KR}),
we have that the map
$\xib \mapsto S^{\Xcb}(\yb^*, \xib)$---obtained by 
fixing the first input of $S^{\Xcb}$ to the observation
$\yb^*\in \re^d$---pushes
forward the posterior
$\pi_{\Xb \vert \yb^*}$
to 
$\Gauss( {\bf 0}, {\bf I}_n )$.
By combining these properties, we arrive at a
definition for the analysis map $T$ as
\begin{equation}
\label{eq:an_map}
  T \coloneqq 
  S^{\Xcb}(\yb^*, \cdot)^{-1}
  \circ 
  S^{\Xcb},
\end{equation}
where $S^{\Xcb}(\yb^*, \cdot)^{-1}$ denotes the inverse function of the mapping $\xib \mapsto S^{\Xcb}(\yb^*, \xib)$.
Inverting a triangular map is
a computationally trivial task, since it reduces to a sequence of one-dimensional
root findings.
See 
Algorithm \ref{alg:invertMap} for details.
It is immediate to verify that the map $T$ 
defined in \eqref{eq:an_map} pushes forward 
$\pi_{\Yb,\Xb}$ to %
$\pi_{\Xb \vert \yb^*}$, and thus defines 
a non-deterministic coupling between prior and posterior
distributions.

In practice, we need to build an estimator, 
$\widehat{T}$, for $T$.
We propose to use %
\begin{equation}
\label{eq:est_an_map}
  \widehat{T} \coloneqq 
  \widehat{S}^{\Xcb}(\yb^*, \cdot)^{-1}
  \circ 
  \widehat{S}^{\Xcb},
\end{equation}
where we replace $S$ in 
\eqref{eq:an_map} by
a constrained maximum likelihood estimator $\widehat{S}$.
We describe this estimator %
in Section \ref{sec:est_KR}.
It turns out that the components of $\widehat{S}$
can be computed quickly---in parallel, and via {\it convex} optimization---from a collection of samples
$(\yb^i,\xb^i) \sim \pi_{\Yb,\Xb}$.

We summarize the resulting analysis step with stochastic maps in Algorithm \ref{alg:analysis_stoc_maps}.
The term stochastic maps refers to the use of 
non-deterministic couplings between forecast and filtering distributions.

\begin{algorithm} %
  \caption{
  {\bf (Invert a triangular map)} \\
  Given a map $S$ of the form 
  \eqref{eq:lowerTri} and an
  $\xb \in \re^n$,  
  solve the
  nonlinear system 
  $S(\zb) = \xb$ for
  $\zb$. 
  }
  \label{alg:invertMap}
  \begin{algorithmic}[1]
    \Procedure{InvertMap}{ S, $\xb$ }%
    \For{$k\gets 1$ to $n$} 
    \State $z_k \gets$ Solve  
    $S^k(z_1,\ldots,z_{k-1}, \xi) = x_k$
    for $\xi \in \re$
    \Comment one-dimensional root finding \cite{press2007numerical}
    \EndFor
    \State \Return ${\zb = (z_1,\ldots,z_n)}$
    \EndProcedure
  \end{algorithmic}
\end{algorithm}

\begin{algorithm} %
\linespread{\spalg}\selectfont
  \caption{
  {\bf (Analysis step with stochastic maps)} \\
  Given $M$ samples $\xb^1,\ldots,\xb^M$
  from the forecast distribution, a likelihood function
  $\pi_{\Yb\vert\Xb}$,
   and an observation
  $\yb^*$, 
  generate a particle approximation 
  $\zb^1,\ldots,\zb^M$ of the filtering
  distribution $\pi_{\Xb \vert \yb^*}$.
  }
  \label{alg:analysis_stoc_maps}
  \begin{algorithmic}[1]
    \Procedure{AnalysisStochastic}{
    $\yb^*$, $\pi_{\Yb \vert \Xb}$,   
    $\xb^1, \ldots,\xb^M$ }%
    \For{$i\gets 1$ to $M$} 
    \Comment Generate samples from $\pi_{\Yb,\Xb}$
    \State %
    $\yb^i \gets$ sample from $\pi_{\Yb\vert\Xb}(\cdot \vert \xb^i)$
    \EndFor
    \State 
    $\widehat{S}^{\Xcb} \gets$ 
    estimator of
    $S^{\Xcb}$ from $(\yb^i,\xb^i)_{i=1}^M$
    given by
    \eqref{eq:mle_KR}.
    $S^{\Xcb}$ is 
    defined
    in \eqref{eq:blockS}
    \For{$i\gets 1$ to $M$} 
    \Comment Action of $\widehat{T}$ as defined in
    \eqref{eq:est_an_map}
    \State $\zb^i \gets \widehat{S}^{\Xcb}(\yb^i,\xb^i)$\State $\zb^i \gets$ 
    \Call{InvertMap}{ $\widehat{S}^{\Xcb}({\yb^*}, \cdot)$, $\zb^i$ }
    \EndFor
    \State \Return 
    $\zb^1,\ldots,\zb^M$ 
    \EndProcedure
  \end{algorithmic}
\end{algorithm}

\subsection{Estimation of the Knothe--Rosenblatt rearrangement from samples}
\label{sec:est_KR}
In this section, we address the 
estimation of 
a KR rearrangement $S$ that pushes forward an arbitrary density $\pi$ on $\re^n$ to a standard normal $\eta = \Gauss( {\bf 0}, {\bf I}_n )$, given only
$M$ samples $\zb^1,\ldots,\zb^M$ from $\pi$.
This problem is at the heart of the filtering methodologies proposed in this paper.  
For instance, in the stochastic map algorithm 
of Section \ref{sec:pert_obs}, we need to estimate a KR
rearrangement that pushes forward the joint distribution of state and data, $\pi_{\Xb,\Yb}$, to a standard normal.
In that case, $\zb^1,\ldots,\zb^M$ represent
samples from $\pi_{\Xb,\Yb}$.

We recall a construction first proposed by
\cite{parno2014transport}.
We can build a constrained maximum likelihood estimator for $S$ as follows: let
$\Hcb$ be a {finite} dimensional approximation space for the KR rearrangement, i.e., let 
$\Hcb$ be a cone  
of monotone triangular maps $U:\re^n \ra \re^n$ that can be
described with finitely many parameters.
For each $U$ in $\Hcb$ consider the density given by pulling back $\eta$ by $U$, defined as
$
U\pull \, \eta \coloneqq 
  (U^{-1})\push \, \eta 
$
and given explicitly by
\begin{equation} \label{eq:def_pullback}
U\pull \, \eta(\zb) = 
\eta(U(\zb))\,\det \nabla U(\zb), 
\end{equation}
where $\det \nabla U$ denotes the determinant of the
gradient of the map.
Note that $\pi$ can be written as
$S\pull\,\eta$ by the definition of the KR
rearrangement $S$.
Thus, the family of densities $(U\pull \, \eta)_{U \in \Hcb}$ defines a parametric model for $\pi$---the law of the samples  $\zb^1,\ldots,\zb^M$.
A %
maximum likelihood estimator, 
$\widehat{S}$, for $S$ can then be written as
\begin{equation} \label{eq:mle_KR}
\widehat{S} \in \arg \max_{U \in \Hcb} \,
\frac{1}{M} \sum_{i=1}^M \,\log U\pull \, \eta(\zb^i).  
\end{equation}
Indeed, $\widehat{S}\pull\,\eta$ minimizes an empirical estimate of the
Kullback–Leibler divergence between $\pi$ and the
class of densities given by 
\eqref{eq:def_pullback} for $U\in\Hcb$.

Since $\eta = \Gauss( {\bf 0}, {\bf I}_n )$, 
a simple calculation shows that each component,
$\widehat{S}^k$, of $\widehat{S}$ can be computed independently as \cite{parno2014transport} %
\begin{equation} \label{eq:component_estimator}
\widehat{S}^k \in \arg \min_{U^k \in \Hcb_k} \,
\frac{1}{M} \sum_{i=1}^M \,\left(\frac{1}{2} \, U^k(\zb^i)^2
- \log \partial_k U^k(
\zb^i) \right),
\end{equation}
where $\Hcb_k$ is a function space for the $k$th
component, $U^k$, of $U \in \Hcb$, and where
$\partial_k U^k$ denotes the derivative of $U^k$ with respect to its
$k$th input variable. 
$\Hcb_k$ contains functions $U^k :\re^k \ra \re$ that are monotone with respect to the last input variable, i.e., where each $\xi \mapsto U^k(z_1,\ldots,z_{k-1}, \xi)$ is an increasing map.

In practice, we need to pick a parameterization for
$U^k$ in \eqref{eq:component_estimator}, i.e., we need to specify a choice for $\Hcb_k$.
Different choices %
lead to different filtering algorithms (cf.~Algorithm \ref{alg:analysis_stoc_maps}).
For instance, if we restrict $\Hcb_k$ to contain only affine maps---i.e., $U^k(\zb)= u_0 + \sum_{i \le k} u_i \, z_i$ for some unknown coefficients $(u_i)$---then we 
obtain an EnKF-type algorithm.
(We clarify this connection in Section \ref{sec:rem_pert_obs}.)
If we depart from the linear ansatz, however,
we obtain a whole new class of nonlinear filtering algorithms.
For example, in this paper we experiment with  
{\it separable} parameterizations of the form
$U^k(\zb)=\sum_{i\le k} \ufrakb_i(z_i)$ 
for some collection of nonlinear functions 
$(\ufrakb_i)$, 
\begin{equation}
 \ufrakb_i(z) = \sum_j u_{ij} \,\psi_j(z),
\end{equation}
which can be
expressed in terms of finitely many %
basis 
functions $(\psi_j)$---e.g.,
Hermite polynomials or radial basis functions---and unknown  
coefficients $(u_{ij})$.
These %
parameterizations
are perhaps the simplest way
to increase complexity %
with respect to a linear function, and  
lead to filtering algorithms that can 
outperform the EnKF in several challenging scenarios, 
with only a marginal increase in computational cost 
(see Section \ref{sec:numerics}). %
We describe separable parameterizations %
in Appendix \ref{sec:param_tri}, together with a general way to represent 
monotone 
triangular maps.

For the parameterizations that we consider in this paper, \eqref{eq:component_estimator}
is a finite-dimensional convex program
that can be solved
efficiently---even in high dimensions---via standard convex optimization methods (e.g.,
Newton's method).
If we make certain structural assumptions on the estimator, however, \eqref{eq:component_estimator} admits a closed form solution without any
need for numerical optimization.
For instance, if we let
$U^k(\zb)= \ufrakb(z_1,\ldots,z_{k-1}) + \ufrakb_k(z_k)$
for some nonlinear function 
$\ufrakb$ and affine $\ufrakb_k$, then \eqref{eq:component_estimator} 
reduces to a standard {\it linear regression} problem,
provided that
we consider parameterizations of  $\ufrakb$ that are linear in the unknown coefficients.
We refer to Appendix \ref{sec:param_tri} for additional  details on this topic.

\subsection{Regularization in high dimensions: localization of nonlinear maps}
\label{sec:reg_map}

In this section, we discuss
how to regularize the estimation of
the KR rearrangement $S$ in high dimensions, by imposing
certain functional constraints---e.g., sparsity---on the structure
of the maximum likelihood estimator 
$\smash{\widehat{S}}$ 
defined in
\eqref{eq:mle_KR}.
Regularization is the cornerstone of
filtering algorithms
that
aim to work in high dimensions with
a limited ensemble size: examples include
covariance tapering and inflation in the EnKF, 
or weight
localization in particle filters \cite{evensen2007data,poterjoy2016localized}.
In this work, we do not
regularize the estimation of
a covariance matrix,
but rather the estimation of a more
general nonlinear function: the
KR rearrangement $S$.

Like every estimator, $\widehat{S}$ has a bias---which quantifies how well we can approximate the true
KR rearrangement in the limit of infinite data---and a  variance---which measures the %
variability of the estimate with respect to different
realizations of the samples $\zb^1,\ldots,\zb^M$ from $\pi$.
Both bias and variance of $\widehat{S}$---which is a random function---can be controlled by changing
the complexity of the
approximation space %
for the estimator.
Very rich parameterizations %
can 
yield estimators
with low bias, but potentially unacceptable variance, if the number of samples
$\zb^1,\ldots,\zb^M$
is insufficient.
There is no point in trying to learn arbitrarily nonlinear transformations.
In fact, 
that would be harmful.
On the other hand, overly simple parameterizations that can be fitted with low variance might have an unnecessarily large estimation bias, 
leading to
a waste of computational resources.
Successful estimation requires addressing this bias-variance tradeoff, %
especially when
dealing with
high dimensions and  
low ensemble regimes.

There are many ways to regularize the estimation of
the KR rearrangement.
For instance, one could
augment the average log-likelihood in
\eqref{eq:mle_KR} with 
a convex 
penalty on the coefficients of
$\widehat{S}$---e.g., a ridge or lasso functional \cite{hastie2005elements}---yielding a
penalized maximum likelihood estimator.
In addition, one could constrain the function space for the estimator
by using only low-order nonlinear basis functions (see Appendix \ref{sec:param_tri}).
In this section, we focus on the complementary idea of regularizing the estimation by imposing
{\it sparsity} constraints on $\widehat{S}$, i.e.,
by
dropping some variable dependence from each component
of the map, such that
\begin{equation} \label{eq:sparsityComp}
 \widehat{S}^k(z_1,\ldots,z_k) = 
 \widehat{S}^k(z_{j_1},\ldots, z_{j_q}, z_k)
\end{equation}
for some indices $j_1,\ldots,j_q$ and integer {$q<k-1$}
\cite{spantini2017inference}---effectively %
introducing a 
notion of ``localization'' for 
nonlinear transformations.
We now discuss %
model assumptions that justify
a sparsity pattern like 
\eqref{eq:sparsityComp}.

The sparsity of a KR rearrangement $S$
that pushes forward a density $\pi$ on
$\re^n$ to $\Gauss({\bf 0}, {\bf I}_n)$ depends
on marginal conditional independence properties of 
$\pi$.
As explained in \cite{spantini2017inference}, if $\Zb=(Z_1,\ldots,Z_n)\sim \pi$, then the $k$th component of $S$ is {\it not} a
function of the $j$th input variable ($j<k$) if
$Z_j$ and $Z_k$ are conditionally independent given
the remaining variables with index less than $k$, i.e.,
if 
\begin{equation} \label{eq:marg_cond_ind}
  Z_j \orth Z_k \vert \Zb_{ \{1:k-1\}\setminus \,j}.
\end{equation}
Whether \eqref{eq:marg_cond_ind} holds, perhaps only
approximately, depends on 
properties of $\pi$ such as 
decay of dependence and %
conditional independence.
 These properties are fairly common when filtering 
 spatially distributed %
 processes since they encode an
 intuitive  
 notion
 of local probabilistic interaction between
 the random variables 
 \cite{reich2015probabilistic,ueno2009covariance}.
\paragraph{Decay of dependence and banded estimators} 
Let $d(\cdot,\cdot)$ be %
a distance function on the index set $\{1,\ldots,n\}$, which defines a natural ordering of the variables 
$\Zb$, 
so that larger values of $d( i, j)$ are associated with
a weaker statistical dependence between $Z_i$ and $Z_j$. %
For instance, in data assimilation  $d(\cdot,\cdot)$ could be
derived from a metric
in physical space. 
The first observation is that the $k$th component of the KR rearrangement should depend 
{\it weakly} on variables that are far away from $Z_k$, according to 
$\db(\cdot,\cdot)$.
To exploit this structure, 
let $\Ac_{k,r}$
denote the set of indices of  state variables that are within a distance 
$r$ from $Z_k$, i.e.,
\begin{equation} \label{eq:nbd_vars}
   \Ac_{k,r} \coloneqq \{ i \in (1,\ldots,k)\,\,{\rm s.t.}\,  
   \,d(i, k) \le r \}.
\end{equation}
We propose to %
build an estimator $\widehat{S}$
of $S$
for which
each component
$\widehat{S}^k$ %
depends only on the %
variables in 
$\Ac_{k,r}$, %
\begin{equation}  \label{eq:threshold_est}
\widehat{S}( \zb ) = \left[\begin{array}{l}
\widehat{S}^1(\zb_{\Ac_{1,r}})\\[2pt] 
\vdots \\ 
\widehat{S}^{n}(\zb_{\Ac_{n,r}}) \\
\end{array}\right], %
\end{equation} 
for an ``optimal'' value of $r\ge0$. This $r$ is obtained 
by minimizing some prediction metric (e.g., RMSE) in an offline calibration phase,
as if it were 
the optimal 
localization radius %
for the
EnKF %
\cite{evensen2007data}.

In the case of linear maps,
 \eqref{eq:threshold_est} corresponds to a ``banded'' estimator for the Cholesky factor of the precision matrix of the ensemble 
 $\zb^1,\ldots,\zb^M$
 \cite{wu2003nonparametric,bickel2008regularized}, and has been used for the purpose of regularized covariance estimation in the context of data assimilation \cite{nino2018ensemble}.
Indeed, if $\pi = \Gauss( {\bf 0}, \Sigmab)$ %
and $\Sigmab = \Lb \Lb^\top$ is a Cholesky decomposition of
$\Sigmab \succ 0$, then
$\Sb \coloneqq \Lb^{-1}$
represents a {\it linear} KR rearrangement 
that pushes forward $\pi$ to 
a standard normal.
In this paper, we extend these banded estimators to the
nonlinear/non-Gaussian case, using the
KR rearrangement to generalize 
the notion of a Cholesky factor.

\paragraph{Markov properties and graphical models} 
The sparsity pattern of the estimator $\widehat{S}$ can also be %
justified on the grounds of
conditional independence properties of $\pi$---the so-called Markov properties.
These properties can be collected in a simple undirected graph
$\Gcb = (\Vc, \Ec)$, where each node $i \in \Vc$ is associated with one of the random variables in the collection $\Zb=(Z_1,\ldots,Z_n)$, and where
each 
edge $(i,j) \in \Ec$ encodes a notion of probabilistic interaction between the pair $Z_i,Z_j$.
We say that $\Zb$ (or $\pi$) is a Markov random field (MRF) with respect to $\Gcb$ if for any partition $\Ac,\Sc,\Bc$ of the vertex set---for which
$\Sc$ is a separator set\footnote{$\Sc$ is a separator set for $\Ac$ and $\Bc$ if every path from a node in
$\Ac$ to a node in $\Bc$ goes through $\Sc$.
}
for $\Ac$ and $\Bc$---we have that
$\Zb_{\Ac}$ and $\Zb_{\Bc}$ are conditionally independent given $\Zb_{\Sc}$,
i.e., $\Zb_{\Ac} \orth \Zb_{\Bc} \vert \Zb_{\Sc}$ \cite{lauritzen1996graphical}.
See Figure \ref{fig:partition_badOrder} ({\it left})
for an example.
For any strictly positive $\pi$, we can find 
the sparsest
$\Gcb$ that is compatible with the conditional independence structure of $\pi$ as follows:
let
$(i,j)\notin \Ec$ if and only if 
$\Zb_{i} \orth \Zb_{j} \vert \Zb_{\Vc\setminus\{i,j\}}$
 \cite{koller2009probabilistic}.

A key 
result in \cite{spantini2017inference} 
links the sparsity of the KR
rearrangement $S$ to the sparsity of $\Gcb$.
For a given $\Gcb$, define the
sequence of marginal graphs $\Gcb^n,\ldots,\Gcb^1$ using a simple
recursion: (1) $\Gcb^n\coloneqq \Gcb$, and (2) $\Gcb^{k-1}$ is
obtained from $\Gcb^k$ by removing node $k$ and by
turning its neighborhood $\neigh (k,\Gcb^k)$ into 
a fully connected subset,\footnote{For any graph $\Gcb$, %
we denote by $\neigh (k,\Gcb)$ the
neighborhood of a node $k$ in $\Gcb$.
} i.e., a clique.
By \cite[Theorem 5.1]{spantini2017inference}, each component $S^k$ of the rearrangement 
can only depend on variables in $\neigh (k,\Gcb^k)$, for all
$k=1,\ldots,n$.
See Figure \ref{fig:marginalGraph} for an illustration (adapted from \cite{spantini2017inference}).
Hence, if we know that $\pi$ is Markov with respect
to a sparse graph, then we can translate this information into sparsity of the estimator 
$\widehat{S}$,
possibly increasing the sparsity pattern
in
\eqref{eq:threshold_est}.

If $\pi = \Gauss({\bf 0},\Sigmab)$, then describing the sparsity of a linear KR
rearrangement due to conditional independence reduces to the problem
of characterizing the fill-in for the Cholesky factor of the precision matrix $\Sigmab^{-1}$---a well-known result
in the field of Gaussian graphical models
\cite{koller2009probabilistic}, which has also been exploited in the context of data
assimilation to devise regularized estimators of the ensemble covariance 
\cite{ueno2009covariance}.
Here, we consider the
non-Gaussian generalization provided by
\cite{spantini2017inference}, where the Cholesky factor is replaced by a
nonlinear KR rearrangement.
In particular, we %
note that 
the sparsity of $S$ depends on the ordering of the input variables in the same way as the fill-in of the  Cholesky factor does. %
Different orderings can yield different sparsity patterns.
See Figure \ref{fig:partition_badOrder} ({\it right}) for an example.
Whenever possible,
we then %
look for orderings that promote sparsity 
(cf.~\cite{spantini2017inference} for
additional details).

In many cases, 
$\pi$ is only
{\it approximately} Markov with respect to a
sparse graph $\Gcb$, e.g., there might exist
a second
density 
$\pi'$ which is Markov with respect to $\Gcb$ and 
such that $\Dkl(\pi' \vert\vert \pi) < \varepsilon$, for some $\varepsilon>0$.
In particular,
we know that the filtering
marginals
will not be ``exactly'' Markov with respect to a sparse 
$\Gcb$, 
since conditional independence properties do not persist over time for general models \cite{koller2009probabilistic}.
Yet, when filtering spatially distributed processes (e.g., in data assimilation) it is reasonable to expect that the filtering (or forecast) marginals will be  
well-approximated by sparse MRFs, since the latter
encode probabilistic models with local or
neighboring interactions.
For example, in Section \ref{sec:numerics} we consider the Lorenz-96 model, which arises from the discretization of a one-dimensional PDE on a periodic domain.
In that case, we can expect that the filtering marginal is %
approximately Markov with respect to a cycle
graph with a low-degree of connectivity between the nodes. %
When dealing with two-dimensional PDEs, it is
natural to consider grid graphs \cite{ueno2009covariance,lindgren2011explicit}, or more general lattices on the plane, and so on for higher dimensional domains.
In practice, we can 
then 
sparsify the estimator
$\widehat{S}$ according to a graph $\Gcb$, which implies Markov properties for $\pi$ that are only
approximately satisfied, and treat
the exact geometry of the graph---e.g., its neighborhood structure---as a calibration parameter of the algorithm, as if it were a  localization radius.
This is equivalent to 
regularizing the estimation of the KR rearrangement by
projecting $\pi$ onto a manifold of sparse MRFs
\cite{koller2009probabilistic}.
\begin{figure}[!ht]
\centering
  \newcommand{\wStarGraph}{0.1cm} %
\newcommand{\scaleStarGraph}{.5} %
\newcommand{\scaleMatrix}{.4}
\newcommand{\wMatrix}{0.8cm}

\subfloat[$\Gcb^5 \coloneqq \Gcb$]{
\begin{tikzpicture}[transform shape, scale = \scaleStarGraph]
	\node[default]  (center) 										{$3$};
 	\node[default]	 (belowCenter)		  		[below=of center]		{$2$} 
 		edge[edgeStyle] (center);
 	\node[default] (aboveCenter)			 		[above=of center]		{$5$}
 		edge[edgeStyle] (center);
 	\node[default] (rightCenter)		  			[right=of center]			{$1$}
 		edge[edgeStyle] (center);
	\node[default] (leftCenter)		  			[left=of center]			{$4$}
 		edge[edgeStyle] (aboveCenter)
 		edge[edgeStyle] (center);		
\end{tikzpicture}
} \hspace{\wStarGraph}
\subfloat[$\Gcb^4$]{
\begin{tikzpicture}[transform shape, scale = \scaleStarGraph]
	\node[default]  (center) 										{$3$};
 	\node[default]	 (belowCenter)		  		[below=of center]		{$2$} 
 		edge[edgeStyle] (center);
 	\node[defaultBlank] (aboveCenter)			 		[above=of center]		{\textcolor{white}{$5$}};
 	\node[default] (rightCenter)		  			[right=of center]			{$1$}
 		edge[edgeStyle] (center);
	\node[default] (leftCenter)		  			[left=of center]			{$4$}
 		edge[edgeStyle] (center);		
\end{tikzpicture}
} \hspace{\wStarGraph}
\subfloat[$\Gcb^3$]{
\begin{tikzpicture}[transform shape, scale = \scaleStarGraph]
	\node[default]  (center) 										{$3$};
 	\node[default]	 (belowCenter)		  		[below=of center]		{$2$} 
 		edge[edgeStyle] (center);
 	\node[defaultBlank] (aboveCenter)			 		[above=of center]		{\textcolor{white}{$5$}};
 	\node[default] (rightCenter)		  			[right=of center]			{$1$}
 		edge[edgeStyle] (center);
	\node[defaultBlank] (leftCenter)		  			
	[left=of center]			{\textcolor{white}{$4$}};
\end{tikzpicture}
} \hspace{\wStarGraph}
\subfloat[$\Gcb^2$]{
\begin{tikzpicture}[transform shape, scale = \scaleStarGraph]
	\node[defaultBlank]  (center) 										{\textcolor{white}{$5$}};
 	\node[default]	 (belowCenter)		  		[below=of center]		{$2$} ;
 	\node[defaultBlank] (aboveCenter)			 		[above=of center]		{\textcolor{white}{$4$}};
 	\node[default] (rightCenter)		  			[right=of center]			{$1$}
 		edge[edgeStyle] (belowCenter)	;	
	\node[defaultBlank] (leftCenter)		  			[left=of center]			{\textcolor{white}{$3$}};
\end{tikzpicture}
}
\hspace{\wMatrix}
\subfloat[$\partial_{j}S^k$]{
\begin{tikzpicture}[transform shape, scale = \scaleMatrix ]
\matrix [draw ,column sep=.15cm, row sep=.15cm, ampersand replacement=\& ]
	{
\node[entriesMatrix] { };\& \node[entriesMatrixZeroDashed]{ };\& \node[entriesMatrixZeroDashed]{ }; \& \node[entriesMatrixZeroDashed] { }; \& \node[entriesMatrixZeroDashed] { };   \\
\node[entriesMatrix] { };\& \node[entriesMatrix]{ };\& \node[entriesMatrixZeroDashed]{ }; \& \node[entriesMatrixZeroDashed] { }; \& \node[entriesMatrixZeroDashed] { };   \\
\node[entriesMatrix] { };\& \node[entriesMatrix]{ };\& \node[entriesMatrix]{ }; \& \node[entriesMatrixZeroDashed] { }; \& \node[entriesMatrixZeroDashed] { };   \\
\node[entriesMatrixZeroDashed] { };\& \node[entriesMatrixZeroDashed]{ };\& \node[entriesMatrix]{ }; \& \node[entriesMatrix] { }; \& \node[entriesMatrixZeroDashed] { };   \\
\node[entriesMatrixZeroDashed] { };  \& \node[entriesMatrixZeroDashed]{ };     \& \node[entriesMatrix]{ }; \& \node[entriesMatrix] { }; \& \node[entriesMatrix] { };    \\
	};
\end{tikzpicture}
}
        \caption[]{Here, we assume that $\pi$ is Markov with respect to the leftmost graph $\Gcb$, and show the sequence of marginals graphs $(\Gcb^k)$ 
        defined
        in Section \ref{sec:reg_map}.
        We then represent the sparsity of the corresponding KR rearrangement $S$ using symbolic matrix notation: the $(j,k)$-th entry of the matrix is {\it not} colored if
the $k$th component of the map does not depend on the $j$th input variable.
        }
\label{fig:marginalGraph}
\vspace{-10pt}
\end{figure}

\begin{figure}[!ht]
\centering
  \newcommand{\wStarGraph}{0.4cm} %
\newcommand{\wStarGraphtwo}{1cm} %
\newcommand{\scaleStarGraph}{.5} %
\newcommand{\fontsizelabelsgraph}{\normalsize} %
\newcommand{\scaleMatrix}{.3} %

\subfloat{
\begin{tikzpicture}%
\begin{scope}[transform shape, scale = \scaleStarGraph]
	\begin{pgfonlayer}{bBack}
		\node[defaultBlank]  (center) {};	
		\node[defaultBlank] (dxCenter)	  			[right=of center]			{};		
	\end{pgfonlayer}		
									
 	\node[default]	 (blCenter)		  		[below=of center]		{} 
 		;
 	\node[default] 	(abCenter)	 		[above=of center]		{}
 		;%
 		
	\node[default] 	(sxCenter)  			[left=of center]			{}
 		edge[edgeStyle] (abCenter)
 		edge[edgeStyle] (blCenter);

 	\begin{pgfonlayer}{bBack}
		\node[defaultBlank]  (absxCenter) [above=of sxCenter] {};
		\node[defaultBlank]  (blsxCenter) [below=of sxCenter] {};
	\end{pgfonlayer}

	\node[default] (dxdxCenter) 	  			[right=of dxCenter]			{}
 		;
	\node[default] 	(abdxCenter)  			[above=of dxCenter]			{}
 		edge[edgeStyle] (abCenter)
 		edge[edgeStyle] (dxdxCenter); 	
	\node[default] 	(bldxCenter)  			[below=of dxCenter]			{}
 		edge[edgeStyle] (abdxCenter)
 		edge[edgeStyle] (blCenter)
 		edge[edgeStyle] (abCenter)
 		edge[edgeStyle] (dxdxCenter);

 	\begin{pgfonlayer}{bBack}
		\node[defaultBlank]  (abdxdxCenter) [above=of dxdxCenter] {};
		\node[defaultBlank]  (bldxdxCenter) [below=of dxdxCenter] {};
	\end{pgfonlayer}	 	 
  \end{scope}

	 \begin{pgfonlayer}{background}
		\node [fill=\colorA, inner sep=.15cm, 
		rounded corners, 
				label={[black,font=\fontsizelabelsgraph]above:$\Aset$}, 
			  fit=(blsxCenter)(absxCenter)] (setA) {};
		\node [fill=\colorB, inner sep=.15cm, rounded corners, 
				label={[black,font=\fontsizelabelsgraph]above:$\Bset$}, 
			  fit=(abdxdxCenter)(bldxCenter)] (setB) {};	
		\node [fill=\colorS, inner sep=.15cm, rounded corners, 
				label={[black,font=\fontsizelabelsgraph]above:$\Sset$}, 
			  fit=(abCenter)(blCenter)] (setS) {};
	\end{pgfonlayer}	
\end{tikzpicture}
}
\hspace{\wStarGraphtwo}
\subfloat[$\Gcb'$]{
\begin{tikzpicture}[transform shape, scale = \scaleStarGraph]
	\node[default]  (center) 										{$5$};
 	\node[default]	 (belowCenter)		  		[below=of center]		{$2$} 
 		edge[edgeStyle] (center);
 	\node[default] (aboveCenter)			 		[above=of center]		{$3$}
 		edge[edgeStyle] (center);
 	\node[default] (rightCenter)		  			[right=of center]			{$1$}
 		edge[edgeStyle] (center);
	\node[default] (leftCenter)		  			[left=of center]			{$4$}
 		edge[edgeStyle] (aboveCenter)
 		edge[edgeStyle] (center);		
\end{tikzpicture}
} \hspace{\wStarGraph}
\subfloat[$\partial_{j}S^k$]{
\begin{tikzpicture}
\begin{scope}[transform shape, scale = .01 ]
\matrix [draw ,column sep=.15cm, row sep=.15cm, ampersand replacement=\& ]
	{
\node[entriesMatrix] { };\& \node[entriesMatrixZeroDashed]{ };\& \node[entriesMatrixZeroDashed]{ }; \& \node[entriesMatrixZeroDashed] { }; \& \node[entriesMatrixZeroDashed] { };   \\
\node[entriesMatrix] { };\& \node[entriesMatrix]{ };\& \node[entriesMatrixZeroDashed]{ }; \& \node[entriesMatrixZeroDashed] { }; \& \node[entriesMatrixZeroDashed] { };   \\
\node[entriesMatrix] { };\& \node[entriesMatrix]{ };\& \node[entriesMatrix]{ }; \& \node[entriesMatrixZeroDashed] { }; \& \node[entriesMatrixZeroDashed] { };   \\
\node[entriesMatrix] { };\& \node[entriesMatrix]{ };\& \node[entriesMatrix]{ }; \& \node[entriesMatrix] { }; \& \node[entriesMatrixZeroDashed] { };   \\
\node[entriesMatrix] { };  \& \node[entriesMatrix]{ };     \& \node[entriesMatrix]{ }; \& \node[entriesMatrix] { }; \& \node[entriesMatrix] { };    \\
	};
\end{scope}	
\end{tikzpicture}
} 
        \caption[]{({\it left}) Example of a graphical model for $\Zb$, together with a partition $\Ac,\Sc,\Bc$ of the vertex set, where 
        $\Sc$ is a separator for $\Ac$ and $\Bc$. By
        definition, $\Zb_{\Ac} \orth \Zb_{\Bc} \vert \Zb_{\Sc}$.
        ({\it center}) Re-ordering of the input variables for the
        example of Figure \ref{fig:marginalGraph}, leading to
         a rearrangement $S$ with no sparsity ({\it right}).
        }
        \label{fig:partition_badOrder}
\vspace{-10pt}
\end{figure}

\subsection{Remarks on the stochastic map filter}
\label{sec:rem_pert_obs}
The stochastic map %
algorithm 
of
Section \ref{sec:pert_obs} is a general tool
for Bayesian inference: given $M$ samples
$(\xb^i)$ from the prior, (1) generate 
$M$ samples $\yb^i \sim \pi_{\Yb\vert\Xb}(\cdot \vert \xb^i)$
from the likelihood, (2) use the pairs $(\yb^i,\xb^i)$ to build an estimator $\widehat{T}$ of the
analysis map $T$ as %
\begin{equation} \label{eq:anmap_and_est}
  T \coloneqq 
  S^{\Xcb}(\yb^*, \cdot)^{-1}
  \circ 
  S^{\Xcb},
  \qquad
  \widehat{T} \coloneqq 
  \widehat{S}^{\Xcb}(\yb^*, \cdot)^{-1}
  \circ 
  \widehat{S}^{\Xcb},
\end{equation}
and (3) obtain a particle approximation of the posterior
as $\widehat{T}(\yb^1, \xb^1),\ldots,
\widehat{T}(\yb^M, \xb^M)$. 

A few remarks are in order.
First, we only need to {\it sample} %
the 
likelihood function 
and do not need explicit access to %
the density $\pi_{\Yb\vert\Xb}$, which might %
be
intractable or expensive to evaluate. 
We are thus in the setting of approximate Bayesian computation (ABC) methods \cite{Marin2011} or,  
more broadly,   %
inference in \emph{generative models}. 
Second, we can compute %
$\smash{\widehat{T}}$ 
using %
{\it convex} optimization:
we only 
need to estimate
a few components, $\smash{S^{\Xcb}}$, of the KR 
rearrangement $S$ that pushes forward 
$\pi_{\Yb,\Xb}$ to a standard normal, using
the constrained maximum likelihood estimator
defined in \eqref{eq:mle_KR}.
In particular, the block $\smash{S^{\Ycb}}$  defined in \eqref{eq:blockS} is not 
needed, while 
the components of $\smash{\widehat{S}^{\Xcb}}$ can be
computed {in parallel} using \eqref{eq:component_estimator}. %
Third, 
the estimation of $\smash{S^{\Xcb}}$ in high dimensions can %
be regularized using the {localization} techniques of
Section \ref{sec:reg_map}.
(For the purpose of
posterior sampling, we may only care about the bias-variance tradeoff in the estimation of 
$T$, and not necessarily that of 
$\smash{S^{\Xcb}}$.)
Fourth, if we constrain the estimator 
$\smash{\widehat{S}^{\Xcb}}$ to be linear, then
we recover an EnKF with %
``perturbed observations,'' i.e., the stochastic EnKF \cite{burgers1998analysis}.

\begin{remark}[Connection with the EnKF]
Assume that the samples $(\yb^i, \xb^i)$ have been
centered, so that they have zero mean.
Then, the best {\it linear} estimator of $\smash{S^{\Xcb}}$---according to \eqref{eq:mle_KR}---is %
given by 
\begin{equation}
\widehat{S}^{\Xcb}(\yb, \xb) = 
\Ab \left( \xb - 
\widehat{\Sigmab}_{\Xb,\Yb} \widehat{\Sigmab}_{\Yb}^{-1}
\yb
\right),
\end{equation}
where $\Ab$ is the inverse of the Cholesky factor of the ``residual'' covariance matrix,
\begin{equation} 
  \widehat{\Sigmab}_{\Xb \vert \Yb}
  \coloneqq 
  \widehat{\Sigmab}_{\Xb} - 
  \widehat{\Sigmab}_{\Xb,\Yb}\,
  \widehat{\Sigmab}_{\Yb}^{-1}\,
  \widehat{\Sigmab}_{\Xb,\Yb}^\top, 
\end{equation}
where $\widehat{\Sigmab}_{\Xb}$, 
$\widehat{\Sigmab}_{\Yb}$, and 
$\widehat{\Sigmab}_{\Xb,\Yb}$ are empirical
(maximum likelihood) estimators of 
$\Var(\Xb)$, $\Var(\Yb)$, and $\Cov(\Xb,\Yb)$,
respectively.\footnote{In the %
case
of a linear--Gaussian observation model, e.g., if
$\Yb = \Hb \Xb + \Ec$ for
$\Ec \sim \Gauss({\bf 0}, \Rb)$ independent of $\Xb$, and with $\Hb$ known,
we could %
regularize the estimation of
$\Cov(\Xb, \Yb)$ and $\Var(\Yb)$ 
by letting  %
$\widehat{\Sigmab}_{\Xb,\Yb}\coloneqq 
\widehat{\Sigmab}_{\Xb}\Hb^\top$ and
$\widehat{\Sigmab}_{\Yb} = (
\Hb \widehat{\Sigmab}_{\Xb} \Hb^\top + \Rb)
$.
}
For a given observation $\yb^*$ of the data, we can then
compute a closed form expression for the inverse
of the map $\xib \mapsto \smash{\widehat{S}^{\Xcb}}(\yb^*, \xib)$ and for the estimator of the analysis map,
\begin{equation}
\label{eq:enkf_pert}
  \widehat{T}(\yb,\xb) = \xb -
  \widehat{\Sigmab}_{\Xb,\Yb} \widehat{\Sigmab}_{\Yb}^{-1} \left( \yb - \yb^* \right),
\end{equation}
which corresponds to the forecast-to-analysis update
of the stochastic EnKF \cite{saetrom2011ensembleB}. %
The 
stochastic map filter %
offers a framework to 
generalize this linear ansatz to more general 
nonlinear transformations.
\end{remark}
At any assimilation time, we need to condition the
forecast distribution on $d$ scalar
observations $\yb^* = (y^*_1,\ldots,y^*_d)$
from $\Yb$. In Section \ref{sec:pert_obs}, we 
presented
an algorithm that can assimilate all these observations 
simultaneously. %
A possible %
alternative if the observations are
conditionally independent given the state, 
i.e.,
\begin{equation}
 \pi_{\Yb \vert \Xb} = \prod_{k=1}^d 
 \pi_{Y_k \vert \Xb},
\end{equation}
is to
process each scalar observation $y^*_k$ individually  and sequentially by iterating $d$ times over Algorithm \ref{alg:analysis_stoc_maps}: %
the posterior ensemble associated with $y^*_k$ 
can be used as a prior
ensemble
when assimilating
$y^*_{k+1}$.
Hence, the analysis step can 
be either performed by computing $d$ analysis maps on $\smash{\re^{n+1}}$---each map associated with one scalar observation---or by computing a single map on $\smash{\re^{n+d}}$
(cf.~the EnKF with single versus multiple observations
\cite{houtekamer2001sequential}).
\begin{remark}[Localize the analysis map for a scalar observation with local likelihood]
\label{rem:localization}
When assimilating a scalar observation (or a small batch thereof) it makes sense to consider
an additional form of {\it localization} for
$\smash{\widehat{S}^{\Xcb}}$ 
with respect to those already introduced in Section \ref{sec:reg_map}.
The idea is that the analysis map $T$ should
revert to the identity function along components that
correspond to unobserved variables that are far from the
observed ones, for instance with respect to the metric
$\db(\cdot,\cdot)$ defined in Section \ref{sec:reg_map}. We can enforce this identity
structure in the estimator $\smash{\widehat{T}}$ by
requiring that some components of $\smash{\widehat{S}^{\Xcb}}$ be the identity function.
Specifically, assume that we are observing a
single scalar observation $Y$ which follows a 
{\it local}
likelihood model, i.e., $\pi_{Y\vert\Xb} = \pi_{Y \vert X_\ell}$ for some
scalar state variable $X_\ell$.
After a {\it permutation}, we can assume that 
$X_\ell = X_1$ and that the state variables have 
been reordered %
in terms
of their increasing 
distance from $X_1$.
We can then look for an estimator 
$\smash{\widehat{S}^{\Xcb}}$ with the following
sparsity pattern,\footnote{One can easily 
{\it increase} this sparsity pattern using the localization ideas
of Section \ref{sec:reg_map}.} 
\begin{equation} 
\label{eq:sparse_analysis_KR}
\widehat{S}^{\Xcb}( y, \xb ) = \left[\begin{array}{l}
s(y,\xb_{1:j})\\[2pt] 
x_{j+1} \\
\vdots \\
x_n
\end{array}\right], \qquad
s:\re \times \re^j \ra \re^j,%
\end{equation}
for some index $j$ that represents the analogue of a 
``localization'' radius for EnKF algorithms.
The effect of \eqref{eq:sparse_analysis_KR} is to produce an estimator $\smash{\widehat{T}}$
for the analysis map  that reverts to the identity 
function after $j$ components, thus producing a truly
{\it local} coupling.
Note that the localization in 
\eqref{eq:sparse_analysis_KR} is only justified because we use $\smash{\widehat{S}^{\Xcb}}$ within the expression for $\smash{\widehat{T}}$ in 
\eqref{eq:anmap_and_est}.
We do not care if 
$\smash{\widehat{S}^{\Xcb}}$ is a poor estimator of  the rearrangement $\smash{S^{\Xcb}}$ as a whole,
as long as the resulting $\widehat{T}$ given by
\eqref{eq:anmap_and_est} is
a good estimator of the analysis map.
\end{remark}

\begin{remark}[Transform rather than sample]%
The estimation of $\smash{S^{\Xcb}}$ by $\smash{\widehat{S}^{\Xcb}}$ yields an implicit approximation of the posterior distribution: by definition, the mapping
$\xib \mapsto \smash{S^{\Xcb}}(\yb^*, \xib)$ pushes forward $\pi_{\Xb \vert \yb^*}$ to a standard normal.
In principle, we could then generate approximate posterior samples by pushing forward samples
from a standard normal $\Gauss({\bf 0}, {\bf I}_{n})$ through the inverse of the map
$\xib \mapsto \smash{\widehat{S}^{\Xcb}}(\yb^*, \xib)$.
In practice, however, we do not do this; rather, we
use $\smash{\widehat{S}^{\Xcb}}$ to build an estimator 
$\smash{\widehat{T}}$ of the prior-to-posterior update.
This is in analogy with the EnKF, where the posterior ensemble is given by a (linear) transformation applied to the prior ensemble, rather than by samples from a Gaussian approximation of the posterior.

In numerical experiments---where transformations can only be {\it  approximate}---we observe that pushing forward samples from $\pi_{\Yb,\Xb}$ through $\smash{\widehat{T}}$ often yields far more accurate posterior approximations than pushing standard normal samples through $\widehat{S}^{\Xcb}(\yb^*,\cdot)^{-1}$. This improvement can be attributed to the cancellation of errors in the composition of $\smash{\widehat{S}^{\Xcb}}$ with its partial inverse. As a limiting but nonetheless illustrative example: when the data are uninformative and $S^{\Xcb}$ does not depend on $\yb$, the Gaussian samples mapped through the inverse of the estimated map will be affected by bias and variance in $\smash{\widehat{S}^{\Xcb}}$, while $\widehat{T}$ will be the identity function and generate exact posterior samples. 
{More generally, consider any map $\smash{\widehat{S}^{\Xcb}}$ that renders $\Zb = \smash{\widehat{S}^{\Xcb}}(\Yb,\Xb)$ independent of $\Yb$; one can show that the transformation $\widehat{T}$ built from this map will again generate exact posterior samples.} %
\end{remark}
We conclude this section with a final remark.
We presented the stochastic map filter
using
the KR rearrangement as our building-block coupling, because the latter can be
easily approximated in high dimensions using 
nonlinear transformations and convex optimization.
Yet the algorithm would still make sense if we were
to replace the KR rearrangement with any other 
transport map $S$ that has the ``block'' structure in \eqref{eq:blockS}, 
and that  pushes forward 
$\pi_{\Yb,\Xb}$ to $\Gauss({\bf 0}, {\bf I}_{d+n})$.

\subsection{Deterministic map filter}
\label{sec:square-root}
In this section, we 
explore the use of deterministic couplings 
in 
the analysis step 
of Section \ref{sec:analysis}. %
In particular, we seek a transport map $T:\re^n \ra \re^n$ that pushes forward prior to posterior, i.e., such that 
$T(\Xb) \sim \pi_{\Xb \vert \yb^*}$.
See Figure \ref{fig:det_coupling} for an illustration.
(Note that in the stochastic map filter of
Section \ref{sec:pert_obs}, the map $T$ was defined
on $\re^{n+d}$ rather than on $\re^n$.)

For all monotone triangular maps $U$ on $\re^n$, define $\pi_{U}$ to be the density %
that is proportional to the function  $\xib \mapsto \pi_{\Yb\vert
  \Xb}(\yb^*\vert \xib)\,U\pull\,\eta(\xib)$, which is the product of
the likelihood and the pullback density $U\pull\,\eta$. 
Now, let $S$ be the KR rearrangement that pushes forward the prior
$\pi_{\Xb}$ to $\eta \coloneqq \Gauss({\bf 0}, {\bf I}_n)$. Hence, the pullback density of $\eta$ by
$S$---see the definition in \eqref{eq:def_pullback}---is 
the prior, i.e., $S\pull \,\eta = \pi_{\Xb}$. It follows that $\pi_{S}$ is the posterior density $\pi_{\Xb \vert \yb^*}$.

We define a candidate %
analysis map $T$ as follows:
\begin{equation} \label{eq:map_det}
  T \coloneqq \Tc \circ S,
\end{equation}
where
$\Tc$ is a KR rearrangement that pushes
forward $\eta$ to $\pi_S$. 
It is immediate to verify that \eqref{eq:map_det} pushes forward
$\pi_{\Xb}$ to $\pi_{\Xb \vert \yb^*}$, and thus defines a valid
deterministic coupling between the prior and posterior distributions.

We propose to estimate $T$ by %
\begin{equation} \label{eq:est_an_map_det}
  \smash{\widehat{T}} \coloneqq 
  \widehat{\Tc} \circ \widehat{S},
\end{equation}
where $\widehat{S}$ and $\smash{\widehat{\Tc}}$ are (constrained) maximum likelihood
estimators of two different  KR rearrangements:
the former %
pushes forward 
$\pi_{\Xb}$ to $\eta$, 
while 
the latter 
pushes forward 
$\eta$ to $\pi_{\widehat{S}}$.
Intuitively,
$\pi_{\widehat{S}}$ %
is an approximation of the posterior distribution, which depends on
the estimator $\smash{\widehat{S}}$.
We can compute $\smash{\widehat{S}}$ efficiently via convex optimization from a collection of prior samples
$(\xb^i) \sim \pi_{\Xb}$ using the construction of
Section \ref{sec:est_KR} and the regularization
techniques of Section \ref{sec:reg_map}.
The computation of ${\widehat{\Tc}}$, however,
is less straightforward, as we explain in the next remark. 
In particular, note that we do not have samples from
$\pi_{\widehat{S}}$; instead, given $\widehat{S}$, we can only evaluate this density up to
a normalizing constant, as long as the likelihood
$\pi_{\Yb \vert \Xb}(\yb^*\vert \cdot) $ can be %
evaluated
explicitly.

\begin{remark}[Estimation of the Knothe--Rosenblatt rearrangement from densities]
${\widehat{\Tc}}$ is a maximum likelihood estimator
of a KR rearrangement that pushes forward a
standard normal 
$\Gauss({\bf 0}, {\bf I}_n)$
to a density $\pi \coloneqq \pi_{\widehat{S}} $  on $\re^n$, i.e., ${\widehat{\Tc}}$ is defined 
analogously to \eqref{eq:mle_KR} as %
\begin{equation} \label{eq:mle_KR_direct}
\widehat{\Tc} \in \arg \max_{U \in \Hcb} \,
\frac{1}{N} \sum_{i=1}^N \,\log U\pull \, \pi(\zb^i), 
\end{equation}
for a collection of $N$ samples $(\zb^i)_{i=1}^N$ from
$\Gauss({\bf 0}, {\bf I}_n)$ and for some
approximation space $\Hcb$.
There is an important difference between
\eqref{eq:mle_KR} and \eqref{eq:mle_KR_direct}:
the density $\eta$ in \eqref{eq:mle_KR} is
always
log-concave, while $\pi$ in \eqref{eq:mle_KR_direct} need not be.
As a result, 
\eqref{eq:mle_KR_direct} is in general {\it not}
a convex problem \cite{Kim2013}.
Moreover, we cannot compute
the components of ${\widehat{\Tc}}$ independently as in
\eqref{eq:component_estimator}, since 
$\pi$ need not factorize as a product of its marginals.
Hence, the computation of ${\widehat{\Tc}}$ is
inherently harder than that of ${\widehat{S}}$, but
still feasible. %
The numerical
solution of \eqref{eq:mle_KR_direct} in the
context of Bayesian inference was pioneered by
\cite{el2012bayesian},
which uses gradient-based optimization (e.g., BFGS or Newton-CG \cite{wright1999numerical}) to minimize
an equivalent reformulation of \eqref{eq:mle_KR_direct},
\begin{equation} \label{eq:mle_KR_direct_expl}
\widehat{\Tc} \in \arg \min_{U \in \Hcb} \,
 - \frac{1}{N} \sum_{i=1}^N \,
 \left( \log \, \pi(U(\zb^i))
 + \sum_{k=1}^n \log \partial_k U^k(\zb^i)
 \right), 
\end{equation}
over a finite dimensional space of polynomial maps
(see Appendix \ref{sec:param_tri} and
\cite{marzouk2016introduction,bigoni2016monotone}).
Three aspects of \eqref{eq:mle_KR_direct_expl} are
particularly important.
First,
we only need to evaluate $\pi_{\widehat{S}}$
up to a normalizing constant. 
Second, %
the number $N$ of samples %
from $\Gauss({\bf 0}, {\bf I}_n)$ 
need not be limited by the cardinality of the
forecast ensemble, 
in contrast to \eqref{eq:mle_KR}.
Third, we must evaluate the likelihood function $\pi_{\Yb|\Xb}$ for each sample in \eqref{eq:mle_KR_direct_expl}. The latter is a crucial contrast with the stochastic map filter of Section~\ref{sec:pert_obs}, which only requires \textit{sampling} from (i.e., simulating) the likelihood.
\end{remark}

In the special case of a Gaussian likelihood $\pi_{\Yb\vert \Xb}$ and linear 
$\widehat{S}$, the estimator
$\pi_{\widehat{S}}$ of the posterior distribution 
becomes
Gaussian.
The corresponding 
KR rearrangement that pushes
forward $\Gauss({\bf 0}, {\bf I}_n)$ %
to $\pi_{\widehat{S}}$ is %
{\it linear}, meaning that %
without loss of generality
we can solve
\eqref{eq:mle_KR_direct_expl} over the space
of linear maps. %
In this case, \eqref{eq:mle_KR_direct_expl} is
a convex (quadratic) problem with a closed
form expression for %
$\smash{\widehat{\Tc}}$,
and %
the deterministic map filter %
reduces
to a particular 
ensemble square-root filter \cite{tippett2003ensemble}.

If either $\pi_{\Yb\vert \Xb}$ is non-Gaussian or
$\smash{\widehat{S}}$ is nonlinear---e.g., if we seek a
non-Gaussian approximation of the prior
distribution---then 
the linear ansatz for $\smash{\widehat{\Tc}}$ is no longer optimal.
In such cases, 
we might want 
to capture non-Gaussian
structure
by considering
parameterizations of
$\smash{\widehat{\Tc}}$ that depart
{gradually} and incrementally from that of
a linear function, 
seeking
a balance
between %
accuracy
and computational cost 
(see Section \ref{sec:est_KR}).
In particular, we can regularize the
estimation of the rearrangement in
high dimensions %
using %
localization ideas similar\footnote{Most of the discussion of Section \ref{sec:reg_map}
is specific to a KR rearrangement whose
target distribution is a product measure on
$\re^n$---e.g., 
$\Gauss({\bf 0}, {\bf I}_n)$---and thus does not apply to the estimation of $\Tc$.
See \cite{spantini2017inference} for additional
details.
}
to those of Section \ref{sec:reg_map}.
We propose to use a ``banded'' estimator
analogous to 
\eqref{eq:threshold_est} to leverage
decay of dependence in $\pi$, i.e.,
\begin{equation}  
\label{eq:threshold_est_direct} 
\widehat{\Tc}( \zb ) = \left[\begin{array}{l}
\widehat{\Tc}^1(\zb_{\Ac_{1,r}})\\[2pt] 
\vdots \\ 
\widehat{\Tc}^{n}(\zb_{\Ac_{n,r}}) \\
\end{array}\right], %
\end{equation}
for some value of the localization radius
$r\ge0$.
For a Gaussian $\pi$, 
\eqref{eq:threshold_est_direct} corresponds to a
banded estimator for the Cholesky factor of 
the {\it covariance} matrix of $\pi$, 
rather than its precision %
(cf.~Section \ref{sec:reg_map}).

We summarize the %
analysis step with deterministic maps in Algorithm \ref{alg:analysis_det_maps}.
Note that
there are many possible variations for the definition
of the prior-to-posterior update \eqref{eq:map_det}.
Most notably, $\smash{\widehat{\Tc}}$ need not 
approximate
a KR rearrangement: 
in fact any map that pushes forward $\eta$ to $\pi_S$ would work,
e.g.,
maps induced by the flows of ODEs 
\cite{daum2008particle,anderes2012general,heng2015gibbs}
or 
by the composition of many simple functions \cite{liu2016stein,detommaso2018stein}, including deep neural networks \cite{rezende2015variational}.

\begin{remark}[Connection with the
variational mapping particle filter]
The variational mapping particle filter
\cite{pulido2018kernel}
uses the nonparametric variational inference algorithm
proposed by \cite{liu2016stein} to
generate approximate samples from a
non-Gaussian approximation of the 
filtering distribution.
The idea is to minimize precisely
\eqref{eq:mle_KR_direct}---a non-convex functional---by
implementing
a form of functional gradient descent on a reproducing kernel Hilbert space (RKHS) of transformations (with no restrictions to triangular maps).
The non-Gaussian approximation of the
unnormalized
filtering distribution is defined by the product of
the likelihood function with an approximation of
the forecast distribution at time $k$ given by\footnote{For ease of exposition, 
we assume that the dynamic of the latent field is only 
given %
at the observation times.}
\begin{equation}
  \pi_{\Zb_{k+1} \vert \yb_{1:k}} 
  \approx
  \frac{1}{M}\,\sum_{i=1}^M
  \pi_{\Zb_{k+1} \vert \Zb_{k}}(\cdot \vert \zb^i),
\end{equation}
where $(\zb^i)$ represent (approximate) samples from the
previous filtering marginal 
$\pi_{\Zb_{k} \vert \yb_{1:k}}$. This approximation of the forecast density contrasts with the 
{\it regularized} maximum likelihood estimation 
of the KR rearrangement
$S$ in \eqref{eq:map_det} 
from the forecast ensemble, 
using 
\eqref{eq:mle_KR}.
Indeed, it would be %
interesting to
approximate the rearrangement 
$\smash{\widehat{T}}$ 
in the deterministic map algorithm using the
technique of
\cite{liu2016stein}
(or its second-order extension \cite{detommaso2018stein}), after %
``localizing'' %
the RKHS of transformations by imposing a
sparsity pattern like the one in
\eqref{eq:threshold_est_direct}.
Regularization is critical in high dimensions.
\end{remark}
The deterministic map filter %
presented in this
section features an inherently non-convex step.
This is an important distinction from the
stochastic map filter of Section \ref{sec:pert_obs}, which 
relies 
solely
on
convex optimization  
(at the price of computing a higher dimensional
transformation).
In Appendix \ref{sec:detmap_local},
we discuss a common structural
assumption on the state-space model 
that allows us to easily
bypass non-convex optimization in the
deterministic map algorithm, leading to a
scheme with close ties to the
multivariate rank histogram filter \cite{metref2014non}.
\begin{algorithm} %
  \linespread{\spalg}\selectfont
  \caption{
  {\bf (Analysis step with deterministic maps)} \\
  Given $M$ samples $\xb^1,\ldots,\xb^M$
  from the forecast distribution, a likelihood function
  $\pi_{\Yb\vert\Xb}$,
  and an observation
  $\yb^*$, 
 generate a particle approximation 
  $\zb^1,\ldots,\zb^M$ of the filtering
  distribution $\pi_{\Xb \vert \yb^*}$.
  Let $\eta \coloneqq \Gauss({\bf 0}, {\bf I}_n)$.
  }
  \label{alg:analysis_det_maps}
  \begin{algorithmic}[1]
    \Procedure{AnalysisDeterministic}{
    $\yb^*$, $\pi_{\Yb \vert \Xb}$,   
    $\xb^1, \ldots,\xb^M$ }%
    \State 
    $\widehat{S} \gets$ 
    estimator of
    $S$ from $(\xb^i)_{i=1}^M$
    given by \eqref{eq:mle_KR} %
    \Comment $S\push\,\pi_{\Xb} = 
    \eta$ 

    \State Define $\pi_{\widehat{S}}$ as 
    $\pi_{\widehat{S}}(\xib) \propto \pi_{\Yb\vert \Xb}(\yb^*\vert \xib)\,\widehat{S}\pull\,\eta(\xib)$

    \State 
    $\widehat{\Tc} \gets$ 
    estimator of
    $\Tc$ from $\pi_{\widehat{S}}$
    given by \eqref{eq:mle_KR_direct_expl}
    \Comment $\Tc \push \,\eta = 
    \pi_{\widehat{S}}$ 
    \For{$i\gets 1$ to $M$} 
    \Comment Action of $\widehat{T}$ as defined in
    \eqref{eq:est_an_map_det}
    \State 
    $\zb^i \gets \widehat{S}(\xb^i)$
    \State $\zb^i \gets \widehat{\Tc}(\zb^i)$ 
    \EndFor
    \State \Return 
    $\zb^1,\ldots,\zb^M$ 
    \EndProcedure
  \end{algorithmic}
\end{algorithm}

\section{Numerical experiments}
\label{sec:numerics}
In this section, we numerically investigate the performance of the stochastic map
filter over a range of examples.
We show that nonlinear configurations of the new filter can have
tracking performance consistently superior to that of the EnKF. We also show that
the stochastic map filter provides better approximations of the
true (Bayesian) filtering distribution.
We focus on the stochastic map filter because it can be implemented
using only convex optimization, regardless of any
nonlinear/non-Gaussian structure in the state-space model.
Numerical experiments with earlier versions of the deterministic map
filter can be found in related work by the authors \cite[Ch.~6]{spantini2017inference}, for the special case of
conditionally independent and local observations (see Appendix \ref{sec:detmap_local}).
Many aspects of our problem setup and our performance assessment follow \cite{lei2011moment}, as detailed below.

\subsection{Common aspects of the problem configurations}
\label{sec:commonsetup}

First we discuss aspects of the numerical experiments that are shared
among the test problems of Sections~\ref{sec:lorenz63}--\ref{sec:bickel_hardcase} and Appendix~\ref{sec:lorenz_heavycase}.

\paragraph{Prior models} 
We consider two dynamical models of increasing dimension: the
Lorenz-63 and the Lorenz-96 models, which are widely used
testbeds for filtering in chaotic dynamical systems. 
The Lorenz-63 model is a three-dimensional system introduced
in~\cite{lorenz1963deterministic} that describes the natural convection of a heated
fluid. The state at time $t$ is a three-dimensional vector $\Zb(t) = (Z_{1}(t),
Z_{2}(t), Z_{3}(t))$ whose dynamics are given
by the ODE system,
\begin{equation} \label{eq:system_lorenz63}
\frac{dZ_{1}}{dt} = -\sigma Z_{1} + \sigma Z_{2}, \;\;\; \frac{dZ_{2}}{dt} = - Z_{1} Z_{3} + \rho Z_{1} - Z_{2}, \;\;\; \frac{dZ_{3}}{dt} = Z_{1}Z_{2} - \beta Z_{3},
\end{equation}
where $(\beta,\rho,\sigma)$ are fixed parameters. In our experiments,
we set $\beta = 8/3$, $\rho = 28$, and $\sigma = 10$, which produces chaotic solutions tending to the well-known Lorenz attractor.

The Lorenz-96 model is a popular testbed for numerical weather
prediction algorithms, reproducing coarse features of the mid-latitude
atmosphere \cite{lorenz1996predictability,majda2012filtering}.  The
state at time $t$ is a 40-dimensional vector,
$\Zb(t)=(Z_1(t),\ldots,Z_{40}(t))$.
The state dynamics are defined by a set of nonlinear ODEs that
represent the spatial discretization of a time-dependent PDE,
i.e., 
\begin{equation} \label{eq:system_lorenz}
  \frac{{\rm d}Z_j}{{\rm d}t} =
  (Z_{j+1} - Z_{j-2}) Z_{j-1} - Z_j + F,
  \qquad
  j=1,\ldots,40,
\end{equation}
with periodic boundary conditions and constant forcing parameter $F$
\cite{reich2015probabilistic}.  In our experiments we use $F=8$, which
leads to a fully chaotic dynamic \cite{majda2012filtering}.

We integrate both ODE systems, \eqref{eq:system_lorenz63} and \eqref{eq:system_lorenz}, using a fourth-order explicit Runge-Kutta method, with constant stepsizes $\Delta t = 0. 05$ and $\Delta t = 0. 01$, respectively. For the Lorenz-63 model only, we also wish to construct a ``reference'' Bayesian filtering solution via a consistent formulation of the particle filter. In this case, we add a small amount of Gaussian noise to the state at each integration step, $\bm{\varepsilon}_{\Delta t} \sim \Gauss(0,10^{-4}\mathbf{I}_{3})$, for the purpose of delaying degeneracy of the particle filter. 
We note that the EnKF and the stochastic map filter remain stable for the Lorenz-63 configurations below
\emph{without} this additive noise. We do not add any noise to the dynamics of the Lorenz-96 model. Therefore, the Markov transition kernel for the Lorenz-96 model is 
intractable.

\paragraph{Likelihood model} 
The state of each system is observed indirectly 
every $\Delta t_{\rm obs}$ 
time units.
Larger values of $\Delta t_{\rm obs}$ are associated with increasingly  non-Gaussian forecast statistics. We index the state at discrete observation times by $(\Zb_{k})_{k \geq 1}$ where $\Zb_{k} = \Zb(k\Delta t_{\text{obs}})$ (e.g., see Figure~\ref{fig:dataAssg}).
For any observation step $k$,  the likelihood model is specified by
\begin{equation}
  \Yb_k = \Hbrm \Zb_k + \Ecb_k,
\end{equation}
where $\Hbrm \in \re^{d \times n}$ is a linear operator that selects $d \leq n$ components of the state at uniform intervals (e.g., every other component, or every fourth component), while $\Ecb_k$ is an additive noise that is independent of $\Zb_k$. In
Sections~\ref{sec:lorenz63} and \ref{sec:bickel_hardcase}, we consider configurations using Gaussian observational noise with zero mean and covariance $\theta^2\mathbf{I}_{d}$. In Section~\ref{sec:lorenz_heavycase}, we employ \emph{heavy-tailed} Laplace
observational noise with zero mean and covariance $2\theta^2\mathbf{I}_{d}$ (i.e., scale parameter $\theta$ for each
dimension).

\paragraph{Simulation setup}
Given a random initial condition $\Zb_{0}$ drawn from 
$\mathcal{N}(\mathbf{0},\mathbf{I}_{n})$, 
we generate a sequence of ``true'' hidden states $(\zb^*_k)_k$ 
and a sequence of synthetic observations 
$(\yb_k^*)_k$, %
with the same model used for filtering. We are thus in the usual ``identical twin experiment'' setting \cite{reich2015probabilistic},
wherein 
the filtering algorithms must
estimate the true state
given only these synthetic observations. 
{In Appendix~\ref{sec:Inverse_crime}, we depart from this setting to demonstrate robustness of the results to 
a mismatch between the discretizations of the data-generating model and the model used for filtering.
}

The initial ensemble for the filtering 
algorithms
is generated through
a 
\emph{spin-up} phase. %
First, we draw $M$ i.i.d.\ random samples of the initial condition,
$\mathcal{N}(\mathbf{0},
\mathbf{I}_{n})$. 
Then, we run the stochastic EnKF---without localization---for $2000$ assimilation steps. 
Using the resulting
ensemble 
at $k=2000$, 
we then apply the filter to be tested (e.g., an EnKF with optimally tuned localization, or %
the stochastic map filter) for an additional $4000$ 
assimilation steps.
 At each step, we use
the 
ensemble mean 
$\bar{\zb}_{k} \coloneqq \frac{1}{M} 
{\sum_{i=1}^{M}} \zb_{k}^{i}$ 
as a point estimate of the
true state $\zb^*_k$. Lastly, to eliminate the transient effect of switching between filtering algorithms, we analyze the assimilation quality only using the last $2000$ assimilation cycles.
We also performed studies with
up to $10^4$ assimilation cycles in this last phase,
but
found 
no significant difference in the results. 

\paragraph{Performance metrics}
We evaluate the tracking performance of the filter via the root-mean-squared error (RMSE), defined at any assimilation step $k$ as $\text{RMSE}_k \coloneqq \|\bar{\zb}_{k} - \zb^*_{k}\|_{2}/\sqrt{n}$. To measure the concentration of the ensemble members, we compute the ensemble spread,  
defined at any assimilation step $k$ as $[\text{tr}(\widehat{\Sigmab}_{k})/n]^{1/2}$, where $\widehat{\Sigmab}_{k}$ denotes the ensemble covariance matrix 
at step $k$. 
In Sections~\ref{sec:lorenz63}--\ref{sec:bickel_hardcase} and \ref{sec:lorenz_heavycase}, we report the \emph{time average} of each of these %
metrics (e.g., over the 2000 test assimilation steps) as a function of the ensemble size $M$.
In addition, we compute the coverage probability of the intervals given
by the empirical $2.5\%$ and $97.5\%$ quantiles of each marginal of
the ensemble, i.e., the frequency with which the $i$th component of
$\zb^*_{k}$ is contained in the $i$th marginal interval. 
For each component, 
this frequency is of the form $k/T$ where $k \in \{0,\dots,T\}$ denotes the number of intervals that contain the corresponding component of the true state, and 
$T$ represents the total number of assimilation times.
We then average these coverage probabilities over the $n$ components
of the state. 
Lastly, evaluations of the ensemble quality based on the \emph{continuous ranked probability score} \cite{gneiting2007probabilistic} are presented in Appendix~\ref{sec:CRPS_results}.

\paragraph{Algorithm parameters}
The results below investigate the performance of the stochastic map filter (Algorithm~\ref{alg:analysis_stoc_maps}) as nonlinearities are gradually introduced in the prior-to-posterior transformation $\widehat{T}$. 

Each component function $U^{k}$ of the map $S^{\Xcb}$ is allowed to depend on the data $y$ and on a subset of the state variables $z_{1},\dots,z_{k}$ given by $\zb_{\Ac_{k,r}}$. 
In particular, our numerical experiments consider a separable representation $U^{k}(y,\zb_{\Ac_{k,r}}) = \ufrakb^k_{0}(y) + \sum_{i \in \Ac_{k,r}} \ufrakb^k_{i}(z_{i})$ for $k=1,\dots,n$. 
For the linear version of the stochastic map filter, labeled as ``linear'' in the figures, $\ufrakb^k_{0}$ and $\ufrakb^k_{i}$ are restricted to be linear functions. 
To increase the complexity of the map, we add radial basis functions (RBFs) to the approximation spaces for $\ufrakb^1_{0}$, $\ufrakb^k_{0}$, and $\ufrakb^k_{i}$ for $i < k$. For $\ufrakb^1_{1}$, we use a monotone parameterization based on integrals of radial basis functions (sigmoids), and we keep $\ufrakb_k^k$ linear for $k > 1$.
See Appendix~\ref{sec:param_tri} for more details on this representation. While many other nonlinear representations of $U^k$ are possible, this particular separable form was chosen because it gradually adds degrees of freedom starting from a linear representation and allows \eqref{eq:component_estimator} to be solved in closed form for $k > 1$.
We label results for maps with linear terms plus $p$ RBFs/sigmoids in each component's functions as ``linear + $p$ RBFs.'' The figures below thus compare the filtering performance of four ensemble methods: the stochastic EnKF, linear maps, and linear maps with $p\in\{1,2\}$ RBFs.

For a fair comparison, all algorithms process the observations at a given assimilation time sequentially (see Section~\ref{sec:rem_pert_obs})
and do not evaluate the likelihood function, which we assume is intractable. The stochastic EnKF computes the Kalman gain and the linear map in~\eqref{eq:enkf_pert} by estimating the cross-covariance 
$\smash{\widehat{\Sigmab}_{\Xb,\Yb}}$ and observation variance 
$\smash{\widehat{\Sigmab}_{\Yb}}$ using 
data simulated from the likelihood model for each forecast sample. 
Since we do not require
direct access to the observation operator $\Hbrm$ or to the covariance of $\Ecb_{k}$, these algorithms are %
generalizable to non-additive and non-Gaussian likelihood models. {We provide examples with a nonlinear observation model and} non-Gaussian observation noise in Appendices~\ref{sec:lorenz_heavycase} {and~\ref{sec:lorenz_nonlinearobs}, respectively.}

As described in Section~\ref{sec:rem_pert_obs}, the linear map algorithm is 
related
to the stochastic EnKF. The only difference between these two algorithms lies
in the way they exploit structure for the purpose of regularization.
While stochastic EnKFs exploit decay of correlation between the variables with increasing spatial distance, %
regularization in the stochastic map filter enforces sparsity in the map $S^{\Xcb}$, based on conditional independence, decay of correlation, and local likelihood structure (see Sections~\ref{sec:reg_map} and~\ref{sec:rem_pert_obs}). In the stochastic map filter, the sparsity structure is parameterized by the distance $r$ that defines the set of input state variables for each component in~\eqref{eq:nbd_vars}, and by the number of components $j \leq n$ where $S^{\Xcb}$ departs from the identity map (see Remark~\ref{rem:localization}). In the stochastic EnKF, covariance localization is applied by multiplying the forecast covariance elementwise with a Gaspari--Cohn tapering function that has an optimally tuned localization radius~\cite{houtekamer2001sequential}.
While the linear maps and the stochastic EnKF behave similarly with larger ensemble sizes---where less regularization is needed---the results may differ for small $M$.

All algorithms considered in this section are implemented with multiplicative prior inflation and localization. %
For the inflation, we increased the spread of the forecast ensemble that is used to estimate $S^{\Xcb}$. We inflated these samples by multiplying their deviations from the sample mean by a scalar $\zeta \geq 1$.
The parameters (i.e., inflation parameter $\zeta$ and localization parameters $r$ and $j$) are tuned for each ensemble size and problem configuration to minimize the time-averaged RMSE. Only the tuned results are presented below, {unless otherwise indicated.} Lastly, for the $p \in \{1,2\}$ stochastic map filter we also investigated the sensitivity of the results with respect to the scaling parameter $\gamma$ 
of the RBFs (described in Appendix~\ref{sec:param_tri}) and found no significant dependence. Thus, the results are presented below for a fixed value of $\gamma = 2.0$ across all of the experiments.

\subsection{Lorenz-63} \label{sec:lorenz63}

\begin{figure}[!ht]
  \centering
  \includegraphics[width=0.45\linewidth]{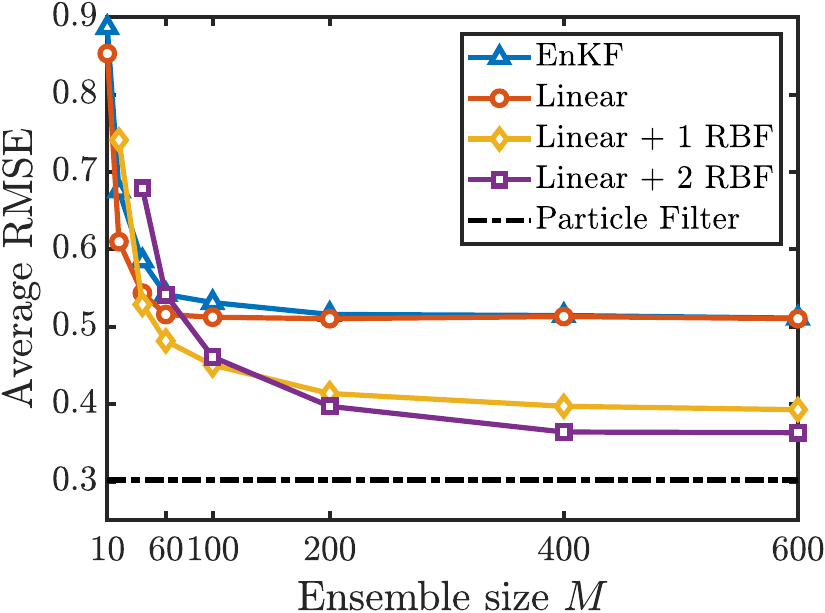}
  \hspace{1cm}
  \includegraphics[width=0.45\linewidth]{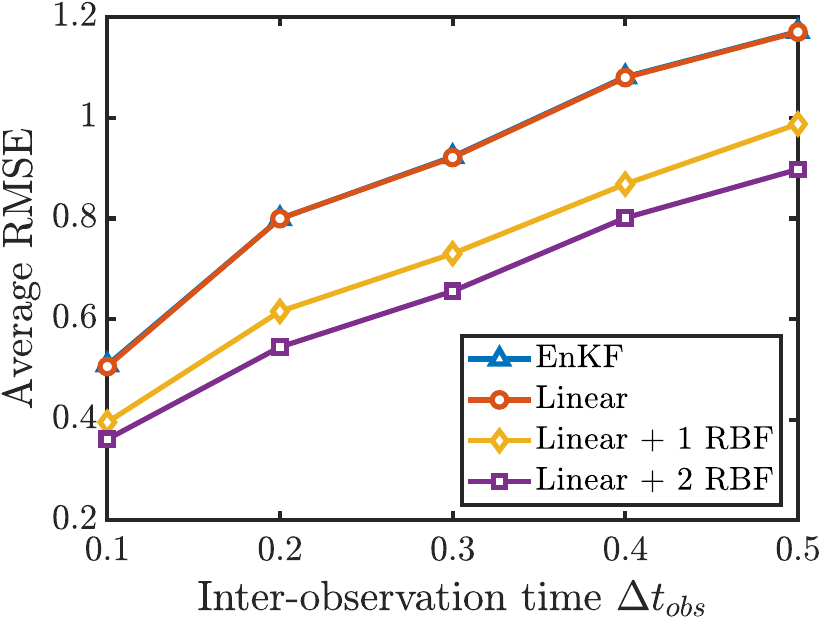}
  \caption{Average RMSE of the Lorenz-63 model for $\Delta t_{\rm obs} = 0.1$ as a function of $M$ ({\it left}), and for $M = 1000$ as a function of the inter-observation time $\Delta t_{\rm obs}$ ({\it right}). \label{fig:lorenz63_RMSE}}
  \vspace{-15pt}
\end{figure}

We first consider the Lorenz-63 system without any form of localization applied to $S^{\Xcb}$.
The idea is to 
isolate the effect of adding nonlinear terms to the prior-to-posterior update.
 In this case, 
 the states of the system are fully observed (i.e., $n = d$) with Gaussian observational noise and variance
 $\theta^2 = 4$. 
The time between observations is set to 
$\Delta t_{\rm obs} = 0.1$.
 The resulting average RMSE, for a range of $M$, is plotted on the left of Figure~\ref{fig:lorenz63_RMSE}.

For very small $M$, linear maps
have the most robust performance.
However, for
sufficiently large ensemble sizes, we observe a consistent improvement in RMSE by adding nonlinearities to the approximation space of the map. Specifically, 
the stochastic map filter with $p=1$ RBFs 
yields better tracking performance
 for $M \geq 40$, while the filter with $p=2$ RBFs provides improved results for $M \geq 200$. 
In both cases, the improvement in average RMSE 
is more than $20\%$.  The average RMSE values obtained with increasing map nonlinearity
 (for large enough $M$)
 also tend towards the
 RMSE of our ``reference'' solution---obtained with a consistent SIR particle filter using  $M=10^6$ samples (with an average relative effective sample size of 0.89), which 
 we plot in Figure~\ref{fig:lorenz63_RMSE} as a dotted line. 

While this experiment did not involve map localization or covariance tapering in the EnKF, the difference between the stochastic EnKF and the linear map results for small $M$ can be attributed to the local likelihood structure that we exploit. When assimilating scalar observations $Y = y^*$ with a local likelihood model (i.e., $Y$ only depends on one component of the state, say $Z_{1}$ after a permutation), the states $Z_{k}$ for $k = 2,\dots,n$ are conditionally independent of $Y$ given $Z_{1}$. This Markov property results in sparsity of the KR rearrangement from $\pi_{\Xb, Y}$ to $\Gauss({\bf 0}, {\bf I}_{n+1})$, and hence the map $S^{\Xcb}$ (see Section~\ref{sec:reg_map}). In particular, the components $U^{k}$ do not depend on $y$ for $k \geq 2$. We note that this sparsity is exact, and thus different from any approximate conditional independence that could further be exploited to regularize the estimation of $S^{\Xcb}$.

In a second experiment, we investigate the performance of the stochastic map filter as a function of
the time $\Delta t_{\rm obs}$ between observations, for a fixed ensemble size.
The idea is to test the filter for a sequence of increasingly non-Gaussian forecast distributions.
The average RMSE of the various algorithms for $\Delta t_{\rm obs} \in [0.1, 0.5]$ 
and for $M = 1000$ is shown in the right panel of Figure~\ref{fig:lorenz63_RMSE}. The stochastic map filters demonstrate a consistently lower RMSE over the full range of inter-observation times. 
This shows that map filters of increasing complexity remain stable (and continue to provide more accurate tracking) even as the state evolution becomes more nonlinear.

\begin{figure}[!ht]
  \centering
  \includegraphics[width=0.45\linewidth]{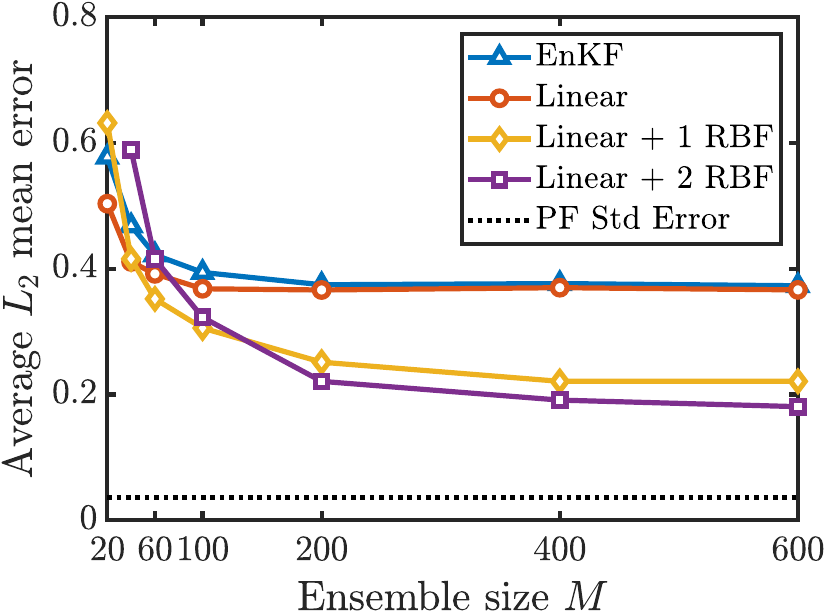}
  \hspace{1cm}
  \includegraphics[width=0.45\linewidth]{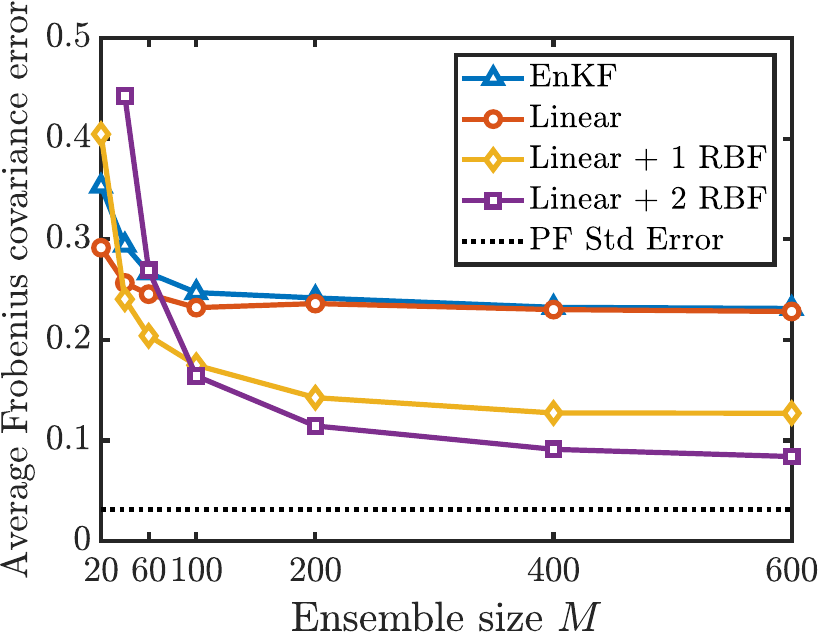}
  \caption{Normalized $L_{2}$ error of the posterior mean ({\it left}) and Frobenius error of the posterior covariance matrix ({\it right}) for the Lorenz-63 model. \label{fig:lorenz63_posterior_stats}
  }
  \vspace{-15pt}
\end{figure}

For this low-dimensional problem, we also investigate how the stochastic map filter captures the true filtering distribution  (i.e., the sequence of posteriors $\pi_{\Xb_k \vert \yb^*_{1:k}} $) over time. 
We first use
an SIR particle filter with $10^6$ samples to estimate the posterior mean $\zb_{k}^{\text{PF}}$ and posterior covariance matrix $\Sigmab_{k}^{\text{PF}}$ 
at each step $k$, with $\Delta t_{\rm obs} = 0.1$. 
In Figure~\ref{fig:lorenz63_posterior_stats}, we plot, for each algorithm,
the normalized $L_{2}$ error in the posterior mean, defined as $\| \bar{\zb}_{k} - \zb_{k}^{\text{PF}} \|_{2}/\sqrt{n}$, and the normalized  error in the posterior covariance, defined as $\| \widehat{\Sigmab}_{k} - \Sigmab_{k}^{\text{PF}} \|_{F}/n$, averaged over 2000 assimilation times $k$.
We observe a significant improvement in the
approximation of 
these posterior statistics, just  
by 
introducing  a few nonlinearities %
in the prior-to-posterior transformation.
For a sufficiently large $M$, the improvement
can be more than $50\%$.

To characterize the variability of the particle filter estimates used above, we repeat the assimilation exercise with the reference $M=10^6$ SIR filter for  100 times. We compute the sample standard deviations of these replicated filtering results (e.g., of the $L_{2}$ norm of the posterior mean and the Frobenius norm of the posterior covariance matrix at each assimilation time) 
and plot their maxima over time using dotted lines in Figure~\ref{fig:lorenz63_posterior_stats}. 
For comparison, the standard errors of the same norms of posterior moments estimated with the stochastic map filter, for $M \in [20,600]$, are 4 to 10 times smaller than the standard errors of the $M=10^6$ SIR filter.

\subsection{Lorenz-96 with long inter-observation time} \label{sec:bickel_hardcase}

\begin{figure}[!ht]
  \centering
  \includegraphics[width=.45\linewidth]{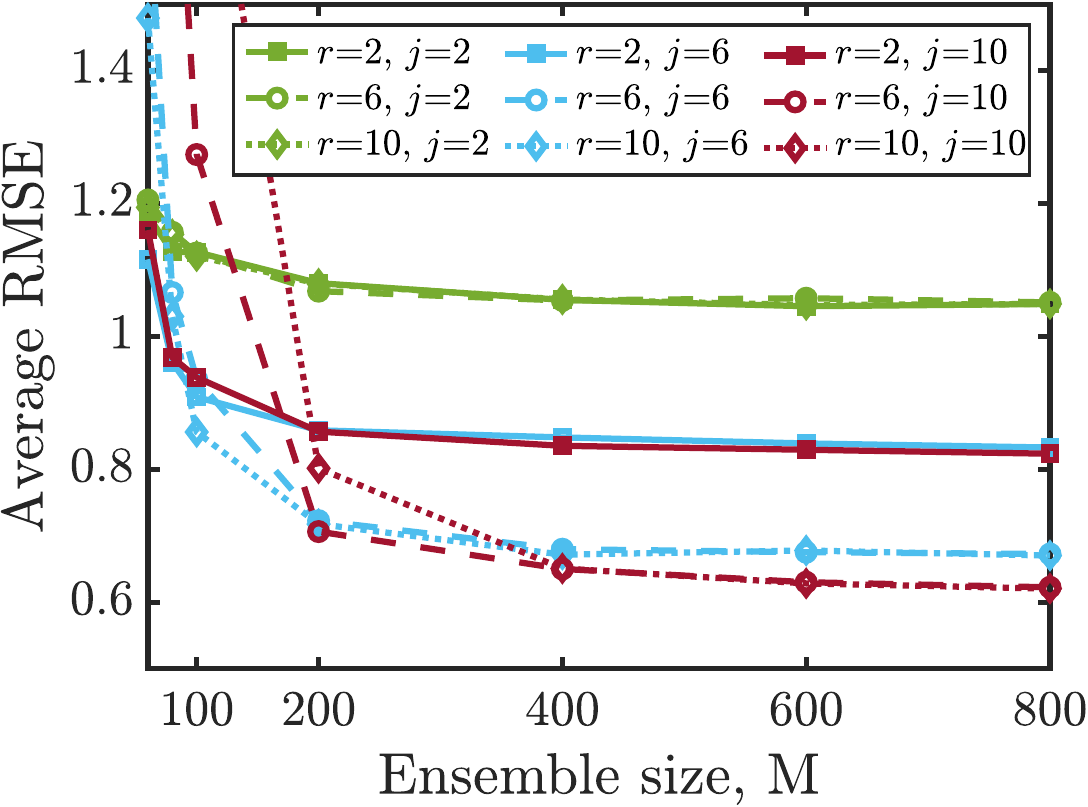}
  \hspace{1cm}
  \includegraphics[width=.45\linewidth]{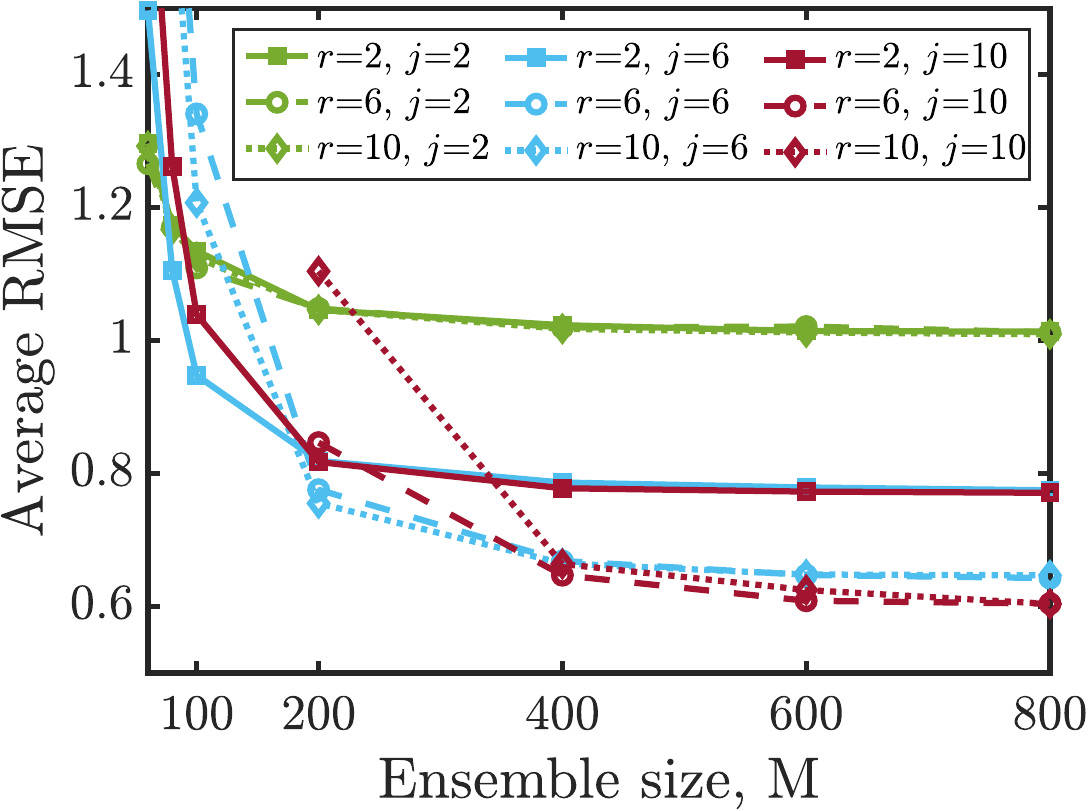}
  \caption{{Average RMSE (over $2000$ assimilation cycles) for the ``hard'' Lorenz-96 configuration of Section~\ref{sec:bickel_hardcase} using `linear + 1 RBF' maps (\emph{left}) and `linear + 2 RBF' maps (\emph{right}) for different localization parameters: the number of non-identity components $j$ in the map $S^{\Xcb}$, and the neighborhood size $r$ defining the input dependence of each map component.} \label{fig:lorenz96_hard_meanRMSE_localization_parameters}}
  \vspace{-12pt}
\end{figure}

Here, we  follow the ``hard case'' 
setup
of~\cite{lei2011moment, bengtsson2003toward} for the Lorenz-96 model, with sparse observations in space and time, i.e., $d = 20$ (observing every other component of the state) and $\Delta t_{\rm obs} = 0.4$. Each observation has independent additive Gaussian noise with variance $\theta^2 = 0.5$. 
For reference, we note that a $\Delta t_{\rm obs}=0.05$ would
correspond to roughly $6$ hours in a global weather
model \cite{lorenz1996predictability,
majda2012filtering}.
The large time interval between observations makes the
forecast distribution highly non-Gaussian. 
As a result, this configuration typically requires ensemble sizes greater than the state dimension $n$ to track $\zb_{k}^{*}$. Furthermore, with a high-dimensional state vector it is necessary to tune the map's localization parameters to achieve stable filtering performance over time. As described in Section~\ref{sec:commonsetup}, we tune the localization parameters $r$ and $j$ for the stochastic map filter and the localization radius for the EnKF, and the inflation parameter $\zeta$ for both algorithms. {Figure~\ref{fig:lorenz96_hard_meanRMSE_localization_parameters} shows the sensitivity of the average RMSE for the nonlinear filters to the localization parameters. For large ensemble sizes, RMSE improves when increasing the number of non-identity map components $j$ and the neighborhood size $r$. For each setting of these parameters, the filter with $p = 2$ RBFs (right) offers a slight benefit over the filter with $p = 1$ RBFs (left). %
We emphasize that increasing $j$, $r$, and $p$, for any $p \geq 1$, \emph{all} comprise ways of increasing the nonlinearity of the map: not just via the choice of approximation space for $\ufrak$, but also by expanding {where} in the map---and in which variables---nonlinearity appears.} %

\begin{figure}
  \centering
  \includegraphics[width=.45\linewidth]{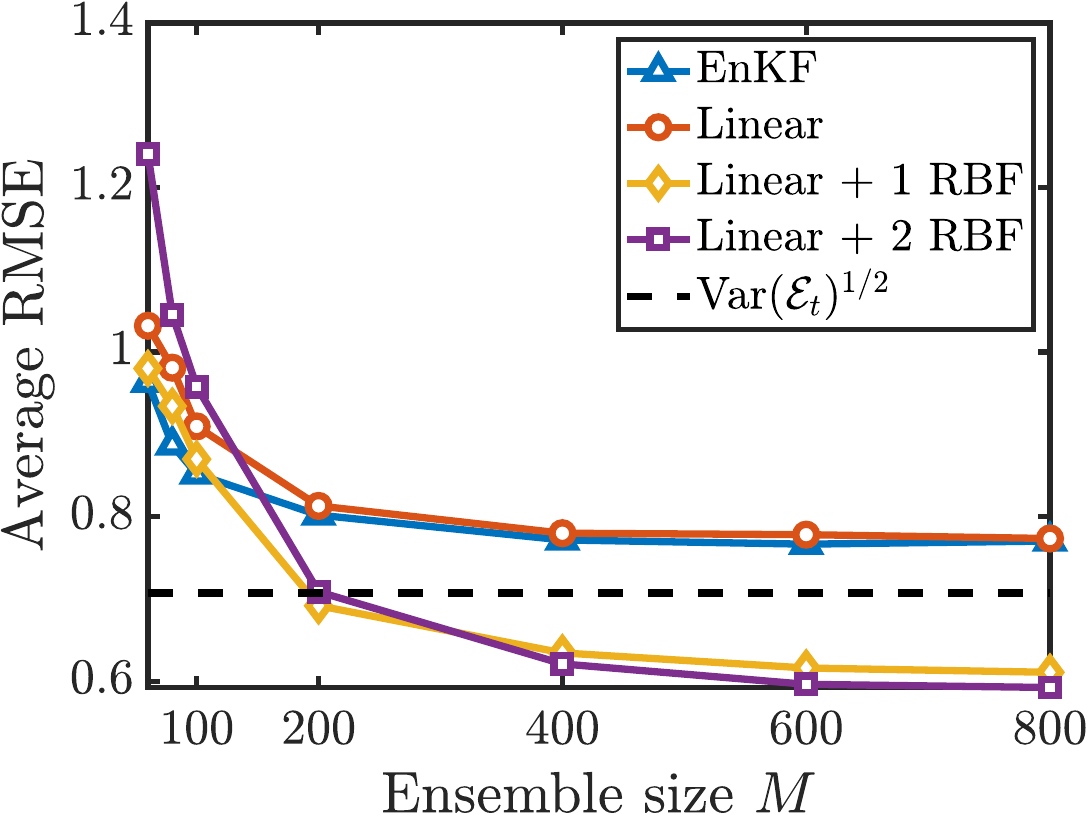}%
  \hspace{1cm}
  \includegraphics[width=.45\linewidth]{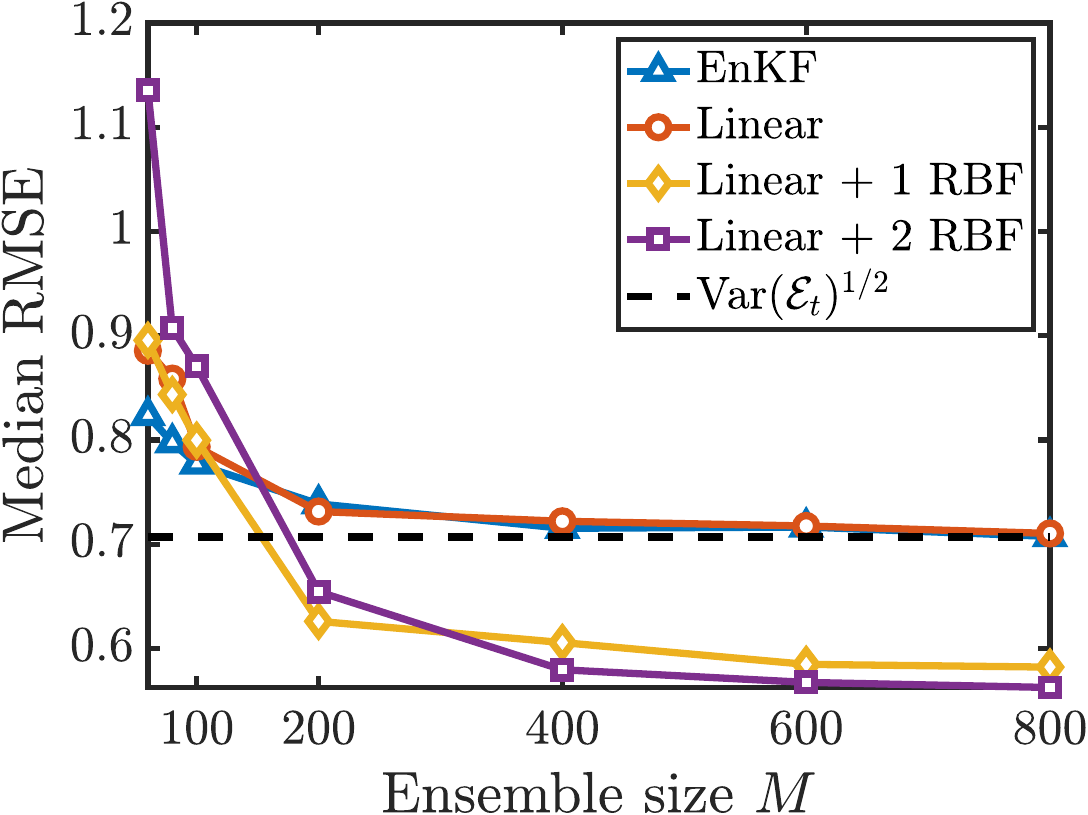}%
  \caption{Average ({\it left}) and median ({\it right}) RMSE (over $2000$ assimilation cycles) for the ``hard'' Lorenz-96 configuration of Section~\ref{sec:bickel_hardcase}, with $\Delta t_{obs} = 0.4$, observing every other component of the state ($d=20$) with additive Gaussian observational noise. Dashed line is the standard deviation of the observational noise. \label{fig:lorenz96_hard_meanRMSE}}
  \vspace{-12pt}
\end{figure}

Figure~\ref{fig:lorenz96_hard_meanRMSE} shows the average and median values of $\text{RMSE}_k$ (over 2000 terminal assimilation cycles $k$) for the optimal setting of the localization parameters as a function of the ensemble size {$M \in [60,800]$}. By increasing the nonlinearity of the map $\widehat{S}^{\Xcb}$ and hence $\widehat{T}$, the stochastic map filter can reduce the bias of the EnKF given sufficient samples. For example, the plateaus of RMSE in Figure~\ref{fig:lorenz96_hard_meanRMSE} for the EnKF reflect the limitation of the EnKF's affine %
transformation: the RMSE initially decreases, then stagnates with increasing $M$. Nonlinear maps reduce the values of these plateaus (by roughly 25\% in this example) and provide smaller RMSE even 
at intermediate ensemble sizes. By introducing relatively few nonlinear basis functions (in $\widehat{S}^{\Xcb}$) whose parameters can be \emph{reliably} learned, the stochastic map filter can better capture the non-Gaussian forecast statistics and improve the tracking of $\zb_{k}^{*}$.

Figure~\ref{fig:lorenz96_hard_meanRMSE} also exposes the bias-variance tradeoff in learning the prior-to-posterior map ${T}$, which corresponds to a link between $M$ and the complexity of the map (i.e., the number of the basis functions in each component of $S^{\Xcb}$). For smaller ensemble sizes, a richer class of nonlinear maps yields estimators $\widehat{T}$ with less bias but higher variance. The EnKF and linear maps, on the other hand, yield more biased but potentially lower-variance estimators that offer more stable tracking performance for small $M$. In the present experiments, as $M$ increases, the map class yielding the minimum RMSE evolves quickly from linear to $p=1$ and then to $p=2$. {Similarly, in Figure~\ref{fig:lorenz96_hard_meanRMSE_localization_parameters}, the expressiveness of the best nonlinear map class, as captured by larger localization parameters, increases with $M$.} 
Overall, the stochastic map filter provides a flexible framework for adjusting the complexity of the map to the given ensemble size in order to extract a good estimator of $T$.

\begin{figure}[!ht]
  \centering
  \includegraphics[width=0.45\linewidth]{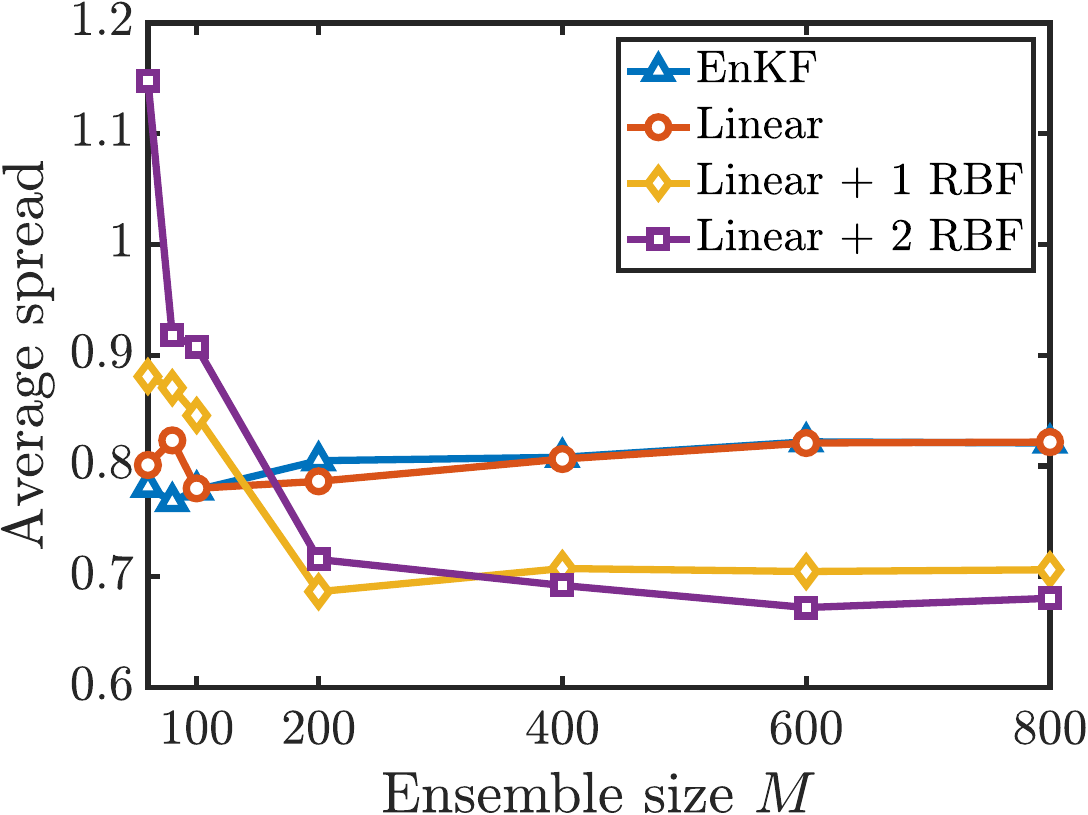}%
  \hspace{1cm}
  \includegraphics[width=0.45\linewidth]{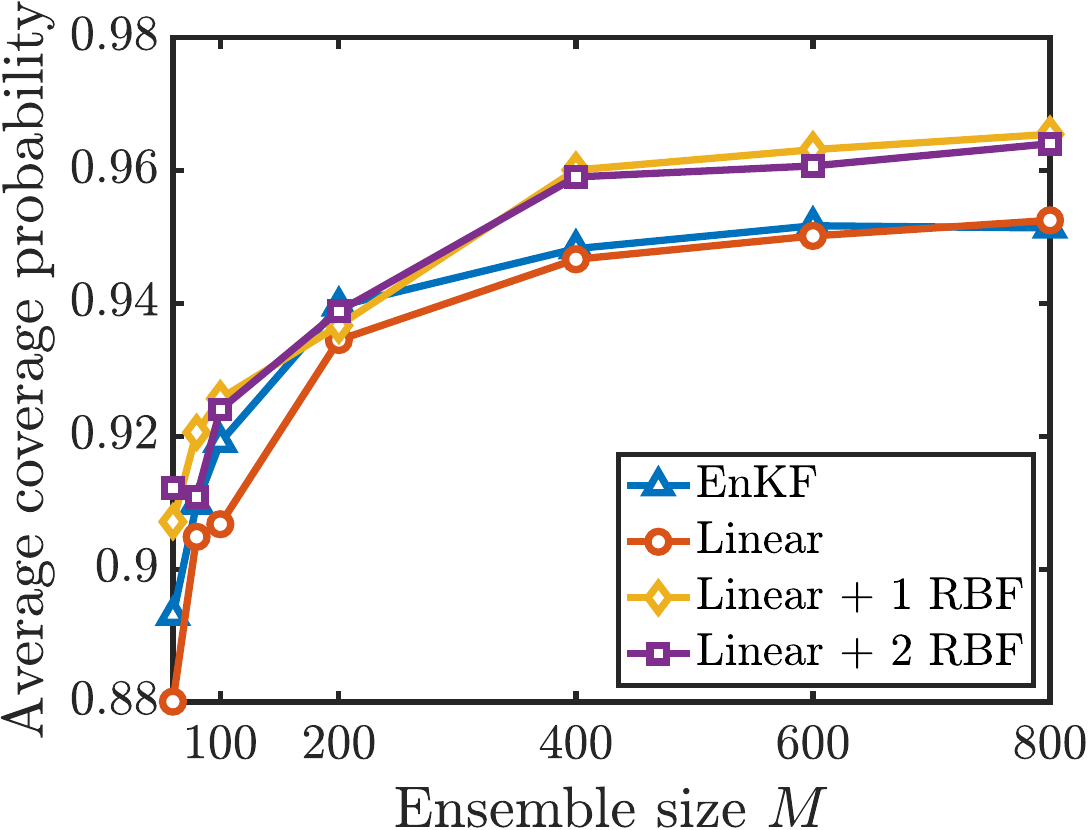}%
  \caption{Average ensemble spread ({\it left}) and average coverage probability of the $[2.5\%, 97.5\%]$ empirical intervals of each state marginal ({\it right}) for the ``hard'' Lorenz-96 configuration. (See precise definitions of these quantities in Section~\ref{sec:commonsetup}.) \label{fig:lorenz96_hard_spread_covprob}}
  \vspace{-10pt}
\end{figure}

In addition to improving the RMSE, the nonlinear stochastic map filter reduces the ensemble spread while simultaneously increasing the probability of covering the true state. We present these results for increasing $M$ in Figure~\ref{fig:lorenz96_hard_spread_covprob}.

\section{Discussion}
\label{sec:discussion}
We have introduced a %
class of nonlinear filtering algorithms
based on the construction of stochastic or deterministic couplings
between continuous distributions.
These algorithms generalize the
EnKF to nonlinear updates of tunable {\it and}
arbitrary complexity.
The key computational task in applying these algorithms is the 
numerical estimation of a KR
rearrangement via convex optimization. Our goal in using nonlinear updates 
is to reduce the intrinsic bias of the EnKF,
while retaining its robustness in
high dimensions. 
We run our
numerical experiments on 
various %
configurations of 
the Lorenz-96 model, %
which is a common testbed for numerical
weather prediction systems; we also include 
a comparison with the true (Bayesian) filtering distributions for the
lower-dimensional Lorenz-63 model.
Through these experiments, we show %
how filtering with nonlinear updates {can} significantly 
 outperform the
EnKF %
in a range of highly non-Gaussian settings, for a marginal increase in ensemble
requirements and computational effort.
In what follows, we briefly discuss
limitations and possible %
extensions of this approach to filtering.

\paragraph{Nonlinear parameterizations}
Highly nonlinear updates do not automatically guarantee %
better filtering performance.
Trying to estimate %
complex transformations
from very few samples
could
yield unacceptable variance.
The bias and variance of a transport map estimator must be balanced, possibly by 
introducing additional information through regularization.
In this paper, we consider several 
regularization
techniques. 
For instance, we define a hierarchy of 
{nonlinear (separable)} parameterizations for the transport maps
{that allow for a gradual departure from the linear \emph{ansatz}, which is key to reducing the bias while managing the variance of the estimator.}
We select
the ``best''
parameterization 
according to 
a prediction error criterion.
We notice several intuitive trends.
Increasingly nonlinear
transformations typically demand
larger ensembles,  
while for 
a fixed parameterization of the maps, the tracking error
plateaus beyond a 
certain ensemble size.
Similarly, for any given problem, %
we expect there to be a 
sufficiently small ensemble size for which the linear ansatz---as in EnKF techniques---performs %
best.
Moreover,
for nonlinear filtering problems that are %
well approximated by %
Gaussian
distributions, we %
do not expect major gains from using nonlinear updates, even for
large ensemble sizes.
Nonlinear updates are mostly needed to capture non-Gaussian structure.
The complexity of the transformations should then be {adapted} to the number of ensemble members and to the non-Gaussianity of the problem.
{Certainly many avenues exist for increasing this complexity. Here we focused on separable parameterizations, motivated in part by computational simplicity, but much more so by the fine-grained control over complexity %
 that they enable. General non-separable maps could be considered as well; see~\cite{baptista2020adaptive} for a recent contribution in this direction.} 
In future work, we wish to automate
this adaptation rather than running 
an offline calibration phase,
i.e., we plan to develop %
{\it nonparametric} %
extensions of the
proposed filtering algorithms. 
{For any of these choices, we emphasize that the estimate of the resulting \emph{composed} transformation $T$---and its expressivity in acting on the forecast distribution---is what ultimately 
affects the filtering approximation.}

\paragraph{Localization and other form of regularization}
A related regularization idea is that of exploiting 
decay of correlation  %
and
approximate conditional independence %
in the filtering
distribution.
This structure is common when filtering spatiotemporal processes.
In this paper,
we introduced  simple 
{\it localization}
ideas
for nonlinear updates by imposing specific sparsity patterns in the transport maps composing the update.
In future work, one could explore
a much wider range of regularization techniques.
For instance, it is natural to consider
LASSO-type %
estimators %
in order
to ``discover''---rather than to enforce---sparsity of the nonlinear transformations.
Or one might exploit low-rank structure in the change
from forecast to filtering distribution in order to reduce the effective dimension
of the nonlinear update, following
the line of work in \cite{spantini2014optimal,cui2014likelihood}.
Accounts of low-rank structure in transport maps can be found in
\cite[Ch.~5]{spantini2017low} and \cite{bigoni2019greedy}; {for an initial
application to filtering, see~\cite{le2021low}}. 
In general, we believe that %
practical high-dimensional %
filtering 
relies crucially on the development of
structure-exploiting
regularization techniques.

\paragraph{Consistency} 
From a theoretical point of view, it would be interesting to study the
behavior of the proposed algorithms in a simultaneous limit of
increasing ensemble size {\it and} increasing complexity of the
rearrangements.
In particular, a key goal should be to understand conditions under which it
is possible to establish some form of \emph{consistency} for the
filters, much in the same spirit as the convergence theory for SMC
methods \cite{crisan2002survey}.  This question is meaningless for the
EnKF, which is inherently inconsistent due to the fixed linear
structure of the update.
It would also be interesting to understand the stochastic map filter
in a continuous-time filtering setting, and thus to explore its
connections with the feedback particle filter
\cite{yang2013feedback}.
This link might offer further insight into
conditions that ensure consistency.

\paragraph{Further extensions}
The ideas presented in this paper can readily be extended to
\textit{smoothing}, i.e., characterizing the conditional distribution
of {past} states, given all available measurements.  In this context,
one can first estimate a (non-singular) Markov transition kernel
between observation times via an upper-triangular KR rearrangement,
and then use this approximation to propagate the filtering ensemble
backwards in time.  The resulting scheme can be interpreted as a Monte
Carlo approximation of the transport map backward-smoothing algorithm
introduced in \cite[Section~7]{spantini2017inference}. 
{It would then be interesting to explore connections with the multi-level generalizations presented in~\cite{houssineau2018multilevel}.}
Relatedly, it would be useful to extend the present framework to problems of sequential \emph{joint
parameter and state} estimation.

Finally, we note that the analysis step of the stochastic map
filter is an instance of \emph{approximate Bayesian computation}, as
it requires only sampling from the likelihood and prior in order to
construct a prior-to-posterior transformation. Realizing and expanding
this approach for generic problems of {simulation-based inference~\cite{cranmer2020frontier}} 
represents an exciting avenue for future work.

\section*{Acknowledgments}
AS, RSB, and YMM gratefully acknowledge support from the AFOSR Computational Mathematics program and AFOSR MURI award FA9550-15-1-0038, and from the US Department of Energy, Office of Advanced Scientific Computing Research, AEOLUS project. RSB also acknowledges support from an NSERC PGSD-D Fellowship. AS acknowledges support from the Deutsche Forschungsgemeinschaft (DFG), through the CRC 1294 ``Data Assimilation'' Project, for a visit to the Institute for Mathematics of the University of Potsdam. The authors also thank Daniele Bigoni, Jana de Wiljes, Matthias Morzfeld, Sebastian Reich, and Xin Tong for many insightful discussions and suggestions.

\appendix

\section{Parameterization and computation of monotone triangular maps}
\label{sec:param_tri}
In this section, we first discuss possible parameterizations for a monotone triangular map $U$ that is meant
to approximate the KR rearrangement %
between a pair of densities 
on $\re^n$ (Section \ref{sec:est_KR}). 
Then, we link specific structures in the
parameterization of $U$ to computational simplifications
for an estimator of the
rearrangement.%

The KR rearrangement between a pair of positive
densities is a triangular bijection on
$\re^n$
such that each slice \eqref{eq:componentMap} is
strictly increasing and absolutely continuous 
\cite{bogachev2005triangular}.
Hence, it makes sense to consider candidate 
rearrangements $U$ whose $k$th components, $U^k$, can 
be written as
\begin{equation} \label{eq:monotone}
  U^k(z_1,\ldots,z_k)= a_k(z_1,\ldots,z_{k-1}) + 
  \int_0^{z_k} \exp \left ( b_k(z_1,\ldots,z_{k-1}, t) \right )\, {\rm d} t,
\end{equation}
for some {\it arbitrary} functions
$(a_k,b_k)$,
and for all
$k=1,\ldots,n$
\cite{ramsay1998estimating,marzouk2016introduction}.
For any choice of $(a_k,b_k)$, the
rearrangement $U$ is a strictly
monotone (increasing) triangular map, since
$\partial_k \,U^k = \exp(b_k)>0$ for all $k$.
In fact, \eqref{eq:monotone} provides 
a rather general way to represent monotone 
triangular functions.

The estimator in 
\eqref{eq:component_estimator}
relies on the definition of
an %
approximation space, $\Hcb_k$,
for every component of the rearrangement.
A viable strategy is to use the %
monotone
representation in \eqref{eq:monotone} and to 
constrain each function
$(a_k,b_k)$ %
to lie
in a finite dimensional space, i.e., %
\begin{equation} \label{eq:linearExp}
  a_k(\zb) = \sum_i a_{i} \,\psi_i^a(\zb),\qquad
  b_k(\zb) = \sum_j b_{j} \,\psi_j^b(\zb),
\end{equation}
for a collection of finitely many 
basis 
functions 
$\psi_i^a, \psi_j^b$---e.g.,
multivariate 
Hermite polynomials, Hermite functions,
splines, or 
radial basis functions---and
unknown  
real valued coefficients $(a_{i},b_{j})$.\footnote{We may substitute the exponential 
in
\eqref{eq:monotone} %
with a
square function or even a sum of squares \cite{jainipolynomialICML2019}.
This offers the computational advantage of 
precomputing any
integral in \eqref{eq:monotone} once and
for all, thanks to
parameterizations of
$b_k$ that 
are 
linear in the
unknown coefficients (see \cite{bigoni2016monotone}).
}
The minimization in 
\eqref{eq:component_estimator}
can then be cast
in terms of 
the unknown
coefficients, which fully
parameterize the maps in
$\Hcb_k$.

As %
explained
in Section \ref{sec:reg_map},
the
``complexity'' of each $\Hcb_k$ is 
a
key ingredient for
regularizing the estimation
of the rearrangement; there is no point
in trying to learn arbitrarily nonlinear maps
from a given finite ensemble.
For instance, the %
EnKF restricts the search to linear maps.
In general, it is possible to depart from the linear ansatz
in a {\it gradual} way, 
by imposing %
structural assumptions
of increasing complexity 
on the
rearrangement.
For example, we can consider maps %
whose components %
are separable as sums of
\emph{univariate} nonlinear
functions $(\ufrakb_i)$ 
that each
admit a linear parameterization like
\eqref {eq:linearExp} or
\eqref{eq:monotone} (if monotone),
i.e.,
\begin{equation} \label{eq:sep_param}
U^k(z_1,\ldots,z_k)=
\ufrakb_1(z_1) + \ldots + \ufrakb_k(z_k).
\end{equation}
Next, 
we can consider separable
parameterizations 
in terms of
bivariate functions, i.e.,
$U^k(z_1,\ldots,z_k)=\sum_{i\le k} \ufrakb_i(z_i, z_k)$,   
and so on.
In essence, the complexity of the parameterization
should be adapted (and calibrated) to the cardinality of the ensemble and to the amount of
prior information that is injected in the estimation
(e.g., see the localization ideas of Section \ref{sec:est_KR}).

In the numerical examples of this paper, we 
explore
the separable parameterization in 
\eqref{eq:sep_param},
since we 
wish to
work with ensemble
sizes comparable to those needed by the EnKF.
We propose a specific strategy to parameterize each
$\ufrakb_i$ in \eqref{eq:sep_param}, although there are plenty of potential %
alternatives.
For $i<k$, $\ufrakb_i$ is simply a univariate function, not necessarily monotone.
In this case, we parameterize $\ufrakb_i$ 
as a linear combination of $p$  
radial basis functions (RBFs) plus a  
global linear term, i.e.,
\begin{equation}
\label{eq:param_nm_univ}
  \ufrakb_i(z) = 
  u_{i0} \,z + \sum_{j=1}^p u_{ij}  \,
  \Gauss( z ; \xi_j, \sigma_j^2 ), 
\end{equation}
for a collection of centers 
$\xib \coloneqq (\xi_1,\ldots,\xi_p)$,
scale parameters 
$\sigmab \coloneqq (\sigma_1,\ldots,\sigma_p)$,
and unknown coefficients $(u_{ij})$.
The number $p$ of RBFs defines the degree
of nonlinearity of $\ufrakb_i$.
For $p=0$, $\ufrakb_i$ reverts to a linear
function.
The centers $\xib$ are chosen from $p$ 
(uniform) 
empirical quantiles of the marginal
distribution of the ensemble along the $i$th
coordinate direction, i.e.,
$\xi_j \coloneqq \hat{q}_{j/(p+1)}$ for $j=1,\ldots,p$, where 
for any $\alpha \in (0,1)$, $\hat{q}_{\alpha}$
denotes 
the $\alpha$-th empirical quantile.
The scale
parameters 
$\sigmab$ are defined to be
proportional to the %
inter-distance between the centers $\xib$, i.e.,
$\sigma_j \coloneqq \gamma\,
  (\xi_{j+1} - \xi_{j-1})/2
$ for $j=1,\ldots,p$, where 
$\xi_0 \coloneqq \xi_1$, 
$\xi_{p+1} \coloneqq \xi_p$, 
and where $\gamma > 0$ is a
tuning parameter.
In our experiments, we consider 
$\gamma \in [1,3]$.
The parameterization of
$\ufrakb_k$
in \eqref{eq:sep_param} is
a bit different, since the monotonicity
constraint on $U^k$ requires
$\ufrakb_k$ to be an {increasing} 
map.
One option is to use the monotone representation
in 
\eqref{eq:monotone} just 
for $\ufrakb_k$.
In practice, however, $\ufrakb_k$ is %
a univariate function and thus we can seek alternative
ways of imposing monotonicity, e.g., by
expressing $\ufrakb_k$ as a nonnegative
combination of {monotone} basis
functions.
For a monotone $\ufrakb_k$, its
derivative $\partial \ufrakb_k$ is always 
nonnegative.
For $p>0$, 
we propose to parameterize  $\partial \ufrakb_k$ 
with $p+2$ positive basis functions,
i.e., 
$\partial \ufrakb_k = \smash{\sum_{j=0}^{p+1}} 
\,u_{kj} \,\psi_j'$, where 
$\psi_j'(z) \coloneqq  \Gauss(z; \xi_j, \sigma_j^2)$
for $j=1,\ldots,p$, while 
\begin{equation}
  \psi_0'(z) \coloneqq
\frac{1}{2}-\frac{1}{2} \erf\left( 
\frac{z - \xi_0}{ 
\sqrt{2}\,\sigma_0} \right),
\quad {\rm and} \quad 
\psi_{p+1}'(z) \coloneqq
\frac{1}{2}+\frac{1}{2} 
\erf\left( \frac{z - \xi_{p+1}}{ 
\sqrt{2}\,\sigma_{p+1}} \right),
\end{equation}
for a collection of $p+2$ centers
$(\xi_i)$
and scale parameters $(\sigma_i)$.
The centers and scale parameters are chosen 
similarly to those in \eqref{eq:param_nm_univ}, i.e.,  
from uniform empirical quantiles of the marginal distribution of the ensemble along the $k$th direction.
Intuitively, $\partial \ufrakb_k$ is the linear
combination of $p$ RBFs $(\psi'_1,\ldots,\psi'_p)$ 
plus a pair of positive 
functions $(\psi'_0,\psi'_{p+1})$ that revert to constants in the tails.
The idea is to promote robustness 
by constraining %
$\ufrakb_k$ to be linear in the tails.
Non-negativity of $\partial \ufrakb_k$ is
imposed by the set of 
linear inequalities $u_{kj}\ge 0$, for
$j=0,\ldots,p+1$.
The resulting
 monotone parameterization
of $\ufrakb_k$ is obtained by integrating
$\partial \ufrakb_k$, i.e.,
$\ufrakb_k(z)\coloneqq \int_0^z \partial 
\ufrakb_k(x)\,{\rm d}x =
c + \smash{\sum_{j=0}^{p+1}} 
\,u_{kj} \,\psi_j(z)$, for an
unknown constant $c$ and antiderivatives 
$\psi_j \coloneqq \int \psi_j'$ given by
\begin{equation}  \label{eq:integrals}
   \left\{
      \begin{array}{lr}
        \psi_0(z) =  
        \frac{1}{2}
        \left( 
        \left( z - \xi_0 \right) \left(
       1 -
        \erf\left( 
        \Delta_0 \right) 
        \right)
         -
        \sigma_0 \,
        \sqrt{2/\pi}
        \,%
        \exp(- \Delta_0^2)
        \right)
        \\[10pt]
        \psi_j(z) = 
        \frac{1}{2} \left( 
        1+ 
         \erf\left( \Delta_j \right) \right)
         & \text{for } 1 \le j \le p \\[10pt]
        \psi_{p+1} = 
        \frac{1}{2}
        \left( 
        \left(
        z - 
        \xi_{p+1}
        \right)
        \left( 1
        + 
        \erf\left( 
        \Delta_{p+1} \right)
        \right)
         +
        \sigma_{p+1} \,
        \sqrt{2/\pi}
        \,%
        \exp(- \Delta_{p+1}^2)
        \right)

        \end{array}
        \right. 
\end{equation}
where
$ 
\Delta_j \coloneqq (z-\xi_j)/(\sqrt{2}\,\sigma_j)
$ for all $j$.
For $p=0$, we set $\ufrakb_k$ to be an affine function.
For the parameterization of $U^k$ given by
\eqref{eq:param_nm_univ} and 
\eqref{eq:integrals},
the minimization in
\eqref{eq:component_estimator} 
is a linearly constrained
convex program that can be solved efficiently via
numerical optimization, e.g., using the 
projected Newton's method \cite{bertsekas1982projected}.
Moreover, if $\ufrakb_k$ is parameterized by an
affine function, 
then
\eqref{eq:component_estimator} reduces to
a standard  {\it linear regression} problem with
a closed form solution.
We %
analyze this fact in the %
 more
general case of parameterizations %
that are separable {\it only} in the  
last (monotone) variable, i.e.,
\begin{equation}  
U^k(z_1,\ldots,z_k)=
\ufrakb(z_1,\ldots,z_{k-1}) +
\ufrakb_k(z_k),
\end{equation}
where $\ufrakb(\zb) = \sum_j u_j \, \psi_j(\zb)$ 
for a collection of arbitrary basis
functions $(\psi_j)$ 
and where $\ufrakb_k(z) = c + \alpha z$.
In this case, we can rewrite
\eqref{eq:component_estimator} as
\begin{equation}  
\label{eq:regression_form}
\min_{(\tilde{u}_j), \tilde{c}, \alpha} \,
\frac{\alpha^2}{2} \,
\left\{ 
\frac{1}{M} \sum_{i=1}^M \,%
\left( \sum_j \tilde{u}_j \,
\psi_j(z^i_1,\ldots,z^i_{k-1}) + \tilde{c} + z^i_k \right)^2
\right\}
- \log \alpha,
\end{equation}
where 
$\zb^i = ( z^i_1,\ldots,z^i_n)$, 
$\tilde{c}\coloneqq c/\alpha$,
and $\tilde{u}_j \coloneqq u_j / \alpha$ for all $j$.
The minimization of \eqref{eq:regression_form} with respect to the new
variables $(\tilde{c}, \tilde{u}_j)$ can be
done independently of $\alpha$, and
corresponds to a standard regression problem,
i.e.,
\begin{equation}  
\kappa^* \coloneqq 
\min_{(\tilde{u}_j), \tilde{c}} \, 
\frac{1}{M} \sum_{i=1}^M \,%
\left( \sum_j \tilde{u}_j \,
\psi_j(z^i_1,\ldots,z^i_{k-1}) + \tilde{c} + z^i_k \right)^2,
\end{equation}
while the optimal $\alpha$ is given by $1/\sqrt{\kappa^*}$.

\section{Deterministic map filter with conditionally independent and local observations}
\label{sec:detmap_local}
In many problems of interest, we have
conditionally independent and local observations,
i.e.,
\begin{equation} \label{eq:local_lik}
  \pi_{\Yb \vert \Xb} = 
  \prod_{k=1}^d \pi_{Y_k \vert X_{\ell_k}},
\end{equation}
for a subset of $d$ state variables
$(X_{\ell_k})$---possibly up to a whitening transformation of the state 
\cite{houtekamer2001sequential,metref2014non}.
In Section \ref{sec:rem_pert_obs},
we saw that \eqref{eq:local_lik} implies
that we can process each of the
$d$ scalar observations individually and
sequentially.
For the deterministic map filter, %
there is a strong incentive in processing
scalar observations one at a time, because we can avoid the explicit solution of a non-convex problem like
\eqref{eq:mle_KR_direct_expl}.

Without loss
of generality, let us 
assimilate a 
scalar observation,
$y^*$,
of the
{\it first} state variable,
possibly up to a {permutation}.
The corresponding likelihood function
is given by
$\pi_{Y\vert X_1}$. 
We want to define a
structure-exploiting analysis map $T$,
similar to \eqref{eq:map_det}.
Let $S$ be a
KR rearrangement 
that pushes
forward $\pi_{\Xb}$ to a 
standard normal $\eta \coloneqq \Gauss({\bf 0}, {\bf I}_n)$---as in 
\eqref{eq:map_det}---and
let $\eta_1 \coloneqq \Gauss(0,1)$ denote the marginal
of $\eta$ along the first variable.
By the definition of KR rearrangement,
the pullback of $\eta_1$ by the
first component, $S^1$, of 
$S$
is equivalent to the prior marginal along the observed component, i.e., 
$(S^1)\pull \,\eta_1 = \pi_{X_1}$.
Moreover,
for all univariate monotone maps $\ufrak$,
define $\pi_{\ufrak}$ to be the density %
that is
proportional to the function 
$\xi  \mapsto \pi_{Y\vert X_1}(y^*\vert \xi)\,\ufrak\pull\,\eta_1(\xi)$, which is the product of
likelihood and pullback %
$\smash{\ufrak\pull\,\eta_1}$.
Note that $\pi_{S^1}$ is equivalent to the
posterior marginal $\pi_{X_1 \vert y^*}$.
Let $\mfrak$ be the increasing rearrangement on
$\re$---really a one-dimensional KR rearrangement---that pushes forward 
$\eta_1$ to $\pi_{S^1}$, and notice that the
posterior distribution factorizes as
\begin{equation}
\label{eq:factorization_post}
  \pi_{\Xb \vert y^*} = 
  \pi_{X_1 \vert y^*} \,
  \pi_{\Xb_{2:n} \vert X_1},
\end{equation}
because of the local structure of the
likelihood function.
The map $\mfrak$ characterizes
the posterior marginal $\pi_{X_1 \vert y^*}$, 
while $S$ gives access to the
prior 
conditional $\pi_{\Xb_{2:n} \vert X_1}$.
To see this, notice that
for all $z \in \re$, the map
\begin{equation}
  z_2,\ldots,z_n \mapsto \left[\begin{array}{l}
  S^2(z,z_2)\\ 
  \vdots \\ 
  S^n(z,z_2,\dots z_n)
  \end{array}\right]
\end{equation}
pushes forward the conditional 
$\pi_{\Xb_{2:n} \vert X_1}(\cdot \vert z)$ to
a standard normal on $\re^{n-1}$ (cf. Section \ref{sec:KR}).
Hence, we can define the analysis map $T$ as follows,
\begin{equation}
\label{eq:update_detmap_single}
T \coloneqq \mapSi^{-1} \circ \Mc \circ S,
\end{equation}
where
$\mapSi$ coincides with $S$ except for the
first component, which is set to the
identity function,
\begin{equation}
  \mapSi( \zb ) \coloneqq \left[\begin{array}{l}
  z_1 \\[2pt]
  S^2(z_1,z_2)\\ 
  \vdots \\ 
  S^n(z_1,z_2,\dots z_n),
  \end{array}\right],
\quad
{\rm and} \quad
\Mc(\zb) \coloneqq 
\left[\begin{array}{l}
\mfrak(z_1)\\[2pt] 
z_2\\ 
\vdots \\ 
z_n
\end{array}\right]. 
\end{equation}
It is immediate to verify
that $T$ pushes forward $\pi_{\Xb}$ to
$\pi_{\Xb \vert y^*}$.
In particular, the map $\Mc$ is responsible for
incorporating information from the 
measurement $y^*$ only in the observed component of the state---this is 
a {\it local assimilation} step---while
the map $\mapSi^{-1}$ is
responsible for
{\it propagating information} %
to the remaining components. %
In practice, we just showed that, 
when assimilating a local scalar observation,
the map $\Tc$ in
\eqref{eq:map_det}
can be rewritten as 
$\Tc = \smash{\mapSi^{-1}} \circ \Mc$.

We propose to estimate $T$ by 
$\widehat{T}$,
\begin{equation}
\label{eq:est_detmap_single}
\widehat{T} \coloneqq \mapSih^{-1} \circ
\widehat{\Mc} \circ \widehat{S},  
\end{equation}
where $\widehat{S}$ is the usual maximum likelihood
estimator of the KR rearrangement $S$ (Section \ref{sec:est_KR}),
$\smash{\mapSih}$ is an estimator
of $\smash{\mapSi}$ obtained (for free) by selecting
the corresponding components of  $\widehat{S}$,
while $\smash{\widehat{\Mc}}$ is %
defined
by an estimator, $\widehat{\mfrak}$, of the increasing
rearrangement  $\mfrak$  that pushes forward 
a standard normal 
to $\pi_{\widehat{S}^1}$. The latter is
an approximation 
of the posterior marginal,
$\pi_{X_1 \vert y^*}$, which depends only on the
first component of $\smash{\widehat{S}}$.

In principle, we could
define $\widehat{\mfrak}$ to be
the solution of a non-convex minimization problem like %
\eqref{eq:mle_KR_direct_expl},
for some parameterization of the rearrangement, which is now just a one dimensional function, and thus
easier to represent than a 
general 
rearrangement on $\re^n$.
In practice, however, we can
simply target a different characterization of the increasing rearrangement in terms of CDFs.
If $F: \re \ra [0,1]$ is the CDF of
$\Gauss(0,1)$ and 
$F_{\pi_{\widehat{S}^1}}: \re \ra [0,1]$ is the CDF of
$\pi_{\widehat{S}^1}$, then we can
express $\mfrak$ as
\begin{equation} \label{eq:increas_rearr}
  \mfrak = F_{\pi_{\widehat{S}^1}}^{-1} \circ F,
\end{equation}
and thus $\widehat{\mfrak}$ can be any estimator
of \eqref{eq:increas_rearr}. There are many options here, from parametric to fully nonparametric estimators.
We propose the following strategy.
For a collection of $N$ grid points
$(x_1,\ldots,x_N)$
on $\re$, let
$f_i$ denote an approximation of 
$F_{\pi_{\widehat{S}^1}}(x_i)$ obtained 
via numerical
integration %
(recall that we
can evaluate the density $\pi_{\widehat{S}^1}$ 
everywhere
up to a normalizing constant).
Notice that 
$\smash{\mfrak^{-1}} = 
\smash{F^{-1}} \circ F_{\pi_{\widehat{S}^1}}$ and let
$\xi_i \coloneqq \smash{F^{-1}}(f_i)$ for all
$i=1,\ldots,N$.
We can think of $\xi_i$ as an approximation of
$\mfrak^{-1}(x_i)$.
Hence, the pairs
$(\xi_i, x_i)_{i=1}^N$ yield approximate
input-output evaluations of the
rearrangement $\mfrak$, and can be used
to define a {\it convex} regression problem
for the estimator
$\widehat{\mfrak}$,
\begin{equation}  
\widehat{\mfrak}\in \arg \min_{\ufrak \in \Hcb} \,
\frac{1}{N} \sum_{i=1}^N 
\left( 
\, \ufrak(\xi_i) - x_i
\right)^2,
\end{equation}
for some space
$\Hcb$ of univariate monotone maps (Appendix \ref{sec:param_tri}).

We can further use %
localization 
ideas %
from
Section \ref{sec:est_KR}
and Remark \ref{rem:localization} to regularize the estimation of $T$ in 
high dimensions.

\begin{remark}[Connection with the multivariate rank 
histogram filter]
Intuitively, %
the
multivariate rank histogram
filter %
implements a 
nonparametric estimator of a
deterministic 
update 
similar to
\eqref{eq:update_detmap_single}, exploiting the
factorization \eqref{eq:factorization_post}
of the posterior density. %
The idea is to (1) apply Bayes' rule locally---along the observed component---combining the histogram 
approximation of 
$\pi_{X_1}$ %
suggested 
by \cite{anderson2010non} with the
likelihood function $\pi_{Y\vert X_1}$ to yield
a piecewise constant approximation of the posterior marginal 
$\pi_{X_1\vert y^*}$, and (2)
propagate information to the unobserved
components by computing a sequence of 
(nonparametric)
{\it particle} approximations of the
conditionals 
$\pi_{X_k \vert \Xb_{1:k-1}}$, for $k\ge 2$.
Recall that the conditionals $\pi_{X_k \vert \Xb_{1:k-1}}$ 
fully 
characterize %
the KR
rearrangement $S$ in \eqref{eq:update_detmap_single}.
In high dimensions, it is increasingly challenging to deploy
particle approximations of the 
conditionals
$\pi_{X_k \vert \Xb_{1:k-1}}$, and thus \cite{metref2014non} resorts
to a mean-field approximation of the posterior density, %
\begin{equation}
  \pi_{\Xb \vert y^*} \approx \pi_{X_1 \vert y^*}\,\prod_{k=2}^n  
  \pi_{X_k \vert X_1},
\end{equation}
which %
relies 
only 
on the characterization of bivariate conditionals.
\end{remark}

\section{Continuous ranked probability score evaluations}
\label{sec:CRPS_results}
To assess the predictive performance of our ensembles, we compute the continuous ranked probability score (CRPS)~\cite{brocker2012evaluating, gneiting2007probabilistic}. The CRPS for the $i$th marginal component of the state at step $k$ compares the ensemble's empirical CDF $\widehat{F}_{k,i}(z)$ with a Heaviside function centered at the true state, $H(z - (\zb^{*}_{k})_{i})$, as given by $\text{CRPS}_{k,i} = \int (\widehat{F}_{k,i}(z) - \mathbbm{1}((\zb^{*}_{k})_{i} \leq z))^2 dz$. We average $\text{CRPS}_{k,i}$ over state components $i$ and assimilation steps $k$.

\begin{figure}[!ht]
  \centering
  \includegraphics[width=0.45\linewidth]{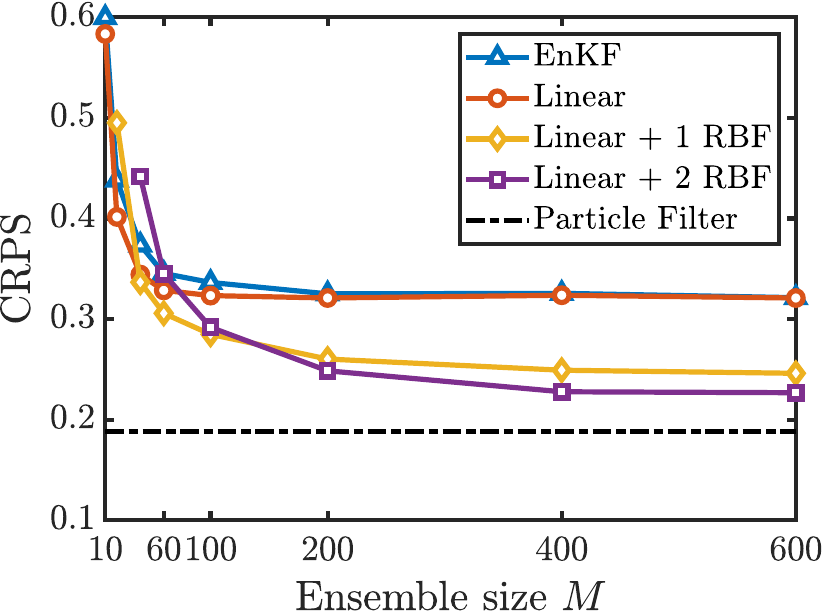}
  \hspace{1cm}
  \includegraphics[width=0.45\linewidth]{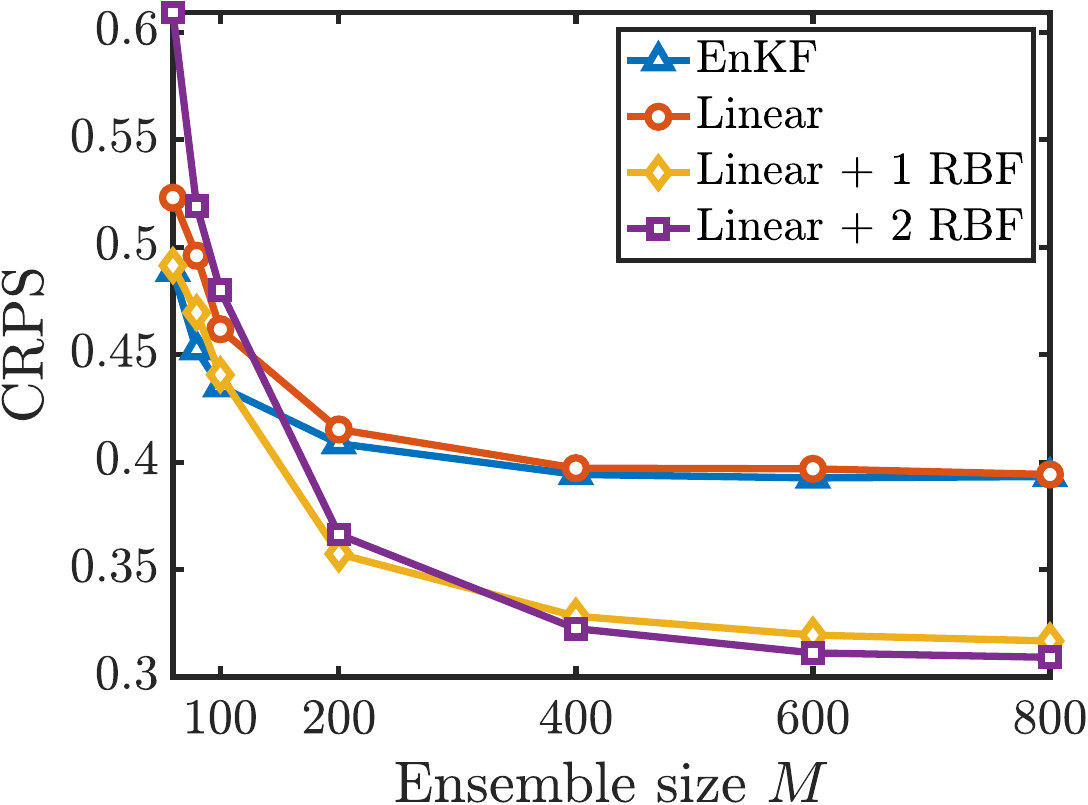}%
  \caption{Average continuous rank probability score (CRPS) of the analysis ensemble for our Lorenz-63 experiments ({\it left}) and the ``hard'' Lorenz-96 configuration ({\it right}). \label{fig:lorenz_models_crps}}
  \vspace{-10pt}
\end{figure}

For the Lorenz-63 model in Section~\ref{sec:lorenz63}, we compute the CRPS and report the average, over all three state components and $2000$ assimilation cycles, for increasing $M$ in Figure~\ref{fig:lorenz_models_crps} (left). For $M \geq 60$ samples, the nonlinear stochastic map filter yields lower values of CRPS, which provides a simultaneous diagnostic of calibration and sharpness as defined in~\cite{gneiting2007probabilistic}. With increasing nonlinearity in the prior-to-posterior transformation, the average CRPS for the Lorenz-63 model approaches that of a particle filter using $10^{6}$ samples.

For the Lorenz-96 configuration in Section~\ref{sec:bickel_hardcase} we compute the average CRPS (over the $n=40$ state components and $2000$ assimilation cycles) and report the results in Figure~\ref{fig:lorenz_models_crps} (right). With $M \geq 100$ samples, the nonlinear stochastic map filter yields a lower CRPS than its linear counterparts. Adding more nonlinearity to the transformation allows the score to improve even further with increasing $M$.

\section{Effect of time discretization}
\label{sec:Inverse_crime}
\begin{figure}[!ht]
  \centering
  \includegraphics[width=0.45\linewidth]{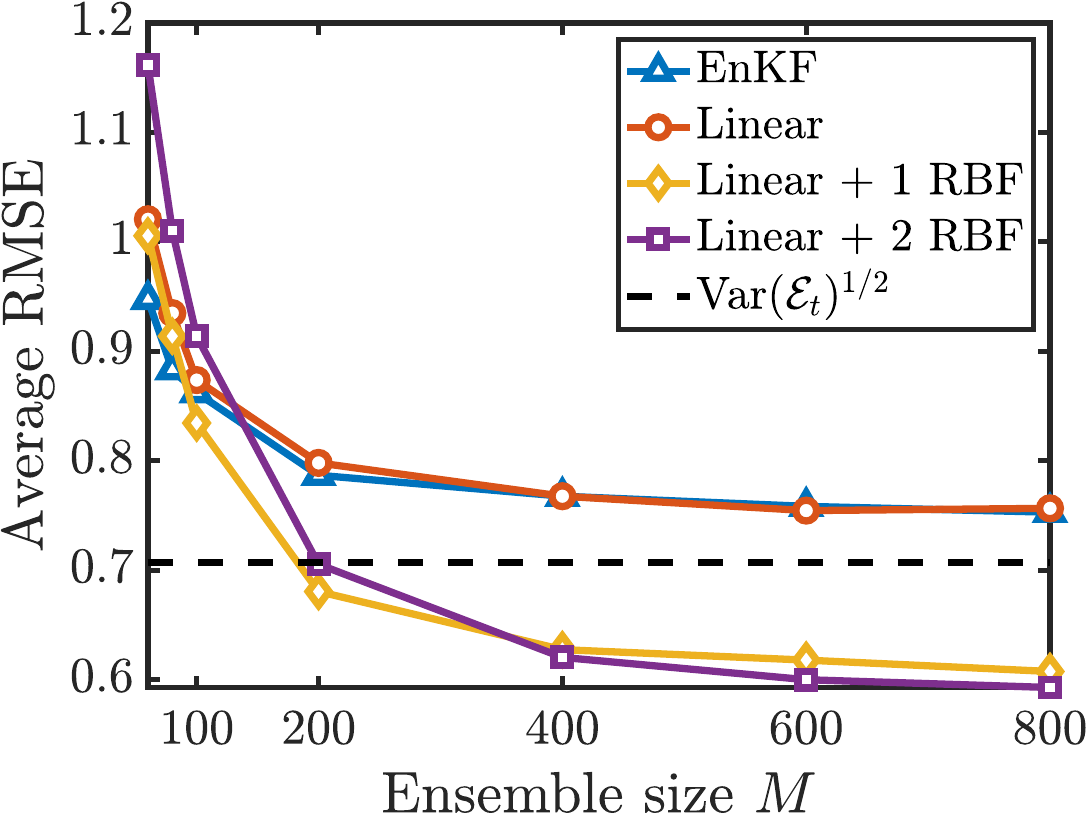}
  \hspace{1cm}
  \includegraphics[width=0.45\linewidth]{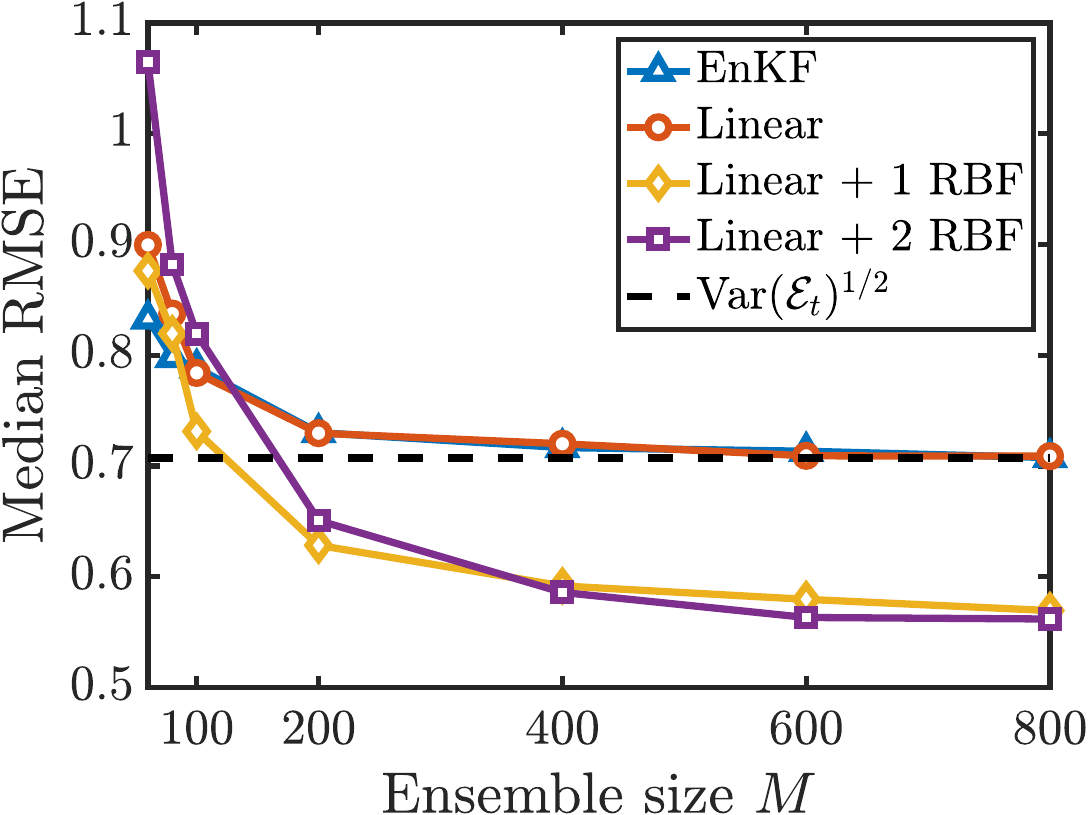}
  \caption{Average ({\it left}) and median ({\it right}) RMSE (over $2000$ assimilation cycles) for the ``hard'' Lorenz-96 configuration. Problem setup is identical to Section~\ref{sec:bickel_hardcase}, except that we use a finer numerical discretization ($\Delta t = 0.001$) to generate the observations.} \label{fig:lorenz96_hard_RMSE_IC}
\end{figure}

\begin{figure}[!ht]
  \centering
  \includegraphics[width=0.45\linewidth]{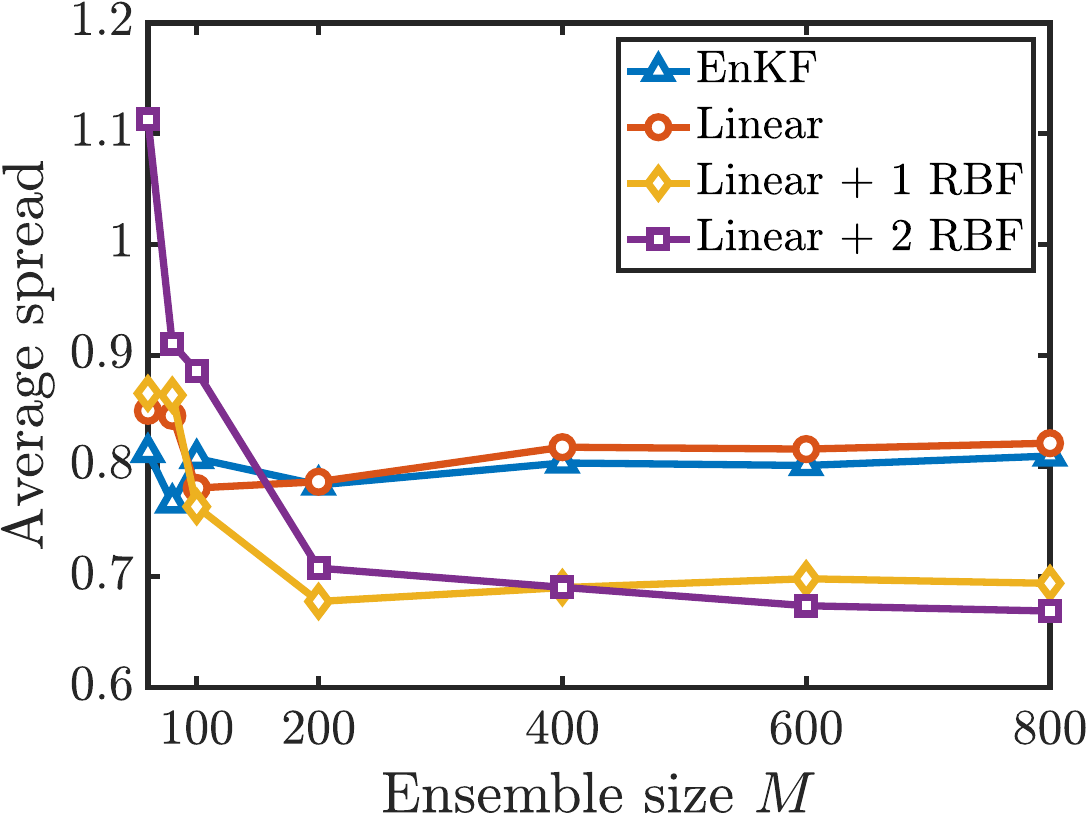}
  \hspace{1cm}
  \includegraphics[width=0.45\linewidth]{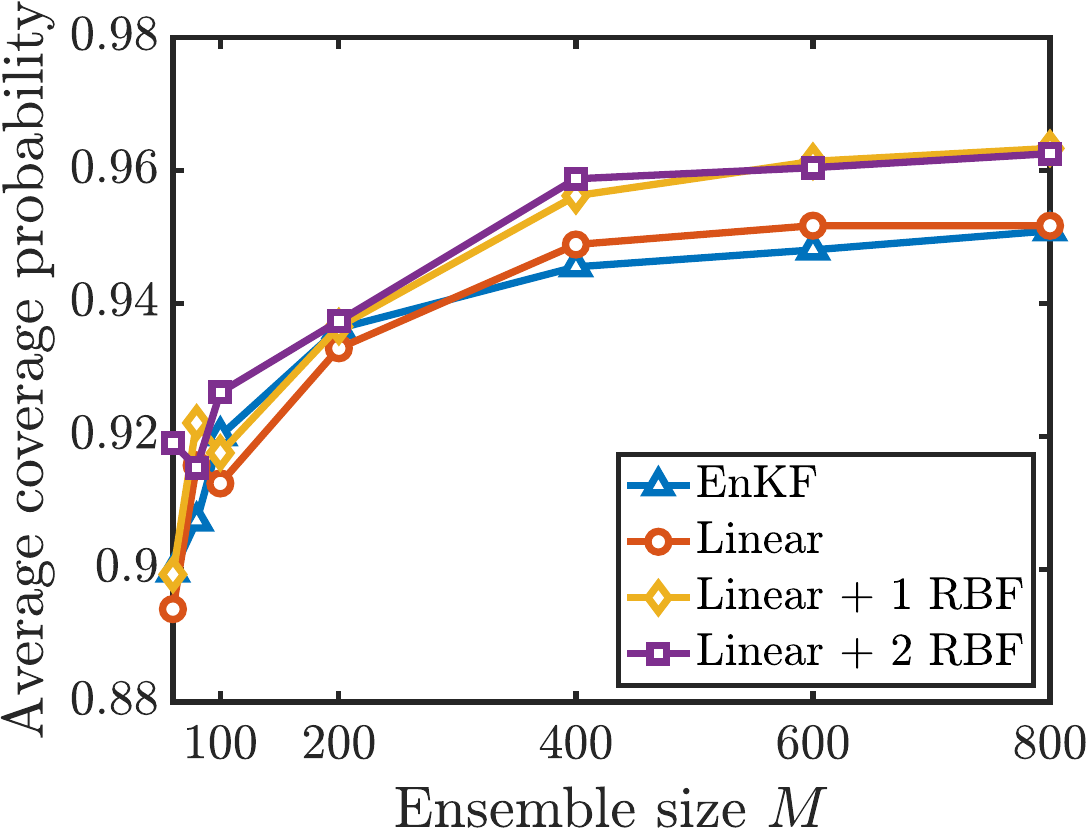}
  \caption{Average spread ({\it left}) and marginal coverage probabilities ({\it right}) (over $2000$ assimilation cycles) for the ``hard'' Lorenz-96 configuration. Problem setup is identical to Section~\ref{sec:bickel_hardcase}, except that we use a finer numerical discretization ($\Delta t = 0.001$) to generate the observations.} 
\label{fig:lorenz96_hard_spread_covprob_IC}
\end{figure}

In this section, we repeat the Lorenz-96 filtering experiment of Section~\ref{sec:bickel_hardcase}, but use a smaller stepsize $\Delta t = 0.001$ when performing numerical integration of \eqref{eq:system_lorenz} to generate the sequence of true hidden states and the synthetic observations $(\yb_k^*)_k$. We maintain the larger stepsize $\Delta t = 0.01$ for numerical integration within each forecast step, prior to assimilating data. This mismatch ensures that there is no ``inverse crime''~\cite{kaipio2007statistical} wherein the same model is used to synthesize and to invert the data. Figure~\ref{fig:lorenz96_hard_RMSE_IC} plots the average and median RMSE over $2000$ assimilation cycles for different ensemble sizes and map parameterizations. This figure matches the trends in Figure~\ref{fig:lorenz96_hard_meanRMSE}, which used identical twin experiments (i.e., the same stepsize $\Delta t = 0.01$ both to generate the observations and to run the filtering algorithms). Figure~\ref{fig:lorenz96_hard_spread_covprob_IC} plots the average spread and marginal coverage probabilities of the ensemble; again, it closely follows the trends in Figure~\ref{fig:lorenz96_hard_spread_covprob}. 
We believe this insensitivity is in part due to the underlying chaotic dynamics of \eqref{eq:system_lorenz}, as well as the noise in the observations; both prevent the model from recovering the true state exactly. Chaotic dynamics, in particular, may prevent any spurious reduction in error that might otherwise come from using common step sizes. 
For the filtering problems in Sections~\ref{sec:lorenz63}, \ref{sec:lorenz_heavycase}, and \ref{sec:lorenz_nonlinearobs} we thus expect a similar insensitivity to the time discretization used to generate the observations, and we omit these comparisons for brevity.

\section{Numerical results for Lorenz-96 model with heavy-tailed observational noise} 
\label{sec:lorenz_heavycase}

A second challenging configuration of the Lorenz-96 model is obtained
with sparse and \emph{heavy-tailed} observations.  In this setting, we
consider $\Delta t_{\rm obs} = 0.1$ and $d = 10$ (observing every
fourth component of the state), and additive independent
Laplace (i.e., double-exponential)-distributed observational
noise. The noise has variance $2\theta^2$, where $\theta = 1$ is the
scale parameter of the Laplace distribution. The large variance of the noise and limited number of observations make filtering difficult; for instance, large RMSE values are obtained when using the EnKF with ensemble sizes smaller than $M = 100$.

\begin{figure}[!ht]
  \centering
  \includegraphics[width=.45\linewidth]{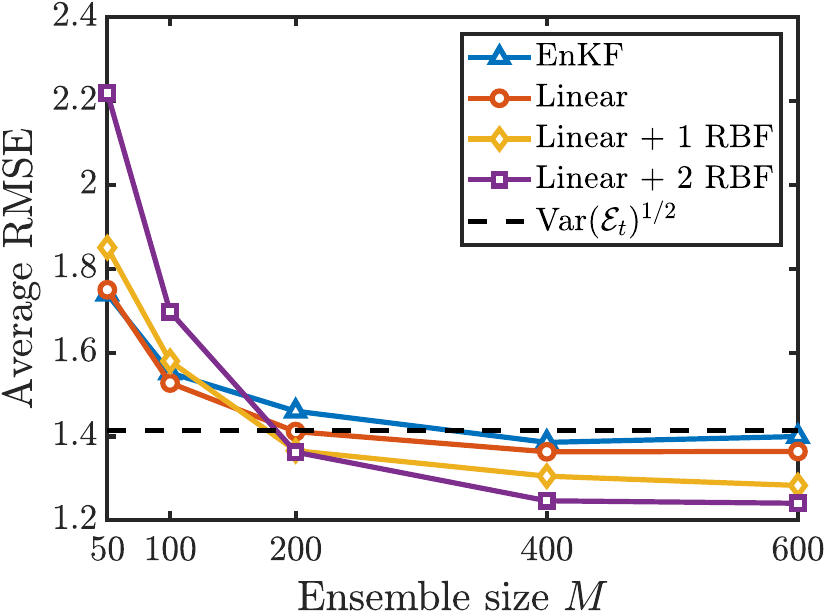}
  \hspace{1cm}
  \includegraphics[width=.45\linewidth]{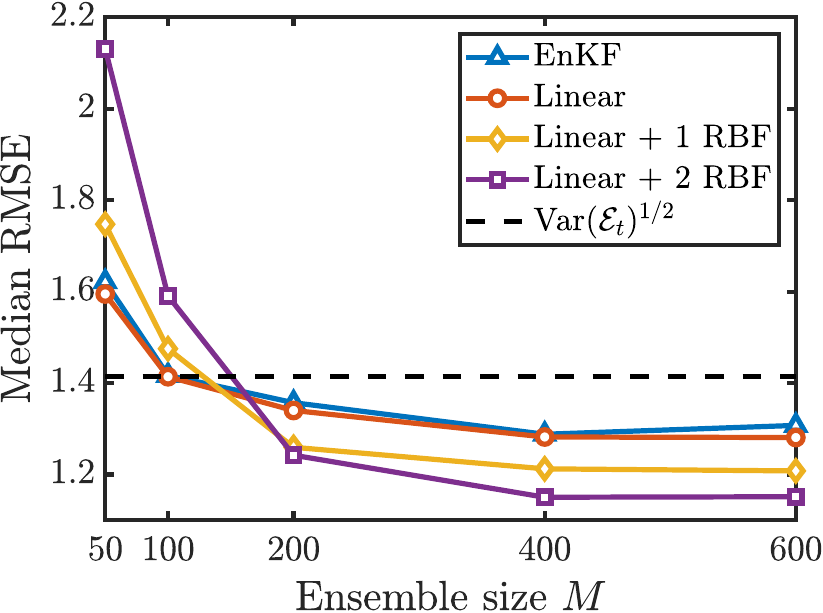}

  \caption{Average ({\it left}) and median ({\it right}) RMSE (over $2000$ assimilation cycles) for the heavy-tailed (Laplace observational noise) Lorenz-96 configuration of Section~\ref{sec:lorenz_heavycase}, with $\Delta t_{obs} = 0.1$ and $d = 10$ observations. Dashed line is the standard deviation of the observational noise.\label{fig:lorenz96_heavy_meanRMSE}}
\end{figure}

\begin{figure}[!ht]
  \centering
  \includegraphics[width=0.45\linewidth]{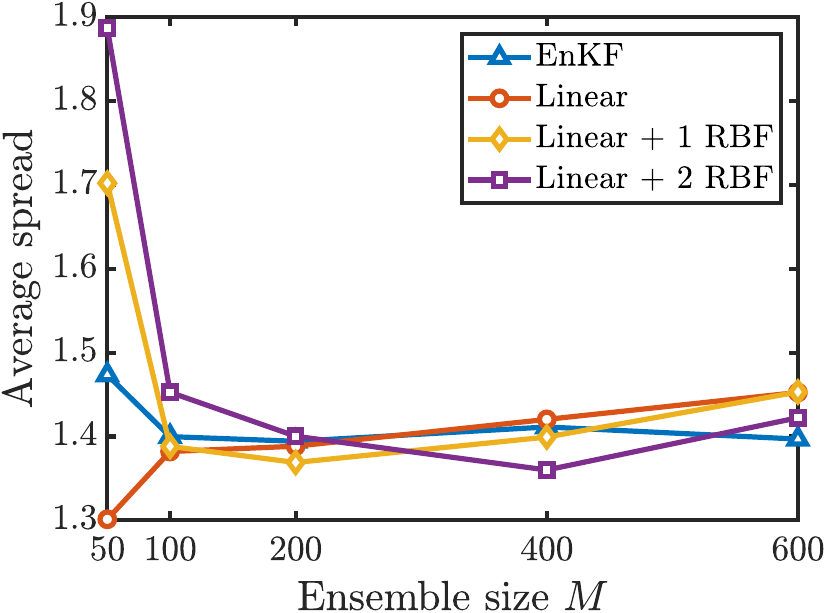}
  \hspace{1cm}
  \includegraphics[width=0.45\linewidth]{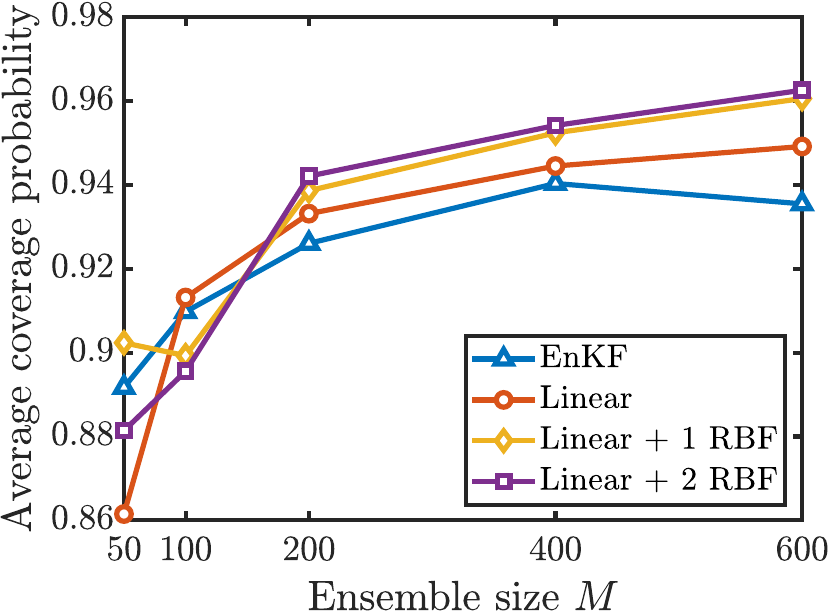}
  \caption{Average ensemble spread ({\it left}) and average coverage probability of the $[2.5\%, 97.5\%]$ empirical intervals of each state marginal ({\it right}) for the heavy-tailed Lorenz-96 configuration. \label{fig:lorenz96_heavy_spread_covprob}}
\end{figure}

Similar to the results in Section~\ref{sec:bickel_hardcase}, we observe improved tracking performance with increasing nonlinearity in the map $S^{\Xcb}$, provided $M$ is not too small. Figure~\ref{fig:lorenz96_heavy_meanRMSE} shows the average and median RMSE for the EnKF and different stochastic map parameterizations. As $M$ increases, the bias-variance tradeoff shifts in favor of more complex maps. Furthermore, as illustrated in Figure~\ref{fig:lorenz96_heavy_spread_covprob}, higher order maps also increase the average coverage probability of the ensemble across $2000$ assimilation steps without increasing the ensemble spread. These results suggest that, by increasing the complexity of the prior-to-posterior transformation, ensemble members from the stochastic map filter better represent the true state $\zb_{k}^{*}$ and its uncertainty for problems with non-Gaussian forecast and analysis distributions.

\section{Numerical results for Lorenz-96 model with nonlinear observations}
\label{sec:lorenz_nonlinearobs}
{In this experiment, we evaluate the applicability of the stochastic map filter to more general likelihood models by filtering the Lorenz-96 system with \emph{nonlinear observations} of the state. We begin with the ``hard case'' configuration of the Lorenz-96 system in Section~\ref{sec:bickel_hardcase}, with $\Delta t_{\text{obs}} = 0.4$ (long time between observations) and $d = 20$ observations at each assimilation step (observing every other component of the state). But we modify the observation model:
now each component of $\Yb_k \in \mathbb{R}^{d}$ results from a nonlinear square root transformation of the corresponding state component, perturbed by additive Gaussian noise. In other words, letting $\{Y_{k,i}\}_{i=1}^d$ denote the components of $\Yb_k$, we let each $Y_{k,i} = \text{sign}(Z_{k,2i-1})\sqrt{|Z_{k,2i-1}|} + \mathcal{E}_{k,i}$, where $\{ \mathcal{E}_{k,i} \}_{i=1}^d \stackrel{\text{iid}}{\sim} \mathcal{N}(0,\theta^2)$ are independent of $\Zb_k$. To maintain a similar signal-to-noise ratio as in the direct observation case, we use an observational noise variance of $\theta^2 = 0.25$. A similar observation model was also studied in~\cite{anderson2020marginal}. In this experiment, we use %
the map parameterization described in Section~\ref{sec:commonsetup}, with component functions $U^k$ that depend linearly on the state variable $z_k$ and nonlinearly on the other variables, %
for all $k$.}%

\begin{figure}[!ht]
  \centering
  \includegraphics[width=0.45\linewidth]{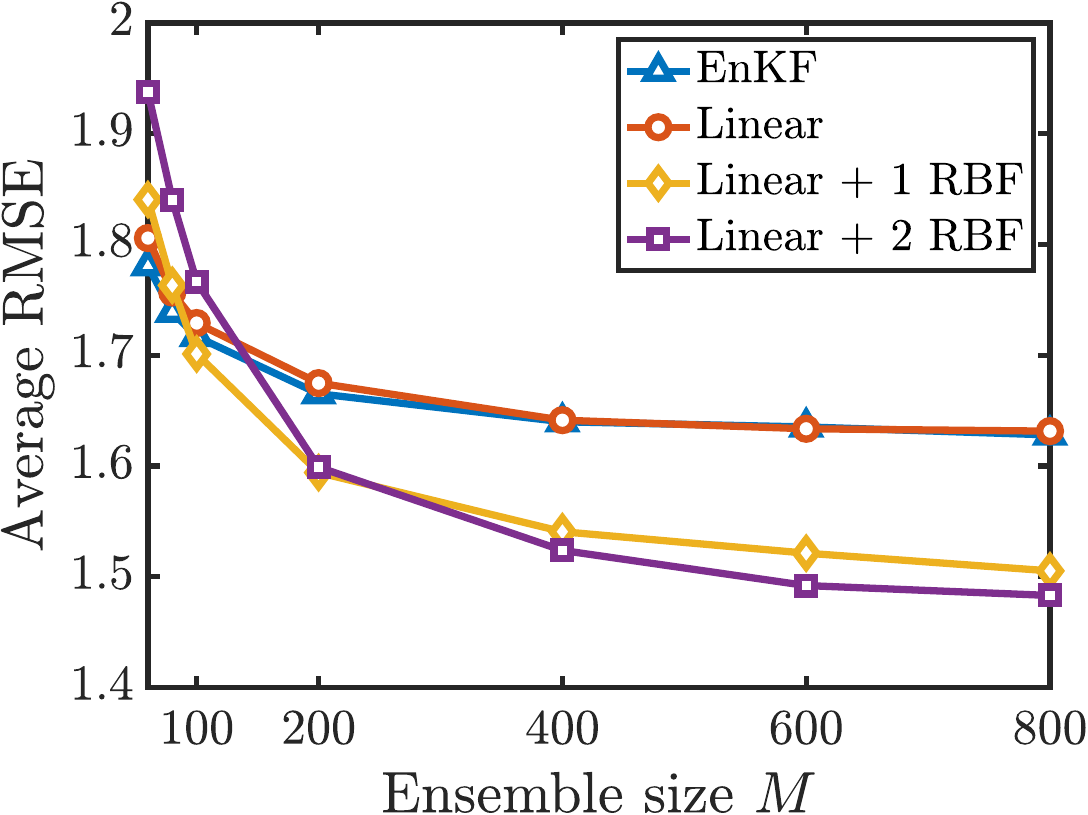}
  \hspace{1cm}
  \includegraphics[width=0.45\linewidth]{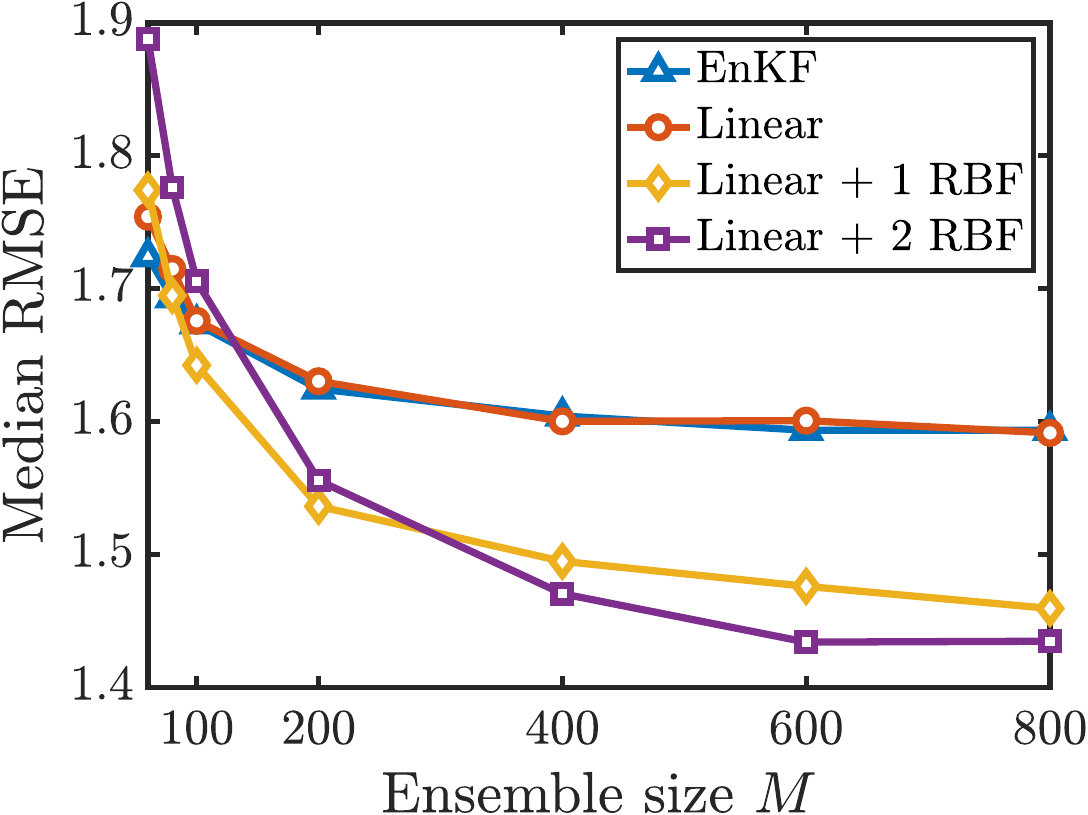}
  \caption{Average ({\it left}) and median ({\it right}) RMSE (over $2000$ assimilation cycles) for the Lorenz-96 configuration of Section~\ref{sec:bickel_hardcase}, with long inter-observation times, modified to have $d = 20$ \emph{square-root} observations of the state.} \label{fig:lorenz96_sqrt_RMSE}
\end{figure}

\begin{figure}[!ht]
  \centering
  \includegraphics[width=0.45\linewidth]{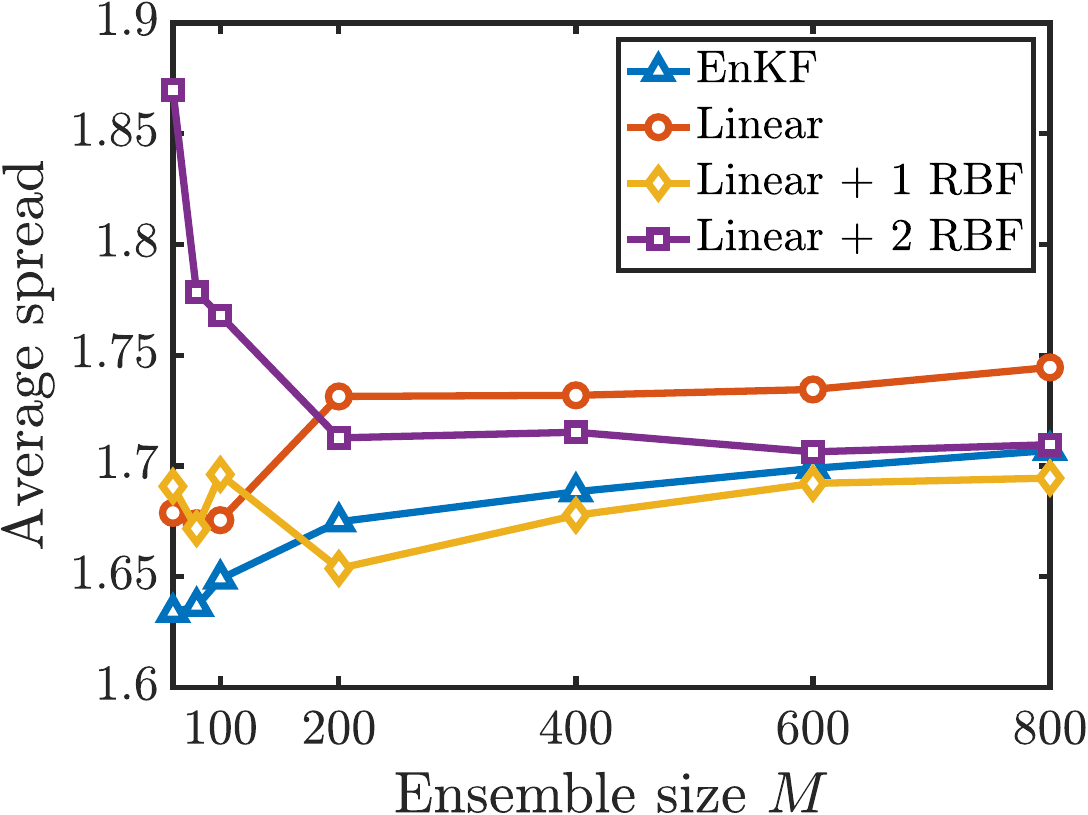}
  \hspace{1cm}
  \includegraphics[width=0.45\linewidth]{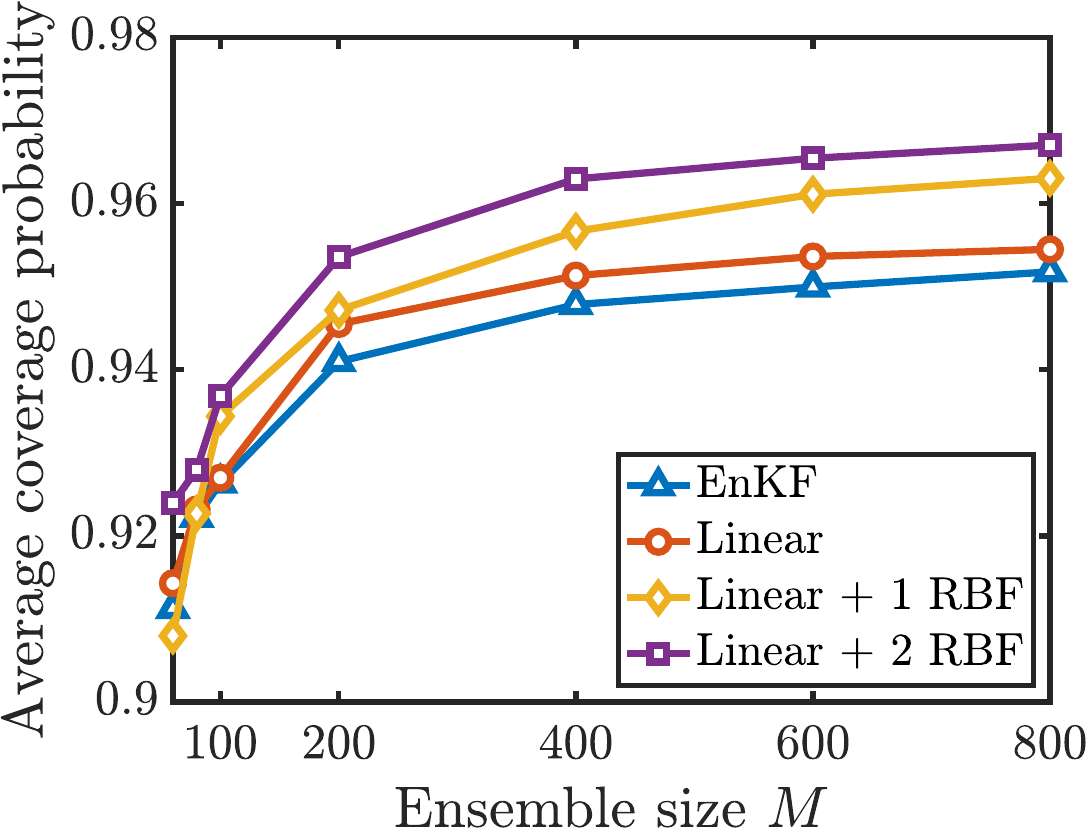}
  \caption{Average ensemble spread ({\it left}) and average coverage probability of the $[2.5\%, 97.5\%]$ empirical intervals of each state marginal ({\it right}) for the Lorenz-96 configuration of Section~\ref{sec:bickel_hardcase}, with long inter-observation times, modified to have $d=20$ \emph{square-root} observations of the state.}
 \label{fig:lorenz96_sqrt_spread_covprob}
\end{figure}

{Figure~\ref{fig:lorenz96_sqrt_RMSE} plots the average and median RMSE for the stochastic EnKF and stochastic map filters, over 2000 assimilation cycles. Similar to the results in Section~\ref{sec:bickel_hardcase}, the stochastic EnKF and linear maps yield the lowest average RMSE at small ensemble size $M$, but their performance plateaus with increasing $M$. Nonlinear maps yield even lower RMSE for progressively larger ensemble sizes. These trends reflect the same bias-variance tradeoff as in our previous examples.
Figure~\ref{fig:lorenz96_sqrt_spread_covprob} shows that the marginal coverage probability increases monotonically with increasing map complexity,
though at the same time more complex maps generally achieve smaller ensemble spread.}

\def\bibfont{\small}
\bibliographystyle{imsart-number}
\bibliography{mapJournalbib}  

\begin{thebibliography}{80}

\bibitem{acevedo2017second}
\begin{barticle}[author]
\bauthor{\bsnm{Acevedo},~\bfnm{Walter}\binits{W.}}, \bauthor{\bparticle{de}
  \bsnm{Wiljes},~\bfnm{Jana}\binits{J.}} \AND
  \bauthor{\bsnm{Reich},~\bfnm{Sebastian}\binits{S.}}
(\byear{2017}).
\btitle{Second-order accurate ensemble transform particle filters}.
\bjournal{SIAM Journal on Scientific Computing}
\bvolume{39}
\bpages{A1834--A1850}.
\end{barticle}
\endbibitem

\bibitem{amezcua2014gaussian}
\begin{barticle}[author]
\bauthor{\bsnm{Amezcua},~\bfnm{Javier}\binits{J.}} \AND
  \bauthor{\bsnm{Van~Leeuwen},~\bfnm{Peter~Jan}\binits{P.~J.}}
(\byear{2014}).
\btitle{Gaussian anamorphosis in the analysis step of the {EnKF}: a joint
  state-variable/observation approach}.
\bjournal{Tellus A: Dynamic Meteorology and Oceanography}
\bvolume{66}
\bpages{23493}.
\end{barticle}
\endbibitem

\bibitem{anderes2012general}
\begin{barticle}[author]
\bauthor{\bsnm{Anderes},~\bfnm{E.}\binits{E.}} \AND
  \bauthor{\bsnm{Coram},~\bfnm{M.}\binits{M.}}
(\byear{2012}).
\btitle{A general spline representation for nonparametric and semiparametric
  density estimates using diffeomorphisms}.
\bjournal{arXiv:1205.5314}.
\end{barticle}
\endbibitem

\bibitem{anderson2010non}
\begin{barticle}[author]
\bauthor{\bsnm{Anderson},~\bfnm{Jeffrey~L}\binits{J.~L.}}
(\byear{2010}).
\btitle{A non-{G}aussian ensemble filter update for data assimilation}.
\bjournal{Monthly Weather Review}
\bvolume{138}
\bpages{4186--4198}.
\end{barticle}
\endbibitem

\bibitem{anderson2020marginal}
\begin{barticle}[author]
\bauthor{\bsnm{Anderson},~\bfnm{Jeffrey~L}\binits{J.~L.}}
(\byear{2020}).
\btitle{A Marginal Adjustment Rank Histogram Filter for Non-{G}aussian Ensemble
  Data Assimilation}.
\bjournal{Monthly Weather Review}
\bvolume{148}
\bpages{3361--3378}.
\end{barticle}
\endbibitem

\bibitem{baptista2020adaptive}
\begin{barticle}[author]
\bauthor{\bsnm{Baptista},~\bfnm{Ricardo}\binits{R.}},
  \bauthor{\bsnm{Zahm},~\bfnm{Olivier}\binits{O.}} \AND
  \bauthor{\bsnm{Marzouk},~\bfnm{Youssef}\binits{Y.}}
(\byear{2020}).
\btitle{An adaptive transport framework for joint and conditional density
  estimation}.
\bjournal{arXiv preprint arXiv:2009.10303}.
\end{barticle}
\endbibitem

\bibitem{bengtsson2003toward}
\begin{barticle}[author]
\bauthor{\bsnm{Bengtsson},~\bfnm{Thomas}\binits{T.}},
  \bauthor{\bsnm{Snyder},~\bfnm{Chris}\binits{C.}} \AND
  \bauthor{\bsnm{Nychka},~\bfnm{Doug}\binits{D.}}
(\byear{2003}).
\btitle{Toward a nonlinear ensemble filter for high-dimensional systems}.
\bjournal{Journal of Geophysical Research: Atmospheres}
\bvolume{108}.
\end{barticle}
\endbibitem

\bibitem{bertsekas1982projected}
\begin{barticle}[author]
\bauthor{\bsnm{Bertsekas},~\bfnm{Dimitri~P}\binits{D.~P.}}
(\byear{1982}).
\btitle{Projected {N}ewton methods for optimization problems with simple
  constraints}.
\bjournal{SIAM Journal on control and Optimization}
\bvolume{20}
\bpages{221--246}.
\end{barticle}
\endbibitem

\bibitem{bickel2008regularized}
\begin{barticle}[author]
\bauthor{\bsnm{Bickel},~\bfnm{P.~J.}\binits{P.~J.}} \AND
  \bauthor{\bsnm{Levina},~\bfnm{E.}\binits{E.}}
(\byear{2008}).
\btitle{Regularized estimation of large covariance matrices}.
\bjournal{The Annals of Statistics}
\bvolume{36}
\bpages{199--227}.
\end{barticle}
\endbibitem

\bibitem{bigoni2016monotone}
\begin{barticle}[author]
\bauthor{\bsnm{Bigoni},~\bfnm{D.}\binits{D.}},
  \bauthor{\bsnm{Spantini},~\bfnm{A.}\binits{A.}} \AND
  \bauthor{\bsnm{Marzouk},~\bfnm{Y.}\binits{Y.}}
(\byear{2019}).
\btitle{On the computation of monotone transports}.
\bjournal{In preparation}.
\end{barticle}
\endbibitem

\bibitem{bogachev2005triangular}
\begin{barticle}[author]
\bauthor{\bsnm{Bogachev},~\bfnm{V.~I.}\binits{V.~I.}},
  \bauthor{\bsnm{Kolesnikov},~\bfnm{A.~V.}\binits{A.~V.}} \AND
  \bauthor{\bsnm{Medvedev},~\bfnm{K.~V.}\binits{K.~V.}}
(\byear{2005}).
\btitle{Triangular transformations of measures}.
\bjournal{Sbornik: Mathematics}
\bvolume{196}
\bpages{309}.
\end{barticle}
\endbibitem

\bibitem{box1964analysis}
\begin{barticle}[author]
\bauthor{\bsnm{Box},~\bfnm{George~EP}\binits{G.~E.}} \AND
  \bauthor{\bsnm{Cox},~\bfnm{David~R}\binits{D.~R.}}
(\byear{1964}).
\btitle{An analysis of transformations}.
\bjournal{Journal of the Royal Statistical Society. Series B (Methodological)}
\bpages{211--252}.
\end{barticle}
\endbibitem

\bibitem{bigoni2019greedy}
\begin{binproceedings}[author]
\bauthor{\bsnm{Brennan},~\bfnm{Michael}\binits{M.}},
  \bauthor{\bsnm{Bigoni},~\bfnm{Daniele}\binits{D.}},
  \bauthor{\bsnm{Zahm},~\bfnm{Olivier}\binits{O.}},
  \bauthor{\bsnm{Spantini},~\bfnm{Alessio}\binits{A.}} \AND
  \bauthor{\bsnm{Marzouk},~\bfnm{Youssef}\binits{Y.}}
(\byear{2020}).
\btitle{Greedy inference with structure-exploiting lazy maps}.
In \bbooktitle{Advances in Neural Information Processing Systems}
\bpages{8330--8342}.
\end{binproceedings}
\endbibitem

\bibitem{brocker2012evaluating}
\begin{barticle}[author]
\bauthor{\bsnm{Br{\"o}cker},~\bfnm{Jochen}\binits{J.}}
(\byear{2012}).
\btitle{Evaluating raw ensembles with the continuous ranked probability score}.
\bjournal{Quarterly Journal of the Royal Meteorological Society}
\bvolume{138}
\bpages{1611--1617}.
\end{barticle}
\endbibitem

\bibitem{burgers1998analysis}
\begin{barticle}[author]
\bauthor{\bsnm{Burgers},~\bfnm{Gerrit}\binits{G.}},
  \bauthor{\bparticle{Jan~van} \bsnm{Leeuwen},~\bfnm{Peter}\binits{P.}} \AND
  \bauthor{\bsnm{Evensen},~\bfnm{Geir}\binits{G.}}
(\byear{1998}).
\btitle{Analysis scheme in the ensemble {K}alman filter}.
\bjournal{Monthly weather review}
\bvolume{126}
\bpages{1719--1724}.
\end{barticle}
\endbibitem

\bibitem{chorin2009implicit}
\begin{barticle}[author]
\bauthor{\bsnm{Chorin},~\bfnm{A.~J.}\binits{A.~J.}} \AND
  \bauthor{\bsnm{Tu},~\bfnm{X.}\binits{X.}}
(\byear{2009}).
\btitle{Implicit sampling for particle filters}.
\bjournal{Proceedings of the National Academy of Sciences}
\bvolume{106}
\bpages{17249--17254}.
\end{barticle}
\endbibitem

\bibitem{cranmer2020frontier}
\begin{barticle}[author]
\bauthor{\bsnm{Cranmer},~\bfnm{Kyle}\binits{K.}},
  \bauthor{\bsnm{Brehmer},~\bfnm{Johann}\binits{J.}} \AND
  \bauthor{\bsnm{Louppe},~\bfnm{Gilles}\binits{G.}}
(\byear{2020}).
\btitle{The frontier of simulation-based inference}.
\bjournal{Proceedings of the National Academy of Sciences}
\bvolume{117}
\bpages{30055--30062}.
\end{barticle}
\endbibitem

\bibitem{crisan2002survey}
\begin{barticle}[author]
\bauthor{\bsnm{Crisan},~\bfnm{D.}\binits{D.}} \AND
  \bauthor{\bsnm{Doucet},~\bfnm{A.}\binits{A.}}
(\byear{2002}).
\btitle{A survey of convergence results on particle filtering methods for
  practitioners}.
\bjournal{IEEE Transactions on Signal Processing}
\bvolume{50}
\bpages{736--746}.
\end{barticle}
\endbibitem

\bibitem{cui2014likelihood}
\begin{barticle}[author]
\bauthor{\bsnm{Cui},~\bfnm{T.}\binits{T.}},
  \bauthor{\bsnm{Martin},~\bfnm{J.}\binits{J.}},
  \bauthor{\bsnm{Marzouk},~\bfnm{Y.}\binits{Y.}},
  \bauthor{\bsnm{Solonen},~\bfnm{A.}\binits{A.}} \AND
  \bauthor{\bsnm{Spantini},~\bfnm{A.}\binits{A.}}
(\byear{2014}).
\btitle{Likelihood-informed dimension reduction for nonlinear inverse
  problems}.
\bjournal{Inverse Problems}
\bvolume{30}
\bpages{114015}.
\end{barticle}
\endbibitem

\bibitem{daum2008particle}
\begin{binproceedings}[author]
\bauthor{\bsnm{Daum},~\bfnm{F.}\binits{F.}} \AND
  \bauthor{\bsnm{Huang},~\bfnm{J.}\binits{J.}}
(\byear{2008}).
\btitle{Particle flow for nonlinear filters with log-homotopy}.
In \bbooktitle{SPIE Defense and Security Symposium}
\bpages{696918--696918}.
\bpublisher{International Society for Optics and Photonics}.
\end{binproceedings}
\endbibitem

\bibitem{del2017stability}
\begin{barticle}[author]
\bauthor{\bsnm{Del~Moral},~\bfnm{Pierre}\binits{P.}},
  \bauthor{\bsnm{Kurtzmann},~\bfnm{Aline}\binits{A.}} \AND
  \bauthor{\bsnm{Tugaut},~\bfnm{Julian}\binits{J.}}
(\byear{2017}).
\btitle{On the stability and the uniform propagation of chaos of a class of
  extended ensemble {K}alman--{B}ucy filters}.
\bjournal{SIAM Journal on Control and Optimization}
\bvolume{55}
\bpages{119--155}.
\end{barticle}
\endbibitem

\bibitem{detommaso2018stein}
\begin{binproceedings}[author]
\bauthor{\bsnm{Detommaso},~\bfnm{Gianluca}\binits{G.}},
  \bauthor{\bsnm{Cui},~\bfnm{Tiangang}\binits{T.}},
  \bauthor{\bsnm{Marzouk},~\bfnm{Youssef}\binits{Y.}},
  \bauthor{\bsnm{Spantini},~\bfnm{Alessio}\binits{A.}} \AND
  \bauthor{\bsnm{Scheichl},~\bfnm{Robert}\binits{R.}}
(\byear{2018}).
\btitle{A Stein variational {N}ewton method}.
In \bbooktitle{Advances in Neural Information Processing Systems}
\bpages{9169--9179}.
\end{binproceedings}
\endbibitem

\bibitem{doucet2009tutorial}
\begin{barticle}[author]
\bauthor{\bsnm{Doucet},~\bfnm{A.}\binits{A.}} \AND
  \bauthor{\bsnm{Johansen},~\bfnm{A.~M.}\binits{A.~M.}}
(\byear{2009}).
\btitle{A tutorial on particle filtering and smoothing: Fifteen years later}.
\bjournal{Handbook of nonlinear filtering}
\bvolume{12}
\bpages{3}.
\end{barticle}
\endbibitem

\bibitem{douglas1999applications}
\begin{binproceedings}[author]
\bauthor{\bsnm{Douglas},~\bfnm{R.~J.}\binits{R.~J.}}
(\byear{1999}).
\btitle{Applications of the {M}onge-{A}mpere equation and {M}onge transport
  problem to meteorology and oceanography}.
In \bbooktitle{Monge Amp{\`e}re Equation: Applications to Geometry and
  Optimization}
\bvolume{226}
\bpages{33}.
\bpublisher{American Mathematical Soc.}
\end{binproceedings}
\endbibitem

\bibitem{durbin2012time}
\begin{bbook}[author]
\bauthor{\bsnm{Durbin},~\bfnm{James}\binits{J.}} \AND
  \bauthor{\bsnm{Koopman},~\bfnm{Siem~Jan}\binits{S.~J.}}
(\byear{2012}).
\btitle{Time series analysis by state space methods}
\bvolume{38}.
\bpublisher{Oxford University Press}.
\end{bbook}
\endbibitem

\bibitem{evensen2007data}
\begin{bbook}[author]
\bauthor{\bsnm{Evensen},~\bfnm{G.}\binits{G.}}
(\byear{2007}).
\btitle{Data Assimilation}.
\bpublisher{Springer}.
\end{bbook}
\endbibitem

\bibitem{frei2013bridging}
\begin{barticle}[author]
\bauthor{\bsnm{Frei},~\bfnm{Marco}\binits{M.}} \AND
  \bauthor{\bsnm{K{\"u}nsch},~\bfnm{Hans~R}\binits{H.~R.}}
(\byear{2013}).
\btitle{Bridging the ensemble {K}alman and particle filters}.
\bjournal{Biometrika}
\bvolume{100}
\bpages{781--800}.
\end{barticle}
\endbibitem

\bibitem{gneiting2007probabilistic}
\begin{barticle}[author]
\bauthor{\bsnm{Gneiting},~\bfnm{Tilmann}\binits{T.}},
  \bauthor{\bsnm{Balabdaoui},~\bfnm{Fadoua}\binits{F.}} \AND
  \bauthor{\bsnm{Raftery},~\bfnm{Adrian~E}\binits{A.~E.}}
(\byear{2007}).
\btitle{Probabilistic forecasts, calibration and sharpness}.
\bjournal{Journal of the Royal Statistical Society: Series B (Statistical
  Methodology)}
\bvolume{69}
\bpages{243--268}.
\end{barticle}
\endbibitem

\bibitem{goodfellow2014generative}
\begin{binproceedings}[author]
\bauthor{\bsnm{Goodfellow},~\bfnm{Ian}\binits{I.}},
  \bauthor{\bsnm{Pouget-Abadie},~\bfnm{Jean}\binits{J.}},
  \bauthor{\bsnm{Mirza},~\bfnm{Mehdi}\binits{M.}},
  \bauthor{\bsnm{Xu},~\bfnm{Bing}\binits{B.}},
  \bauthor{\bsnm{Warde-Farley},~\bfnm{David}\binits{D.}},
  \bauthor{\bsnm{Ozair},~\bfnm{Sherjil}\binits{S.}},
  \bauthor{\bsnm{Courville},~\bfnm{Aaron}\binits{A.}} \AND
  \bauthor{\bsnm{Bengio},~\bfnm{Yoshua}\binits{Y.}}
(\byear{2014}).
\btitle{Generative adversarial nets}.
In \bbooktitle{Advances in neural information processing systems}
\bpages{2672--2680}.
\end{binproceedings}
\endbibitem

\bibitem{gordon1993novel}
\begin{binproceedings}[author]
\bauthor{\bsnm{Gordon},~\bfnm{Neil~J}\binits{N.~J.}},
  \bauthor{\bsnm{Salmond},~\bfnm{David~J}\binits{D.~J.}} \AND
  \bauthor{\bsnm{Smith},~\bfnm{Adrian~FM}\binits{A.~F.}}
(\byear{1993}).
\btitle{Novel approach to nonlinear/non-{G}aussian {B}ayesian state
  estimation}.
In \bbooktitle{IEE Proceedings F-radar and signal processing}
\bvolume{140}
\bpages{107--113}.
\bpublisher{IET}.
\end{binproceedings}
\endbibitem

\bibitem{hastie2005elements}
\begin{barticle}[author]
\bauthor{\bsnm{Hastie},~\bfnm{Trevor}\binits{T.}},
  \bauthor{\bsnm{Tibshirani},~\bfnm{Robert}\binits{R.}},
  \bauthor{\bsnm{Friedman},~\bfnm{Jerome}\binits{J.}} \AND
  \bauthor{\bsnm{Franklin},~\bfnm{James}\binits{J.}}
(\byear{2005}).
\btitle{The elements of statistical learning: data mining, inference and
  prediction}.
\bjournal{The Mathematical Intelligencer}
\bvolume{27}
\bpages{83--85}.
\end{barticle}
\endbibitem

\bibitem{heng2015gibbs}
\begin{barticle}[author]
\bauthor{\bsnm{Heng},~\bfnm{J.}\binits{J.}},
  \bauthor{\bsnm{Doucet},~\bfnm{A.}\binits{A.}} \AND
  \bauthor{\bsnm{Pokern},~\bfnm{Y.}\binits{Y.}}
(\byear{2015}).
\btitle{Gibbs flow for approximate transport with applications to {B}ayesian
  computation}.
\bjournal{arXiv:1509.08787}.
\end{barticle}
\endbibitem

\bibitem{houssineau2018multilevel}
\begin{barticle}[author]
\bauthor{\bsnm{Houssineau},~\bfnm{Jeremie}\binits{J.}},
  \bauthor{\bsnm{Jasra},~\bfnm{Ajay}\binits{A.}} \AND
  \bauthor{\bsnm{Singh},~\bfnm{Sumeetpal~S}\binits{S.~S.}}
(\byear{2018}).
\btitle{Multilevel {M}onte {C}arlo for smoothing via transport methods}.
\bjournal{SIAM Journal on Scientific Computing}
\bvolume{40}
\bpages{A2315--A2335}.
\end{barticle}
\endbibitem

\bibitem{houtekamer2001sequential}
\begin{barticle}[author]
\bauthor{\bsnm{Houtekamer},~\bfnm{Peter~L}\binits{P.~L.}} \AND
  \bauthor{\bsnm{Mitchell},~\bfnm{Herschel~L}\binits{H.~L.}}
(\byear{2001}).
\btitle{A sequential ensemble {K}alman filter for atmospheric data
  assimilation}.
\bjournal{Monthly Weather Review}
\bvolume{129}
\bpages{123--137}.
\end{barticle}
\endbibitem

\bibitem{jainipolynomialICML2019}
\begin{binproceedings}[author]
\bauthor{\bsnm{Jaini},~\bfnm{Priyank}\binits{P.}},
  \bauthor{\bsnm{Selby},~\bfnm{Kira~A.}\binits{K.~A.}} \AND
  \bauthor{\bsnm{Yu},~\bfnm{Yaoliang}\binits{Y.}}
(\byear{2019}).
\btitle{Sum-of-Squares Polynomial Flow}.
In \bbooktitle{International Conference on Machine Learning}.
\bnote{arXiv:1905.02325}.
\end{binproceedings}
\endbibitem

\bibitem{kaipio2007statistical}
\begin{barticle}[author]
\bauthor{\bsnm{Kaipio},~\bfnm{Jari}\binits{J.}} \AND
  \bauthor{\bsnm{Somersalo},~\bfnm{Erkki}\binits{E.}}
(\byear{2007}).
\btitle{Statistical inverse problems: discretization, model reduction and
  inverse crimes}.
\bjournal{Journal of computational and applied mathematics}
\bvolume{198}
\bpages{493--504}.
\end{barticle}
\endbibitem

\bibitem{kantorovich1965best}
\begin{bbook}[author]
\bauthor{\bsnm{Kantorovich},~\bfnm{L.~V.}\binits{L.~V.}}
(\byear{1965}).
\btitle{The best use of economic resources.}
\bpublisher{Oxford \& London: Pergamon Press.}
\end{bbook}
\endbibitem

\bibitem{katzfuss2020ensemble}
\begin{barticle}[author]
\bauthor{\bsnm{Katzfuss},~\bfnm{Matthias}\binits{M.}},
  \bauthor{\bsnm{Stroud},~\bfnm{Jonathan~R}\binits{J.~R.}} \AND
  \bauthor{\bsnm{Wikle},~\bfnm{Christopher~K}\binits{C.~K.}}
(\byear{2020}).
\btitle{Ensemble {K}alman methods for high-dimensional hierarchical dynamic
  space-time models}.
\bjournal{Journal of the American Statistical Association}
\bvolume{115}
\bpages{866--885}.
\end{barticle}
\endbibitem

\bibitem{Kim2013}
\begin{binproceedings}[author]
\bauthor{\bsnm{Kim},~\bfnm{Sanggyun}\binits{S.}},
  \bauthor{\bsnm{Ma},~\bfnm{Rui}\binits{R.}},
  \bauthor{\bsnm{Mesa},~\bfnm{Diego}\binits{D.}} \AND
  \bauthor{\bsnm{Coleman},~\bfnm{TP}\binits{T.}}
(\byear{2013}).
\btitle{{Efficient Bayesian Inference Methods via Convex Optimization and
  Optimal Transport}}.
In \bbooktitle{IEEE Symposium on Information Theory}
\bvolume{6}.
\end{binproceedings}
\endbibitem

\bibitem{koller2009probabilistic}
\begin{bbook}[author]
\bauthor{\bsnm{Koller},~\bfnm{D.}\binits{D.}} \AND
  \bauthor{\bsnm{Friedman},~\bfnm{N.}\binits{N.}}
(\byear{2009}).
\btitle{Probabilistic graphical models: principles and techniques}.
\bpublisher{MIT press}.
\end{bbook}
\endbibitem

\bibitem{lauritzen1996graphical}
\begin{bbook}[author]
\bauthor{\bsnm{Lauritzen},~\bfnm{S.~L.}\binits{S.~L.}}
(\byear{1996}).
\btitle{Graphical models}.
\bpublisher{Oxford University Press}.
\end{bbook}
\endbibitem

\bibitem{law2012evaluating}
\begin{barticle}[author]
\bauthor{\bsnm{Law},~\bfnm{Kody~JH}\binits{K.~J.}} \AND
  \bauthor{\bsnm{Stuart},~\bfnm{Andrew~M}\binits{A.~M.}}
(\byear{2012}).
\btitle{Evaluating data assimilation algorithms}.
\bjournal{Monthly Weather Review}
\bvolume{140}
\bpages{3757--3782}.
\end{barticle}
\endbibitem

\bibitem{le2021low}
\begin{binproceedings}[author]
\bauthor{\bsnm{Le~Provost},~\bfnm{Mathieu}\binits{M.}},
  \bauthor{\bsnm{Baptista},~\bfnm{Ricardo}\binits{R.}},
  \bauthor{\bsnm{Marzouk},~\bfnm{Youssef}\binits{Y.}} \AND
  \bauthor{\bsnm{Eldredge},~\bfnm{Jeff}\binits{J.}}
(\byear{2021}).
\btitle{A low-rank nonlinear ensemble filter for vortex models of aerodynamic
  flows}.
In \bbooktitle{AIAA Scitech 2021 Forum}
\bpages{1937}.
\end{binproceedings}
\endbibitem

\bibitem{lei2011moment}
\begin{barticle}[author]
\bauthor{\bsnm{Lei},~\bfnm{Jing}\binits{J.}} \AND
  \bauthor{\bsnm{Bickel},~\bfnm{Peter}\binits{P.}}
(\byear{2011}).
\btitle{A moment matching ensemble filter for nonlinear non-{G}aussian data
  assimilation}.
\bjournal{Monthly Weather Review}
\bvolume{139}
\bpages{3964--3973}.
\end{barticle}
\endbibitem

\bibitem{lindgren2011explicit}
\begin{barticle}[author]
\bauthor{\bsnm{Lindgren},~\bfnm{F.}\binits{F.}},
  \bauthor{\bsnm{Rue},~\bfnm{H.}\binits{H.}} \AND
  \bauthor{\bsnm{Lindstr{\"o}m},~\bfnm{J.}\binits{J.}}
(\byear{2011}).
\btitle{An explicit link between {G}aussian fields and {G}aussian {M}arkov
  random fields: the stochastic partial differential equation approach}.
\bjournal{Journal of the Royal Statistical Society: Series B}
\bvolume{73}
\bpages{423--498}.
\end{barticle}
\endbibitem

\bibitem{liu2016stein}
\begin{binproceedings}[author]
\bauthor{\bsnm{Liu},~\bfnm{Q.}\binits{Q.}} \AND
  \bauthor{\bsnm{Wang},~\bfnm{D.}\binits{D.}}
(\byear{2016}).
\btitle{Stein variational gradient descent: a general purpose {B}ayesian
  inference algorithm}.
In \bbooktitle{Advances in Neural Information Processing Systems}
\bpages{2370--2378}.
\end{binproceedings}
\endbibitem

\bibitem{lorenz1963deterministic}
\begin{barticle}[author]
\bauthor{\bsnm{Lorenz},~\bfnm{Edward~N}\binits{E.~N.}}
(\byear{1963}).
\btitle{Deterministic nonperiodic flow}.
\bjournal{Journal of the atmospheric sciences}
\bvolume{20}
\bpages{130--141}.
\end{barticle}
\endbibitem

\bibitem{lorenz1996predictability}
\begin{binproceedings}[author]
\bauthor{\bsnm{Lorenz},~\bfnm{Edward~N}\binits{E.~N.}}
(\byear{1996}).
\btitle{Predictability: {A} problem partly solved}.
In \bbooktitle{Proc. Seminar on predictability}
\bvolume{1}.
\end{binproceedings}
\endbibitem

\bibitem{majda2012filtering}
\begin{bbook}[author]
\bauthor{\bsnm{Majda},~\bfnm{Andrew~J}\binits{A.~J.}} \AND
  \bauthor{\bsnm{Harlim},~\bfnm{John}\binits{J.}}
(\byear{2012}).
\btitle{Filtering complex turbulent systems}.
\bpublisher{Cambridge University Press}.
\end{bbook}
\endbibitem

\bibitem{mandel2011convergence}
\begin{barticle}[author]
\bauthor{\bsnm{Mandel},~\bfnm{Jan}\binits{J.}},
  \bauthor{\bsnm{Cobb},~\bfnm{Loren}\binits{L.}} \AND
  \bauthor{\bsnm{Beezley},~\bfnm{Jonathan~D}\binits{J.~D.}}
(\byear{2011}).
\btitle{On the convergence of the ensemble {K}alman filter}.
\bjournal{Applications of Mathematics}
\bvolume{56}
\bpages{533--541}.
\end{barticle}
\endbibitem

\bibitem{Marin2011}
\begin{barticle}[author]
\bauthor{\bsnm{Marin},~\bfnm{J.~M.}\binits{J.~M.}},
  \bauthor{\bsnm{Pudlo},~\bfnm{P.}\binits{P.}},
  \bauthor{\bsnm{Robert},~\bfnm{C.~P.}\binits{C.~P.}} \AND
  \bauthor{\bsnm{Ryder},~\bfnm{R.~J.}\binits{R.~J.}}
(\byear{2012}).
\btitle{{Approximate Bayesian computational methods}}.
\bjournal{Statistics and Computing}
\bvolume{22}
\bpages{1167--1180}.
\end{barticle}
\endbibitem

\bibitem{marzouk2016introduction}
\begin{bincollection}[author]
\bauthor{\bsnm{Marzouk},~\bfnm{Y.}\binits{Y.}},
  \bauthor{\bsnm{Moselhy},~\bfnm{T.}\binits{T.}},
  \bauthor{\bsnm{Parno},~\bfnm{M.}\binits{M.}} \AND
  \bauthor{\bsnm{Spantini},~\bfnm{A.}\binits{A.}}
(\byear{2016}).
\btitle{An introduction to sampling via measure transport}.
In \bbooktitle{Handbook of Uncertainty Quantification}.
\end{bincollection}
\endbibitem

\bibitem{metref2014non}
\begin{barticle}[author]
\bauthor{\bsnm{Metref},~\bfnm{S.}\binits{S.}},
  \bauthor{\bsnm{Cosme},~\bfnm{E.}\binits{E.}},
  \bauthor{\bsnm{Snyder},~\bfnm{C.}\binits{C.}} \AND
  \bauthor{\bsnm{Brasseur},~\bfnm{P.}\binits{P.}}
(\byear{2014}).
\btitle{A non-{G}aussian analysis scheme using rank histograms for ensemble
  data assimilation}.
\bjournal{Nonlinear Processes in Geophysics}
\bvolume{21}
\bpages{869--885}.
\end{barticle}
\endbibitem

\bibitem{monge1781memoire}
\begin{bbook}[author]
\bauthor{\bsnm{Monge},~\bfnm{Gaspard}\binits{G.}}
(\byear{1781}).
\btitle{M{\'e}moire sur la th{\'e}orie des d{\'e}blais et des remblais}.
\bpublisher{De l'Imprimerie Royale}.
\end{bbook}
\endbibitem

\bibitem{morzfeld2012random}
\begin{barticle}[author]
\bauthor{\bsnm{Morzfeld},~\bfnm{Matthias}\binits{M.}},
  \bauthor{\bsnm{Tu},~\bfnm{Xuemin}\binits{X.}},
  \bauthor{\bsnm{Atkins},~\bfnm{Ethan}\binits{E.}} \AND
  \bauthor{\bsnm{Chorin},~\bfnm{Alexandre~J}\binits{A.~J.}}
(\byear{2012}).
\btitle{A random map implementation of implicit filters}.
\bjournal{Journal of Computational Physics}
\bvolume{231}
\bpages{2049--2066}.
\end{barticle}
\endbibitem

\bibitem{el2012bayesian}
\begin{barticle}[author]
\bauthor{\bsnm{Moselhy},~\bfnm{T.}\binits{T.}} \AND
  \bauthor{\bsnm{Marzouk},~\bfnm{Y.}\binits{Y.}}
(\byear{2012}).
\btitle{Bayesian inference with optimal maps}.
\bjournal{Journal of Computational Physics}
\bvolume{231}
\bpages{7815--7850}.
\end{barticle}
\endbibitem

\bibitem{nino2018ensemble}
\begin{barticle}[author]
\bauthor{\bsnm{Nino-Ruiz},~\bfnm{Elias~D}\binits{E.~D.}},
  \bauthor{\bsnm{Sandu},~\bfnm{Adrian}\binits{A.}} \AND
  \bauthor{\bsnm{Deng},~\bfnm{Xinwei}\binits{X.}}
(\byear{2018}).
\btitle{An Ensemble {K}alman Filter Implementation Based on Modified Cholesky
  Decomposition for Inverse Covariance Matrix Estimation}.
\bjournal{SIAM Journal on Scientific Computing}
\bvolume{40}
\bpages{A867--A886}.
\end{barticle}
\endbibitem

\bibitem{parno2014transport}
\begin{barticle}[author]
\bauthor{\bsnm{Parno},~\bfnm{M.}\binits{M.}} \AND
  \bauthor{\bsnm{Marzouk},~\bfnm{Y.}\binits{Y.}}
(\byear{2018}).
\btitle{Transport map accelerated {M}arkov chain {M}onte {C}arlo}.
\bjournal{SIAM/ASA Journal on Uncertainty Quantification}
\bvolume{2}
\bpages{645--682}.
\end{barticle}
\endbibitem

\bibitem{poterjoy2016localized}
\begin{barticle}[author]
\bauthor{\bsnm{Poterjoy},~\bfnm{Jonathan}\binits{J.}}
(\byear{2016}).
\btitle{A localized particle filter for high-dimensional nonlinear systems}.
\bjournal{Monthly Weather Review}
\bvolume{144}
\bpages{59--76}.
\end{barticle}
\endbibitem

\bibitem{press2007numerical}
\begin{bbook}[author]
\bauthor{\bsnm{Press},~\bfnm{W.~H.}\binits{W.~H.}},
  \bauthor{\bsnm{Teukolsky},~\bfnm{S.~A.}\binits{S.~A.}},
  \bauthor{\bsnm{Vetterling},~\bfnm{W.~T.}\binits{W.~T.}} \AND
  \bauthor{\bsnm{Flannery},~\bfnm{B.~P.}\binits{B.~P.}}
(\byear{2007}).
\btitle{Numerical recipes: The art of scientific computing. Third edition.}
\bpublisher{Cambridge University Press}.
\end{bbook}
\endbibitem

\bibitem{pulido2018kernel}
\begin{barticle}[author]
\bauthor{\bsnm{Pulido},~\bfnm{Manuel}\binits{M.}} \AND
  \bauthor{\bsnm{vanLeeuwen},~\bfnm{Peter~Jan}\binits{P.~J.}}
(\byear{2018}).
\btitle{Kernel embedding of maps for sequential {B}ayesian inference: The
  variational mapping particle filter}.
\bjournal{arXiv preprint arXiv:1805.11380}.
\end{barticle}
\endbibitem

\bibitem{ramsay1998estimating}
\begin{barticle}[author]
\bauthor{\bsnm{Ramsay},~\bfnm{J.~O.}\binits{J.~O.}}
(\byear{1998}).
\btitle{Estimating smooth monotone functions}.
\bjournal{Journal of the Royal Statistical Society: Series B}
\bpages{365--375}.
\end{barticle}
\endbibitem

\bibitem{reich2013nonparametric}
\begin{barticle}[author]
\bauthor{\bsnm{Reich},~\bfnm{S.}\binits{S.}}
(\byear{2013}).
\btitle{A nonparametric ensemble transform method for {B}ayesian inference}.
\bjournal{SIAM Journal on Scientific Computing}
\bvolume{35}
\bpages{A2013--A2024}.
\end{barticle}
\endbibitem

\bibitem{reich2015probabilistic}
\begin{bbook}[author]
\bauthor{\bsnm{Reich},~\bfnm{S.}\binits{S.}} \AND
  \bauthor{\bsnm{Cotter},~\bfnm{C.}\binits{C.}}
(\byear{2015}).
\btitle{Probabilistic Forecasting and Bayesian Data Assimilation}.
\bpublisher{Cambridge University Press}.
\end{bbook}
\endbibitem

\bibitem{rezende2015variational}
\begin{barticle}[author]
\bauthor{\bsnm{Rezende},~\bfnm{D.~J.}\binits{D.~J.}} \AND
  \bauthor{\bsnm{Mohamed},~\bfnm{S.}\binits{S.}}
(\byear{2015}).
\btitle{Variational inference with normalizing flows}.
\bjournal{arXiv:1505.05770}.
\end{barticle}
\endbibitem

\bibitem{rosenblatt1952remarks}
\begin{barticle}[author]
\bauthor{\bsnm{Rosenblatt},~\bfnm{M.}\binits{M.}}
(\byear{1952}).
\btitle{Remarks on a multivariate transformation}.
\bjournal{The Annals of Mathematical Statistics}
\bpages{470--472}.
\end{barticle}
\endbibitem

\bibitem{saetrom2011ensembleB}
\begin{barticle}[author]
\bauthor{\bsnm{S{\ae}trom},~\bfnm{Jon}\binits{J.}} \AND
  \bauthor{\bsnm{Omre},~\bfnm{Henning}\binits{H.}}
(\byear{2011}).
\btitle{Ensemble {K}alman filtering with shrinkage regression techniques}.
\bjournal{Computational Geosciences}
\bvolume{15}
\bpages{271--292}.
\end{barticle}
\endbibitem

\bibitem{santambrogio2015optimal}
\begin{bbook}[author]
\bauthor{\bsnm{Santambrogio},~\bfnm{F.}\binits{F.}}
(\byear{2015}).
\btitle{Optimal Transport for Applied Mathematicians}
\bvolume{87}.
\bpublisher{Springer}.
\end{bbook}
\endbibitem

\bibitem{snyder2008obstacles}
\begin{barticle}[author]
\bauthor{\bsnm{Snyder},~\bfnm{Chris}\binits{C.}},
  \bauthor{\bsnm{Bengtsson},~\bfnm{Thomas}\binits{T.}},
  \bauthor{\bsnm{Bickel},~\bfnm{Peter}\binits{P.}} \AND
  \bauthor{\bsnm{Anderson},~\bfnm{Jeff}\binits{J.}}
(\byear{2008}).
\btitle{Obstacles to high-dimensional particle filtering}.
\bjournal{Monthly Weather Review}
\bvolume{136}
\bpages{4629--4640}.
\end{barticle}
\endbibitem

\bibitem{spantini2017low}
\begin{bphdthesis}[author]
\bauthor{\bsnm{Spantini},~\bfnm{Alessio}\binits{A.}}
(\byear{2017}).
\btitle{On the low-dimensional structure of Bayesian inference}
\btype{PhD thesis},
\bpublisher{Massachusetts Institute of Technology}.
\end{bphdthesis}
\endbibitem

\bibitem{spantini2017inference}
\begin{barticle}[author]
\bauthor{\bsnm{Spantini},~\bfnm{Alessio}\binits{A.}},
  \bauthor{\bsnm{Bigoni},~\bfnm{Daniele}\binits{D.}} \AND
  \bauthor{\bsnm{Marzouk},~\bfnm{Youssef}\binits{Y.}}
(\byear{2018}).
\btitle{Inference via low-dimensional couplings}.
\bjournal{The Journal of Machine Learning Research}
\bvolume{19}
\bpages{2639--2709}.
\end{barticle}
\endbibitem

\bibitem{spantini2014optimal}
\begin{barticle}[author]
\bauthor{\bsnm{Spantini},~\bfnm{A.}\binits{A.}},
  \bauthor{\bsnm{Solonen},~\bfnm{A.}\binits{A.}},
  \bauthor{\bsnm{Cui},~\bfnm{T.}\binits{T.}},
  \bauthor{\bsnm{Martin},~\bfnm{J.}\binits{J.}},
  \bauthor{\bsnm{Tenorio},~\bfnm{L.}\binits{L.}} \AND
  \bauthor{\bsnm{Marzouk},~\bfnm{Y.}\binits{Y.}}
(\byear{2015}).
\btitle{Optimal low-rank approximations of {B}ayesian linear inverse problems}.
\bjournal{SIAM Journal on Scientific Computing}
\bvolume{37}
\bpages{A2451--A2487}.
\end{barticle}
\endbibitem

\bibitem{tippett2003ensemble}
\begin{barticle}[author]
\bauthor{\bsnm{Tippett},~\bfnm{Michael~K}\binits{M.~K.}},
  \bauthor{\bsnm{Anderson},~\bfnm{Jeffrey~L}\binits{J.~L.}},
  \bauthor{\bsnm{Bishop},~\bfnm{Craig~H}\binits{C.~H.}},
  \bauthor{\bsnm{Hamill},~\bfnm{Thomas~M}\binits{T.~M.}} \AND
  \bauthor{\bsnm{Whitaker},~\bfnm{Jeffrey~S}\binits{J.~S.}}
(\byear{2003}).
\btitle{Ensemble square root filters}.
\bjournal{Monthly Weather Review}
\bvolume{131}
\bpages{1485--1490}.
\end{barticle}
\endbibitem

\bibitem{ueno2009covariance}
\begin{barticle}[author]
\bauthor{\bsnm{Ueno},~\bfnm{G.}\binits{G.}} \AND
  \bauthor{\bsnm{Tsuchiya},~\bfnm{T.}\binits{T.}}
(\byear{2009}).
\btitle{Covariance regularization in inverse space}.
\bjournal{Quarterly Journal of the Royal Meteorological Society: A journal of
  the atmospheric sciences, applied meteorology and physical oceanography}
\bvolume{135}
\bpages{1133--1156}.
\end{barticle}
\endbibitem

\bibitem{van2018particle}
\begin{barticle}[author]
\bauthor{\bparticle{van} \bsnm{Leeuwen},~\bfnm{Peter~Jan}\binits{P.~J.}},
  \bauthor{\bsnm{K{\"u}nsch},~\bfnm{Hans~R}\binits{H.~R.}},
  \bauthor{\bsnm{Nerger},~\bfnm{Lars}\binits{L.}},
  \bauthor{\bsnm{Potthast},~\bfnm{Roland}\binits{R.}} \AND
  \bauthor{\bsnm{Reich},~\bfnm{Sebastian}\binits{S.}}
(\byear{2018}).
\btitle{Particle filters for applications in geosciences}.
\bjournal{arXiv preprint arXiv:1807.10434}.
\end{barticle}
\endbibitem

\bibitem{villani2008optimal}
\begin{bbook}[author]
\bauthor{\bsnm{Villani},~\bfnm{C.}\binits{C.}}
(\byear{2008}).
\btitle{Optimal transport: old and new}
\bvolume{338}.
\bpublisher{Springer Science \& Business Media}.
\end{bbook}
\endbibitem

\bibitem{wackernagel1996multivariate}
\begin{binproceedings}[author]
\bauthor{\bsnm{Wackernagel},~\bfnm{Hans}\binits{H.}}
(\byear{1996}).
\btitle{Multivariate geostatistics: an introduction with applications}.
In \bbooktitle{International Journal of Rock Mechanics and Mining Sciences and
  Geomechanics Abstracts}
\bvolume{33}
\bpages{363A--363A}.
\bpublisher{Springer}.
\end{binproceedings}
\endbibitem

\bibitem{wright1999numerical}
\begin{bbook}[author]
\bauthor{\bsnm{Wright},~\bfnm{S.~J.}\binits{S.~J.}} \AND
  \bauthor{\bsnm{Nocedal},~\bfnm{J.}\binits{J.}}
(\byear{1999}).
\btitle{Numerical Optimization}
\bvolume{2}.
\bpublisher{Springer New York}.
\end{bbook}
\endbibitem

\bibitem{wu2003nonparametric}
\begin{barticle}[author]
\bauthor{\bsnm{Wu},~\bfnm{Wei~Biao}\binits{W.~B.}} \AND
  \bauthor{\bsnm{Pourahmadi},~\bfnm{Mohsen}\binits{M.}}
(\byear{2003}).
\btitle{Nonparametric estimation of large covariance matrices of longitudinal
  data}.
\bjournal{Biometrika}
\bvolume{90}
\bpages{831--844}.
\end{barticle}
\endbibitem

\bibitem{yang2013feedback}
\begin{barticle}[author]
\bauthor{\bsnm{Yang},~\bfnm{Tao}\binits{T.}},
  \bauthor{\bsnm{Mehta},~\bfnm{Prashant~G}\binits{P.~G.}} \AND
  \bauthor{\bsnm{Meyn},~\bfnm{Sean~P}\binits{S.~P.}}
(\byear{2013}).
\btitle{Feedback particle filter}.
\bjournal{IEEE Transactions on Automatic Control}
\bvolume{58}
\bpages{2465--2480}.
\end{barticle}
\endbibitem

\end{thebibliography}

\end{document}